\documentclass[aps,twocolumn,pra,
 amsmath,amssymb,
 aps,
]{revtex4-1}

\usepackage{graphicx}
\usepackage{dcolumn}
\usepackage{bm}
\usepackage{graphicx,lastpage}
\usepackage{graphics}
\usepackage{amsmath}
\usepackage{amssymb}
\usepackage{upgreek}
\usepackage{color}
\usepackage{float}
\usepackage{soul}
\usepackage{ragged2e}
\usepackage{subfig,caption,subcaption}
\usepackage{censor}
\usepackage{enumerate}
\usepackage{textcomp}
\usepackage{braket}
\usepackage{placeins}
\usepackage{ulem}

\begin{document}


\title{Cotunneling assisted nonequilibrium thermodynamics of a  photosynthetic junction }

\author{Debasish Sharma$^a$}
\author{Manash Jyoti Sarmah$^a$}
\author{Mriganka Sandilya$^{b,c}$}
\author{Himangshu Prabal Goswami$^a$}
\email{hpg@gauhati.ac.in}

\affiliation{$^a$Department of Chemistry,$^b$Department of Physics, Gauhati University,
Gopinath Bordoloi Nagar, Jalukbari, Guwahati-781014, Assam, India\\
$^c$Light and Matter Physics Group, Raman Research Institute, Bengaluru- 560080, India}%

\date{\today}

\begin{abstract}
We theoretically investigate a photosystem II-based reaction center modeled as a nonequilibrium quantum junction. We specifically focus on the electron-electron interactions that enable cotunneling events to be captured through quantum mechanical rates due to the inclusion of a negatively charged manybody state. Using a master equation framework with realistic spectral profiles, we analyze the cotunneling assisted current, power, and work. Amplification of the cotunneling assisted current and power occurs over a narrower bias range, reflecting a trade-off where higher flux is compensated by a reduced work window. We further report that the cotunneling-enhanced thermodynamic variables, particularly within specific bias windows, depends on the interplay between cotunneling amplitudes, electron transition rates, and interaction energy. Both attractive and repulsive electronic interactions can enhance cotunneling, but this effect is sensitive to the energy balance between states and the tunneling strength asymmetries.

\end{abstract}

\maketitle

\section{\label{sec:level1}Introduction}

\begin{figure*}[ht!]
    \centering
    \includegraphics[height=6.5cm, width=18cm]{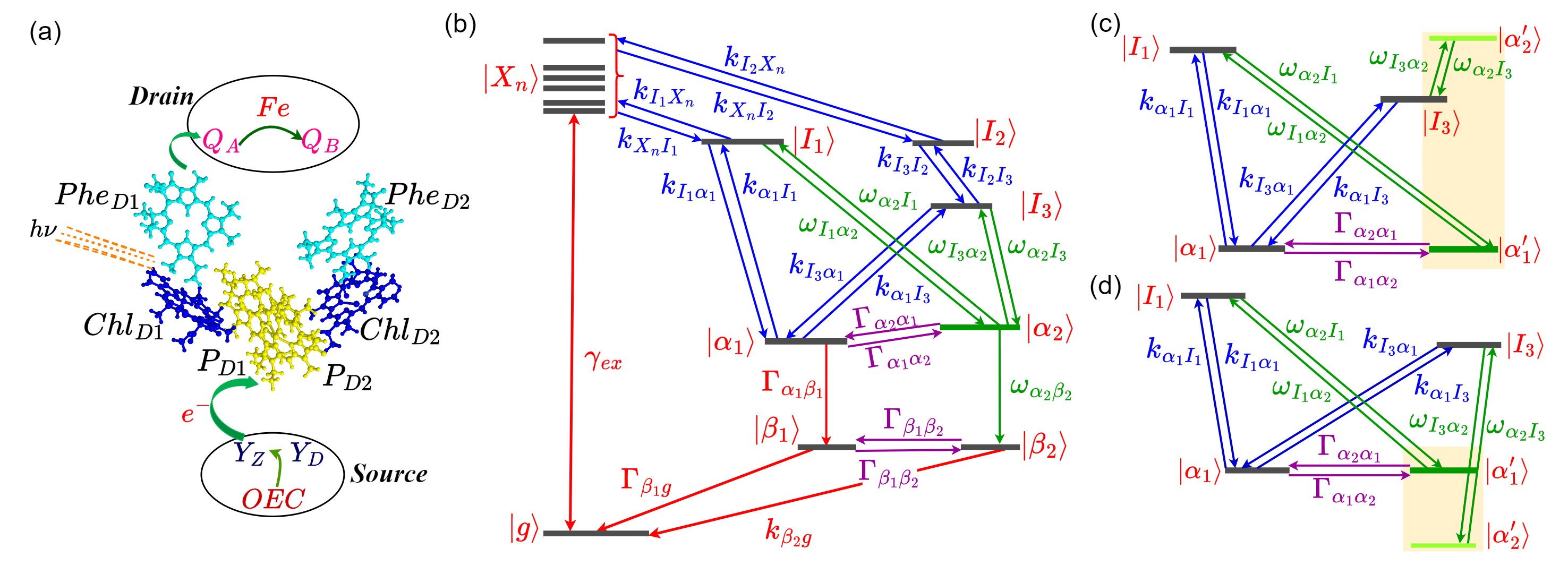}
    \caption{\justifying Model of the PSIIRC as a quantum junction. (a) A schematic representation of the considered model of the PSIIRC, (b) Energy level diagram of the electronic state space model. The single unidirectional arrows represent rates connecting the ground ($\ket{g}$) and empty states ($\ket{\beta_{1}}$, $\ket{\beta_{2}}$) to the excited state manifold. $\Gamma_{\beta_{1} g}$, $k_{\beta_{2} g}$ and $\Gamma_{\alpha_{1}\beta_{1}}$  are the unidirectional transfer rates while $\omega_{\alpha_{2}\beta_{2}}$ is the rate of cotunneling. 
    $\gamma_{ex}$ is the rate of photoexcitation. Coupled blue arrows represent F$\ddot{o}$rster-like rates for electron transfer from the excitons ($\ket {X_n}, n=1...6$) to the charge-separated states ($\ket {I_i},i=1,2,3$ and $\ket{\alpha_{1}}$). The coupled arrows represent the simultaneous two electron transport from two distinct charge-separated states to the negatively charged state ($\ket{\alpha_{2}}$). (c) and (d): Schematic model of the negatively charged state. The negatively charged state comprises two interacting states $\ket{\alpha'_{1}}$ and $\ket{\alpha'_{2}}$, represented by the highlighted region. Simultaneous electron transfer from the charge-separated states $\ket{I_{1}}$ and $\ket{I_{3}}$ populates the two interacting states.}
    \label{fig:1}  
\end{figure*}

Electron transfer within biosystems leverages  tunneling leading to quantum mechanical rates between participating states, enhancing energy conversion efficiency through  exploration over multiple pathways \cite{Gitt,Moser1992,Engel2007}. Control over the flow of electrons through such transfer pathways involve directed configurational changes in the protein environment surrounding the active biomolecular systems or pigments \cite{MARCUS1985265,RENGER20121164,Siegbahn2008}. An interesting example is the Photosystem II, a natural energy converter that produces dioxygen $O_{2}$ through the splitting of water\cite{ch1}. Experimental evidence characterizes the presence of \textit{chlorophyll a} and other accessory pigments such as \textit{pheophytin}, \textit{$\beta$-carotene}, and \textit{plastoquinone} which take an active part in light absorption and subsequent energy transfer\cite{Guskov2009, Tomo2021} through different quantum transport pathways \cite{pnas.1105234108}. Especially in the Photosystem II Reaction Center (PSIIRC), two separate electron transfer pathways exist which can be simultaneously accessed within the same timescale by suitable tuning of the surrounding protein environment\cite{Vladimir}. The protein environment also controls the generation of charge-separated as well as charged states along with the overall dynamics of PSIIRC. The environment establishes transverse and lateral excitonic asymmetry among the PSII pigments,  allowing  efficient electron transfer and charge separation. This asymmetry is achieved through the precise spatial arrangement of the protein subunits and cofactors within the reaction center. Further, the protein environment helps to delocalize triplet states away from $Chl_{D1}$ onto other pigments, preventing selective damage to the D1 branch of the protein \cite{Hayase2023, Kamlowski,D3SC02985A}, the active branch in charge separation\cite{annurev}. Such a photoprotection is essential for maintaining the functionality of the reaction center under extreme conditions\cite{sirohiwal2020protein,D3SC02985A}, highlighting the role of environment-assisted quantum dynamics \cite{Mohseni} in these pigments.

Following a plethora of similar observations, a theoretical understanding of the dynamics of such bioquantum systems is construed using principles from open quantum dynamics. Protein environments are treated as classical, structured, or unstructured quantum mechanical reservoirs \cite{skourtis2010fluctuations,dorfman2013photosynthetic,rouse2024light,PRXEnergy.2.013002,wang2020dissipative,dodin2022noise,poteshman2023network}. From a general perspective, the dynamics of a target quantum system interacting with multiple environments are typically analyzed using approaches such as Markovian or non-Markovian master equations, hierarchical equations of motion, or Green's functions, with the choice of framework determined by the nature of the underlying interactions. \cite{fang2019nonequilibrium,joubert2023quantum,singh2011electronic,yang2020steady,chen2015using,PhysRevResearch.5.013181,zhang2023many,harbola2008superoperator,levi2015quantum,karafyllidis2017quantum,suess2014hierarchy,timm2008tunneling,welack2008single,papp2024computation}.  Be it strong or weak, accurate modeling of system-environment correlations allows the exploration of exotic dynamics assisted by coherences, phonons, polarons, and cotunneling whose signatures can be seen in the current-voltage characteristics \cite{Hong-Guang,Gitt,Zhou,jha2024unraveling}. Inspired by such theories, it has been demonstrated that binding a photosystem I complex between a gold electrode and a gold-coated glass tip enables photocurrent measurement from a single photosynthetic reaction center \cite{Gerster2012}. This breakthrough in single-molecule techniques provided insight into the mechanisms influencing photocurrent in reaction centers \cite{Pillai}. 

The protein environment in PSIIRC is also known to generate triplet states primarily on the D1 branch which leads to low-energy charge-transfer excitations between the D1 and D2 branches thus affecting the current-voltage characteristics\cite{D3SC02985A}. Given that a triplet state can accommodate two electrons, an important question arises regarding how the simultaneous transport of two electrons, i.e. \textit{cotunneling}, influences the current-voltage characteristics. Whilst the role of cotunneling in quantum transport has previously explored \cite{cabrera2023charge,weymann2011dark,bian2022charge,donarini2024transport,hsiao2024exciton}, its role in biosystems is lacking. Asymmetrically biased tunneling amplitudes between the system and reservoirs is known to enhance current output at specific values of electronic interaction energy \cite{carmi2012enhanced,sandilya2024cotunneling}. The PSIIRC surrounded by its protein environment can offer a platform to realize asymmetrically tuned tunneling amplitudes that may be brought about by appropriate configurational changes in the protein environment. Further, experiments suggest both charge-separated and positively charged states contribute during the electron transfer processes in PSIIRC, and hence the possibility of other states that can promote cotunneling cannot be ignored. Motivated by this hypothesis, in this work, we investigate how the presence of electron cotunneling influences the overall thermodynamic variables within the PSIIRC. By leveraging on the parameters associated with the presence of a negatively charged state, we aim to delve into the role of cotunneling in the current and power characteristics of the PSIIRC.

\section{\label{sec:level2}PSIIRC Junction}

The PSIIRC model considered in this work is based on the crystallographic data detailing the arrangement of chromophores involved in charge transfer \cite{ferreira2004architecture, Umena2011}. Structural analysis reveals that the reaction center is composed of four chlorophyll molecules,  a special pair (\textit{P$_{D1}$} and \textit{P$_{D2}$}) and an accessory pair (\textit{Chl$_{D1}$} and \textit{Chl$_{D2}$}); two pheophytins (\textit{Phe$_{D1}$} and \textit{Phe$_{D2}$}) and two quinones (\textit{Q}$_A$ and \textit{Q}$_B$) symmetrically arranged along the D1 and D2 protein branches\cite{Zouni2001}.  Additionally two \textit{Chl} molecules, $Chlz_{D1}$ and $Chlz_{D2}$ are placed oppositely on the periphery of the reaction center. An isolated PSIIRC (Fig.~\ref{fig:1}(a)) positioned between the Oxygen Evolution Complex (OEC) and the $Q_A\to Q_BH_2$ reduction complex (QRC) can be envisioned as a molecular junction with the OEC being a reservoir of electrons (source) and the reduction complex being a sink of electrons (drain). The reservoirs are akin to two electrodes or noninteracting electronic reservoirs \cite{Tao2006}. This configuration positions the PSIIRC molecular junction to be in a non-equilibrium state, enabling an electronic current through the PSIIRC.

 As per existing spectroscopic evidence \cite{Brixner2005, Duan2017}, the quantum states involved in electron transfer between source and drain include the ground state $\ket{g}$, the six excitonic states $\ket{X_{n}}, n =1\ldots6 $ formed by the mixing of the six core chromophores, the four chlorophyll and the two pheophytin molecules viz $P_{D1}, P_{D2}, Chl_{D1}$, $Chl_{D2}$, $Phe_{D1}$ and $Phe_{D2}$. From these excitonic states, electron transfer occurs to two different charge-separated states viz. $\ket{Chl_{D1}^{+}Phe_{D1}^{-}}$ (shorthand notation: $\ket{I_{1}}$), $\ket{P_{D2}^{+}P_{D1}^{-}}$ $\equiv$ $\ket{I_{2}}$. Further electron transfer includes two other charge-separated states, $\ket{P_{D1}^{+}Chl_{D1}^{-}}\equiv \ket{I_{3}}$ and $\ket{P_{D1}^{+}Phe_{D1}^{-}}\equiv\ket{\alpha_1}$, and the positively charged state $\ket{P_{D1}^{+}Phe_{D1}}$ $\equiv$ $\ket{\beta_{1}}$. From a nonequilibrium manybody perspective, these electronic states (except the $\ket{\beta_{1}}$ state) are analogous to $N$-electron manybody states. The positively charged state, $\ket{\beta_{1}}$, corresponds to a $N-1$-electron many body state. The electron flow channel represents a closed circuit when  the $N-1$-electronic state, $\ket{\beta_{1}}$ gains an electron from the source and the PSIIRC is returned to the ground state.

Evidences \cite{novoderezhkin2007mixing, Vladimir} suggest the possibility of simultaneous flow of electrons via two quantum channels: (a) relaxation of the photochemically excited states to the $P_{D1}^+$ $P_{D2}^-$ state with further electron transfer to $Chl_{D1}$ and $Phe_{D1}$ and (b) through the formation of $Chl_{D1}^{+}$ $Phe_{D1}^{-}$ and $P_{D1}^{+}$ $Phe_{D1}^{-}$ from the excited state of $Chl_{D1}$. In the current terminology, the two pathways are, $\ket{X_n}\rightarrow \ket {I_1}\rightarrow \ket{\alpha_1}$ and $\ket{X_n}\rightarrow \ket {I_2}\rightarrow \ket {I_3}\rightarrow \ket{\alpha_1}$. The energy-level diagram illustrating the electron transfer pathways is schematically represented by Fig.~\ref{fig:1}(b). In the scenario, when the two pathways combine to give a single quantum channel of electron transport, the possibility of the involvement of negatively charged states or $N+1,N+2$ manybody states during electron transport cannot be ignored \cite{welack2008single,goswami2015electroluminescence}. In such a case, the negative states could be accessible in the pathway under a proper environment that facilitates enough coupling. This is especially true since it has been recently predicted that the electrostatic field of the protein matrix (the environment) engulfing the PSIIRC molecular junction causes the red-shifting of chlorophylls and blue-shifting of pheophytins, contributing to the creation of additional charge transfer excited states along specific pathways within the PSIIRC\cite{sirohiwal2020protein}. Further, new pathways involving singlet-triplet conversion have also been unraveled where the protein matrix is found to play the deciding role \cite{D3SC02985A}. One can identify the existence of such states by properly assessing the entire configurational space of the protein with quantum molecular dynamic simulations \cite{Brunk}. In the current work, we do not address this latter aspect. To account for such a case in a simplistic manner, we introduce an additional possible negatively charged state into the PSRCII Hamiltonian. Although there might be many such states, we only assume the existence of a negative state $\ket{\alpha_{2}}$, representing a $N+2$ manybody state, which is considered to be generated due to the electrostatic field of the protein matrix. As an initial case study, we further assume that the state $\ket{\alpha_{2}}$ can be populated via the $N$ electron states $\ket{I_{1}}$ and $\ket{I_{3}}$ as shown in Fig.~\ref{fig:1}(b). The existence of the state $\ket{\alpha_{2}}$ is explicable if and only if, two electrons simultaneously get transported from the two $N$-body states within the same timescale, a phenomenon termed as cotunneling in quantum transport \cite{Aghassi,carmi2012enhanced}. To justify the role of simultaneous two-electron transfer in the system's dynamics, we can understand this negatively charged state to be comprising of two different interacting $N+1$ states, $\ket{\alpha'_{1}}$ and $\ket{\alpha'_{2}}$ with an interaction energy $U$ as shown in Fig.~\ref{fig:1}(c).  When $U\ne 0$, there is an interaction (attractive or repulsive) between the two electrons, these occupied states collectively form the negatively charged state. We also introduce another $N$-electron state $\ket{\beta_{2}}$ to which the system jumps from the $\ket{\alpha_{2}}$ after losing two electrons to the drain via cotunneling. The $N$-electronic state , $\ket{\beta_{2}}$ nonriadiatively relaxes back to the ground state. Overall, the PSIIRC model that we study is assumed to have twelve $N$-electron states, one $N+2$ electron state, and a single $N-1$ electron state which is schematically shown in Fig.~\ref{fig:1}(b).  

\section{\label{sec:level3}Cotunneling Assisted Dynamics}

To quantify the quantum dynamics of the PSIIRC, illustrated in Fig.~\ref{fig:1}(b),  we use the master equation framework derivable by considering an open quantum system Hamiltonian of the type, $\hat H =\hat H_q+\hat H_E+\hat H_B$, representing the molecule, environment and molecule-environment interaction parts. These Hamiltonians are discussed the Appendix A. The state vector (or the vectorized density matrix), $\ket{\hat\rho}$, describing the PSIIRC is composed of the manybody state populations (Appendix A). We define it in the following basis, ${\ket{g}, \ket{X_{1}},\ldots, \ket{X_{6}},\ket{I_{1}}, \ket{I_{2}}, \ket{I_{3}}, \ket{\alpha_{1}}, \ket{\alpha_{2}}, \ket{\beta_{1}}, \ket{\beta_{2}}}$. The general form of the time evolution of the density vector follows $\ket{\dot{\hat\rho}_q} = \breve {\cal L}\ket{\hat\rho_q}$ and it maintains detailed balance condition. $\breve {\cal L}$ is the superoperator that contains the effective electron and energy transfer rates between the manybody states. We refer to the appendices for a detailed derivation of the master equation.  $\breve {\cal L}$ can be evaluated analytically and is given by Eq.~\ref{A9}, in Appendix~\ref{SD}. The  exchange rates between the manybody states that contribute to the dynamics are schematically shown in Fig.~\ref{fig:1}(b) and the expressions for each rate are provided in the appendix. Physically, there are several different types of rates based on the involved states in energy or electronic exchange.  The rate of photoexcitation (excitation due to concentrated but incoherent solar radiation) from the ground state to the excitonic states is denoted as $\gamma_{ex}$ while the electron ejection rate from the state $\ket{\beta_{1}}$  is denoted as $\Gamma_{\beta_{1} g}$. The nonradiave relaxation from state $\ket{\beta_2}$ is $k_{\beta_{2} g}$. These rates are the reservoir-induced processes in standard nonequilibrium quantum transport \cite{Konstantin, Creatore} derived under the weak coupling limit. Secondly, there are intra-exciton population transfer rates amongst the six excitons $\ket{X_n}$, denoted by $r_{pq},p\ne q\in \{1\ldots 6\}$ ( not shown in Fig.1b) which are calculated within the framework of exciton relaxation dynamics using modified Redfield Theory \cite{YANG2002355}. These sets of rates are dependent on the re-organization energy ($\lambda_n$) and the line-broadening function $(g(t))$ of the excitons. The analytical expressions are given in Eq.~\ref{D1} of Appendix~\ref{ED}.

From the exciton states $\ket{X_{n}}$, energy is exchanged with intermediate states via resonance energy transfer leading the charge transfer rates  between $\ket{I_{1}}$ and $\ket{I_{2}}$, denoted as $k_{xy},x\ne y $ with $x\in \ket X_n$ and $y\in \{\ket{I_1}$,$\ket{I_2}$\}. These charge transfer rates are estimated using F$\ddot{o}$rster theory which depends on the spectral overlap and the electronic coupling between the states. Fluorescence experiments have revealed highly structured spectral densities \cite{Peterman1998TheNO}, offering a realistic depiction of exciton-phonon interactions. This includes 48 high-frequency vibrational modes alongside low-frequency phonons \cite{novoderezhkin2007mixing}. In modeling the F$\ddot{o}$rster rates,  we employ realistic spectral densities, ensuring that calculations are performed over the complete spectral profile in league with earlier studies \cite{C7SC02983G}. Previous studies have shown that there is no effective exchange between $\ket {X_n}$ and $\ket {I_3}$ due to insufficient spectral overlap or electronic coupling. Likewise, F$\ddot{o}$rster-like rates are used to describe population transfer among the charge-separated states $\ket{I_{1}}, \ket{I_{2}}, \ket{I_{3}}$ and $\ket{\alpha_{1}}$. These F$\ddot{o}$rster-like rates are controlled by the strength of the electronic coupling and the spectral overlap between the states across which the transfer takes place. The analytical expressions are provided in Appendix~\ref{FD} from Eqs.~\ref{C2} - \ref{C4}. Four electron shuttling rates are also present in the dynamics:  $\Gamma_{\alpha_{j}\alpha_{j'}}$ ($j \ne j'$ and $j, j' \in {1, 2}$)  between the $N$ state ($\ket{\alpha_1}$) and the $N+1$ component ($E_{\alpha_1'}$) of the state ($\ket{\alpha_2}$), and  $\Gamma_{\beta{j}\beta{j'}}$ ($j \ne j'$ and $j, j' \in {1, 2}$) between the $N-1$ state ($\ket{\beta_{1}}$) and the $N$ state ($\ket{\beta_{2}}$).  Two unidirectional electron tunneling rates, $\Gamma_{\alpha_j\beta_{j}}$ and $\omega_{\alpha_{j'}\beta_{j'}}$, ($j =1$ and $j' = 2$) allow electronic population transfer from the $\ket {\alpha_k}$-th state to the $\ket{\beta_{k}}$-th state. These rates have been obtained using a second order perturbation on the molecule-electronic reservoir coupling. Lastly, following the fourth order  perturbation  on the molecule-electronic reservoir \cite{carmi2012enhanced} we obtain the cotunneling rates, $\omega_{lm}, ~ l \ne m$ ($l, m \in \{\ket{I_{1}}, \ket{I_{3}}, \ket{\alpha_{2}},\ket{\beta_{2}} \}$). These rates account for the simultaneous population transfer from the charge-separated states $\ket{I_{1}}$ and $\ket{I_{3}}$ to $\ket{\alpha_{2}}$ as well as from $\ket{\alpha_2}$ to $\ket{\beta_2}$.  The cotunneling rates depend on the energies of the involved states, $\ket{I_{1}}$, $\ket{I_{3}}$, $\ket{\alpha'_{1}}$ and $\ket{\alpha'_{2}}$; the cotunneling coupling coefficients, $t_{I_{1}\alpha_{2}}$ and $t_{I_{3}\alpha_{2}}$, representing the strength of tunneling between $\ket{I_{1}}$, $\ket{\alpha_{2}}$ and $\ket{I_{3}}$, $\ket{\alpha_{2}}$ along with the interaction energy, U. The analytical expressions are presented in Eqs.~\ref{D2} and \ref{D3} of Appendix~\ref{CT}

\begin{figure}[h!]
    \centering
    \begin{minipage}{0.493\columnwidth} 
        \centering
        \includegraphics[width=4cm, height=3.5cm]{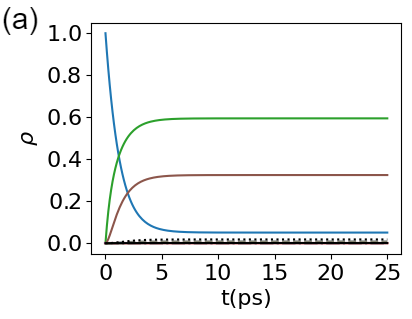}  
    \end{minipage}
    \hspace{-1mm} 
    \begin{minipage}{0.493\columnwidth}
        \centering
        \includegraphics[width=3cm, height=3cm]{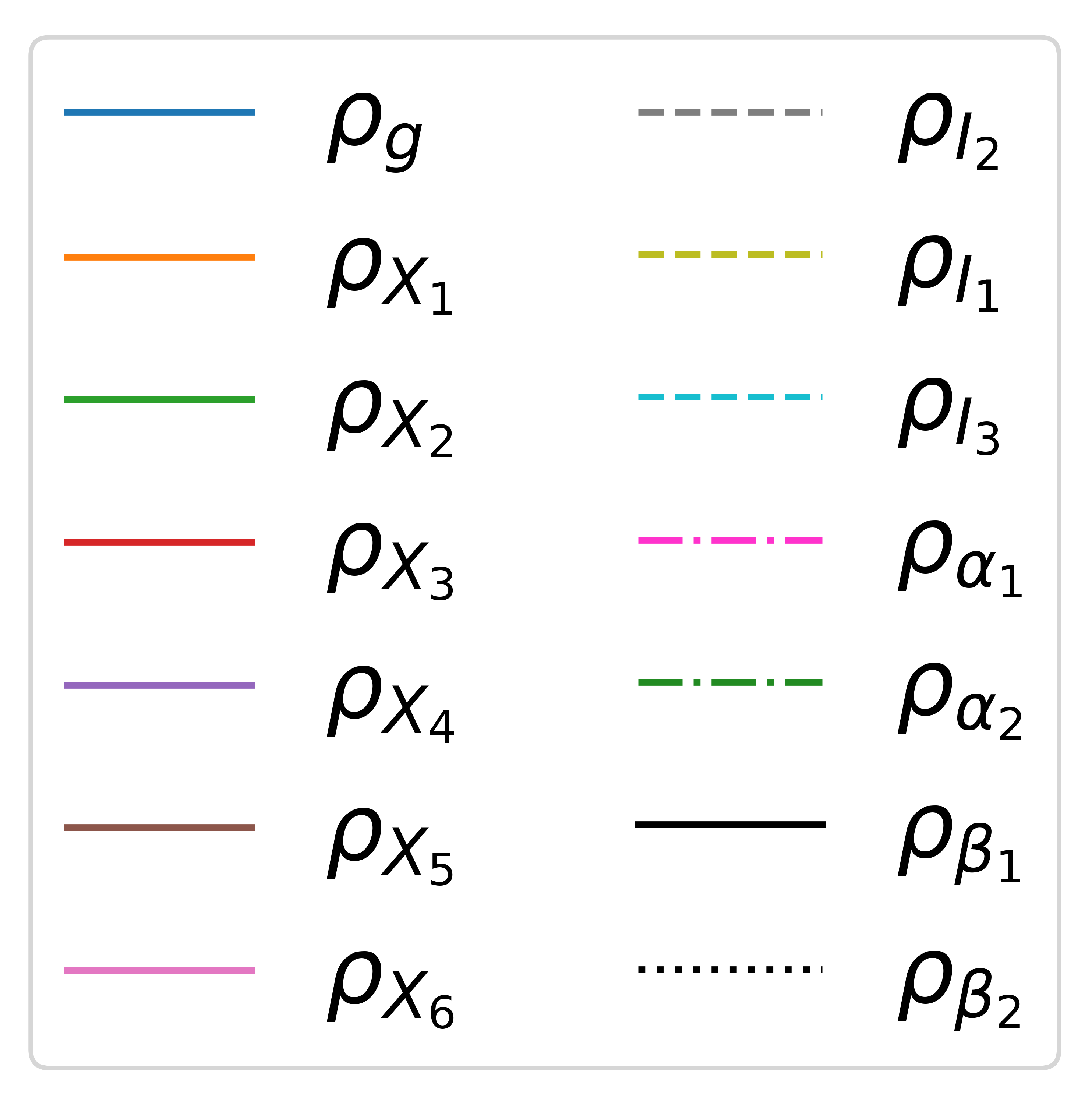}  
    \end{minipage}
    
    \vspace{0.1cm} 

    \begin{minipage}{0.493\columnwidth} 
        \centering
        \includegraphics[width=4cm, height=3.5cm]{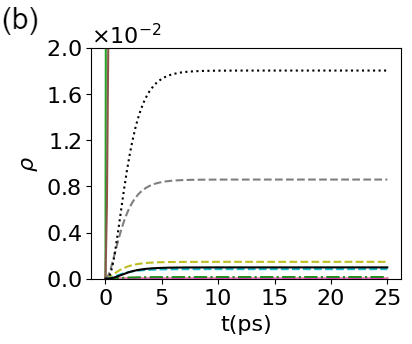} 
    \end{minipage}
    \hspace{-1mm} 
    \begin{minipage}{0.493\columnwidth}
        \centering
        \includegraphics[width=4cm, height=3.5cm]{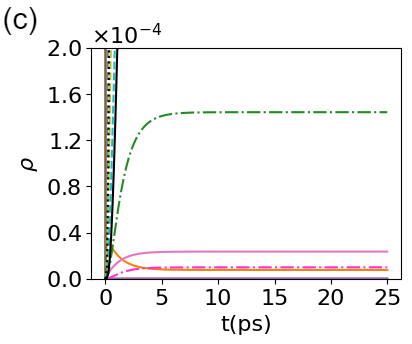}  
    \end{minipage}
    
    \vspace{0.1cm} 
    
    \begin{minipage}{0.493\columnwidth} 
        \centering
        \includegraphics[width=4cm, height=3.5cm]{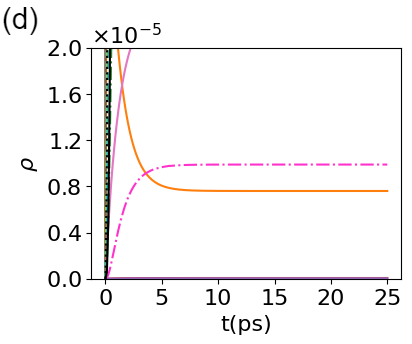}  
    \end{minipage}
    \hspace{-1mm} 
    \begin{minipage}{0.493\columnwidth}
        \centering
        \includegraphics[width=4cm, height=3.5cm]{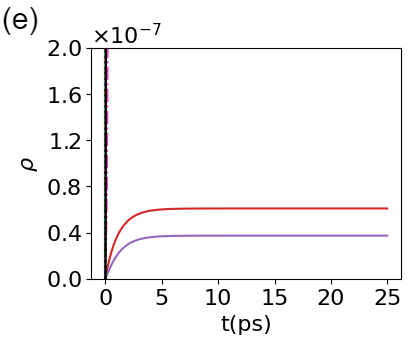}  
    \end{minipage}

    \caption{\justifying Population dynamics of the PSIIRC under the scenario when all the couplings are weak and are in a comparable range, with $E'_{\alpha_{2}} = 15670 cm^{-1}$, $E'_{\alpha_{1}} = 13972 cm^{-1}$, $\Gamma_{\alpha_{2}\alpha_{1}} = 50 cm^{-1}$, $\Gamma_{\alpha_{1}\alpha_{2}} = 5 cm^{-1}$, $\Gamma_{\beta_{1}\beta_{2}} = \Gamma_{\beta_{2}\beta_{1}} = 5 cm^{-1}$, $\gamma_{ex}=75cm^
    {-1}$, $t_{I_{1}\alpha_{2}} = t_{I_{3}\alpha_{2}} = 50 cm^{-1}$ and $U = -1000 cm^{-1}$.}
    \label{fig:2}
\end{figure}

The parameters that we choose to modulate in our study are the energy of the $N+2$ state (through the energies of the two constituent N+1 electronic states), the interaction energy $U$, the cotunneling amplitudes $t_{ab}$, the shuttling rates, $\Gamma_{\alpha_{j}\alpha_{j'}}$, unidirectional rate $\Gamma_{\alpha_j\beta_{j}}$ and the cotunneling rate $\omega_{\alpha_{j'}\beta_{j'}}$. 
To obtain the quantum dynamics from the master equation, 
we assume the values of the absolute shuttling rates and the cotunneling amplitudes to be within the same order as the electronic couplings involved in the charge transfer rates. Solving the Pauli master equation (Eq.~\ref{A9}), gives us the population dynamics for this specific scenario. We choose this condition so that the populations of all the states achieve a steady state within the considered timescale. The time evolution of the populations of the manybody states is shown in Fig.~\ref{fig:2}. Fig.\ref{fig:2} illustrates the population dynamics of the system for a particular instance when $E'_{\alpha_{2}}>E'_{\alpha_{1}}$; $\Gamma_{\alpha_{2}\alpha_{1}} > \Gamma_{\alpha_{1}\alpha_{2}}$; $t_{I_{1}\alpha_{2}} \approx t_{I_{3}\alpha_{2}}$ and $U = -1000 cm^{-1}$.

\section{\label{sec:level4}Thermodynamics of PSIIRC}
\begin{figure}[h!]
    \centering
    \begin{minipage}{0.494\columnwidth}
        \centering
        \includegraphics[width=4.33cm, height=3.5cm]{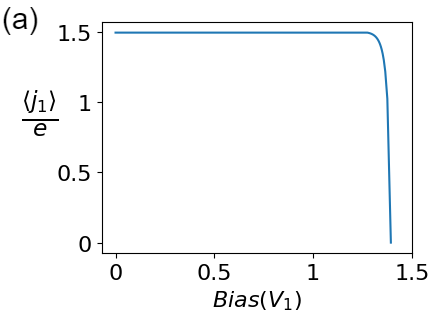}
    \end{minipage}
    \hspace{-1mm}
    \begin{minipage}{0.494\columnwidth}
        \centering
        \includegraphics[width=4.33cm, height=3.5cm]{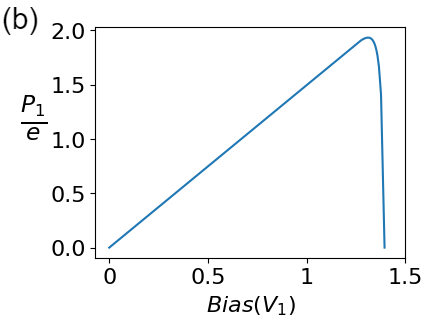}
    \end{minipage}
    
    \vspace{0.1cm} 

    \begin{minipage}{0.494\columnwidth}
        \centering
        \includegraphics[width=4cm, height=3.5cm]{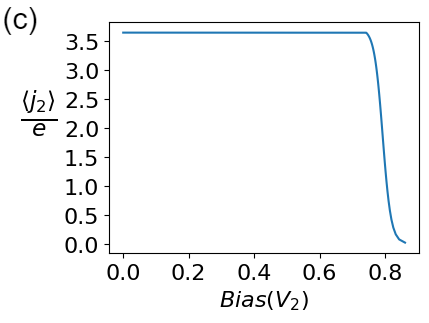}
    \end{minipage}
    \hspace{-1mm}
    \begin{minipage}{0.494\columnwidth}
        \centering
        \includegraphics[width=4cm, height=3.5cm]{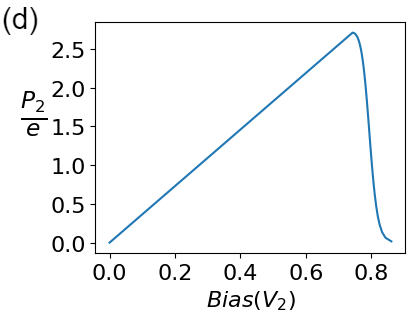}
    \end{minipage}

    \caption{\justifying (a) and (b):$\langle j_{1} \rangle - V_{1}$ and $P_{1}-V_{1}$ graphs across the states $\ket{\alpha_{1}}-\ket{\beta_{1}}$ for a fixed value of $\omega_{\alpha_{2}\beta_{2}}$. (c) and (d):$\langle j_{2} \rangle - V_{2}$ and $P_{2}-V_{2}$ graphs across the states $\ket{\alpha_{2}}-\ket{\beta_{2}}$ for a fixed value of $\Gamma_{\alpha_{1}\beta_{1}}$.} 
    \label{fig:3}
\end{figure}

Typically, from a quantum transport perspective, one is interested in the thermodynamic quantities such as current, power, and work at a particular terminal of quantum junctions \cite{esposito2009nonequilibrium}, either the source or the drain. In the PSIIRC junction, the electronic transport across the states, $\ket {\alpha_{1}}$ to $\ket{\beta_{1}}$  or $\ket{\alpha_{2}}$ to $\ket{\beta_{2}}$ is representative of a junction terminal where a steady-state current  is of interest. The current due to transport from $\ket {\alpha_{1}}\rightarrow $  $\ket{\beta_{1}}$ is termed as the conventional sequential current, $\langle j_{1} \rangle $ and has been well documented \cite{C7SC02983G}. The current due to transport from $\ket{\alpha_{2}} \rightarrow \ket{\beta_{2}}$ is the cotunneling current, $\langle j_{2} \rangle $ arising due to an electronic transport involving $N+2 \rightarrow N$ manybody state. The two electrons are created in the sink.  Mathematically, the steady-state values of the individual currents, $\langle j_{k} \rangle$ for the $k$th transport process can be evaluated using,
\begin{equation}\label{1}
    \langle j_{k} \rangle = k\times e \Gamma_{\alpha_{k}\beta}^{}\times\rho_{\alpha_{k}}^\infty, k =1,2
\end{equation}
where $\rho_{\alpha_{k}}^\infty$ is the steady-state population of the $\ket\alpha_{k}$ state. When $k=2$, the current is twofold since cotunneling involves two simultaneous electrons.
The steady-state output power of the individual processes is obtainable using,
\begin{equation}\label{2}
    P_{k} = \langle j_{k} \rangle V_{k}, k =1,2
\end{equation}
with $V_k$ being the electrochemical work done in the k-th individual electron transport process.
 Following standard thermodynamic consideration in photophysical systems \cite{Ross, Shockley}, the work done is directly obtainable from the effective bias between the energies of the $\ket{\alpha_{1}}-\ket{\beta_{1}}$ and $\ket{\alpha_{2}}-\ket{\beta_{2}}$. It is quantifiable as,
 
\begin{equation}\label{3}
    eV_{k}k 
    =  E_{\alpha_{k}}-E_{\beta_{k}} + k_{B}Tln\bigg[ \frac{\rho_{\alpha_{k}}^\infty}{\rho_{\beta_{k}}^\infty}\bigg], ~k = 1,2
\end{equation}
where $\rho_{\beta_{k}}^\infty$ is the steady-state population of the $\ket{\beta_{k}}$-th state, $k_{B}$ is the Boltzmann constant, T is the temperature with \textit{e} being the electric charge. The steady-state populations $\rho_m^\infty$ can be obtained by setting the master equation $\ket{\dot{\hat\rho}_q} = \breve {\cal L}\ket{\hat\rho_q} = 0$ along with $\sum_m\rho_m=1$. The characteristic current-voltage and power-voltage, i.e $\langle j_{k} \rangle - V_{k}$ and $P_{k}-V_{k}$ curves are then computed using Eqs.~(\ref{1}, \ref{2} and \ref{3}). For the quantum dynamics evaluated for parameters displayed in Fig.~\ref{fig:2}(a), the $\langle j_{1} \rangle - V_{1}$, $P_{1}-V_{1}$ and $\langle j_{2} \rangle - V_{2}$, $P_{2}-V_{2}$ curves are displayed in Fig.~\ref{fig:3}(a), 3(b) and \ref{fig:3}(c), 3(d) respectively. The qualitative shape of all the curves follows the universally accepted characteristic pattern \cite{huang2023high, Photocurrent}. The sequential current, $\langle j_{1} \rangle $  is constant at lower bias. As the bias increases to a certain threshold, a sharp drop in current is observed. This suggests that the system reaches a point where further electron transfer becomes unfavorable,  due to competing rates that do not favour populating the $\ket{\alpha_1}$ state,  leading to a reduction in the effective current. The cotunneling current-bias, $\langle j_{2} \rangle - V_2$, curve also exhibits a similar trend but is larger in magnitude than $\langle j_1\rangle$ and the steep drop in the cotunneling current occurs at a lower voltage when compared to the sequential current. This indicates that the cotunneling current is persistent only over a smaller range of bias between the source and the drain. We can conclude that the increase in the magnitude of $\langle j_2\rangle$ is available over a smaller bias range, typical of a trade-off characteristic behavior: a high value of flux is compensated by squeezing the voltage window. The $P_k - V_k$ curves are shown in Fg. (\ref{fig:3}(b) and (d)) complement the observation and conclusion of the $\langle j_k\rangle-V_k$ curves. A higher value of cotunneling assisted power is available within a smaller span of bias in comparison to sequential power.

To further comprehend the cotunneling assisted characteristics of the PSIIRC junction, we focus on the key parameters that involve cotunneling: the cotunneling specific coupling coefficients ($t_{I_1\alpha_1}$ and $t_{I_3\alpha_2}$), shuttling rates between  $\ket{\alpha_1}$ and $\ket{\alpha_2}$ ($\Gamma_{\alpha_{1}\alpha_{2}}$ and $\Gamma_{\alpha_{2}\alpha_{1}}$), the interaction energy ($U$), and the relative energy of the levels $\ket{\alpha'_1}$ and $\ket{\alpha'_{2}}$ ($E'_{\alpha_1}$ and $E'_{\alpha_2}$). The F$\ddot{o}$rster-like rates and the intra-exciton transfer rates remain constant throughout the calculations. However, adjusting the concerning key parameters alters the cotunneling rates under different parametric scenarios. We analyse a few relevant parameter scenarios by constructing contour diagrams that depict the ratio of cotunneling current to the sequential current, $\langle j_2 \rangle/\langle j_1 \rangle$, along with the ratio of cotunneling power to sequential power, $P_2/P_1$, as functions of the electron ejection rate $\Gamma_{\alpha_1\beta_{1}}$ and the cotunneling rate $\omega_{\alpha_2\beta_{2}}$. Given that the master equation approach is valid only in the weak coupling regime, we maintain that all relevant couplings we vary do not exceed the PSIIRC energy scales. Hence the range of the electron ejection rates is restricted to the domain $\Gamma_{\alpha_{1}\beta_{1}}, \omega_{\alpha_{2}\beta_{2}}\in\{0,~25000cm^{-1}\}$.

\begin{figure}[h!]
    \centering
    \begin{minipage}{0.494\columnwidth}
        \centering
        \includegraphics[width=4cm, height=3.3cm]{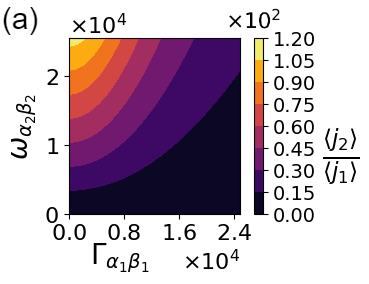} 
    \end{minipage}
    \begin{minipage}{0.494\columnwidth}
        \centering
        \includegraphics[width=4cm, height=3.3cm]{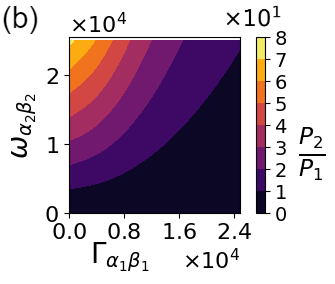}
    \end{minipage}

    \vspace{0.1cm} 

    \begin{minipage}{0.494\columnwidth}
        \centering
        \includegraphics[width=4cm, height=3.3cm]{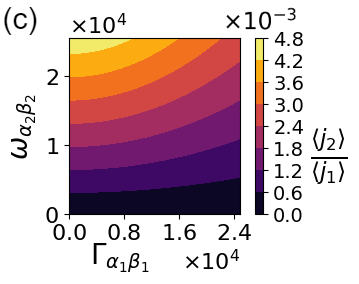}
    \end{minipage}
    \begin{minipage}{0.494\columnwidth}
        \centering
        \includegraphics[width=4cm, height=3.3cm]{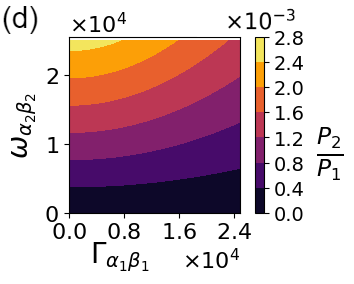}
    \end{minipage}

     \caption{\justifying Contour plots of ${\langle j_{2} \rangle}/{\langle j_{1} \rangle}$ and ${P_{2}}/{P_{1}}$ when all the couplings are weak and comparable, with $t_{I_{1}\alpha_{2}} \approx t_{I_{3}\alpha_{2}}, \Gamma_{\alpha_{2}\alpha_{1}} > \Gamma_{\alpha_{1}\alpha_{2}}$ and $E'_{\alpha_{2}} > E'_{\alpha_{1}}$. For (a), (b): $U = -1000 cm^{-1}$, and (c), (d): $U = 1000 cm^{-1}$.} 
    \label{fig:4}
\end{figure}

First, we consider a scenario where inherent energies of the $\ket{\alpha_2}$ state have an offset, i.e $E'_{\alpha_2} > E'_{\alpha_1}$ (note that $E_{\alpha_2}= E'_{\alpha_2} + E'_{\alpha_1}$+U), the shuttling rates are asymmetrically tuned such that $\Gamma_{\alpha_{2}\alpha_{1}} > \Gamma_{\alpha_{1}\alpha_{2}}$ (physically the rate of transfer from the $\ket{\alpha_{2}}$ state to the $\ket{\alpha_{1}}$ state is larger) while the cotunneling amplitudes are taken to be approximately equal, i.e. $t_{I_1 \alpha_2} \approx t_{I_3 \alpha_2}$. The resultant current and power ratio are shown in Figs.~\ref{fig:4}(a) and \ref{fig:4}(b) for attractive ($U <0$) and repulsive ($U>0$) interaction energies as a function of the sequential ($\Gamma_{\alpha_{1}\beta_{1}}$) and cotunneling ejection rates ($\omega_{\alpha_{2}\beta_{2}}$).  The attractive interaction energy $U$ allows a 100-fold enhancement of the cotunneling current and a 10-fold enhancement of the cotunneling power compared to the sequential counterparts, as seen in Figs.~\ref{fig:4}(a) and \ref{fig:4}(b). This enhancement persists even when the sequential ejection rate, $\Gamma_{\alpha_1 \beta_{1}}$ is larger than $\omega_{\alpha_2 \beta_{2}}$, indicating that the negative $U$ creates a sufficiently stable attractive interaction between electrons in the negatively charged state, which allows simultaneous tunneling of two electrons to the drain in the same timescale. Conversely, when $U$ is positive, the cotunneling contributions diminish, as shown in Figs.~\ref{fig:4}(c) and \ref{fig:4}(d), with the ratio of cotunneling current and power bounded below unity. In this scenario, positive $U$ represents electron-electron repulsion. This leads to an unfavorable negatively charged state where electrons are either transferred back to the $\ket {\alpha_1}$ state or lost, nullifying the cotunneling contributions. Maintaining the same parameter conditions but reversing the asymmetry in the shuttling rates, i.e, considering $\Gamma_{\alpha_2 \alpha_1} < \Gamma_{\alpha_1 \alpha_2}$,  doesn't alter the aforementioned observation (see Fig.~\ref{fig: 12} in Appendix~\ref{CP}).  Although the cotunneling contributions increase more sharply at higher negative $U$ values, this difference can be attributed to the higher electron influx into the $\ket{\alpha_2}$ state, driven by the dominant shuttling rate $\Gamma_{\alpha_1 \alpha_2}$.

\begin{figure}[h!]
    \centering
    \begin{minipage}{0.494\columnwidth}
        \centering
        \includegraphics[width=4cm, height=3.5cm]{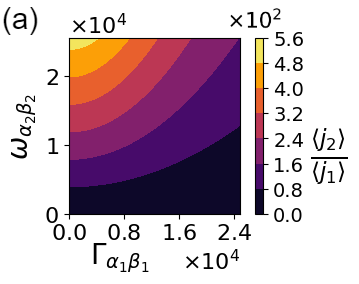} 
    \end{minipage}
    \begin{minipage}{0.494\columnwidth}
        \centering
        \includegraphics[width=4cm, height=3.5cm]{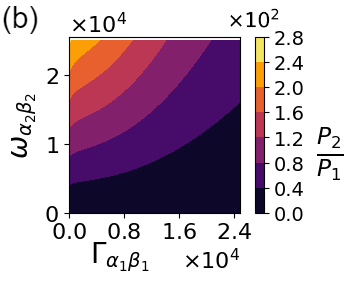}
    \end{minipage}

    \vspace{0.1cm} 

    \begin{minipage}{0.494\columnwidth}
        \centering
        \includegraphics[width=4cm, height=3.5cm]{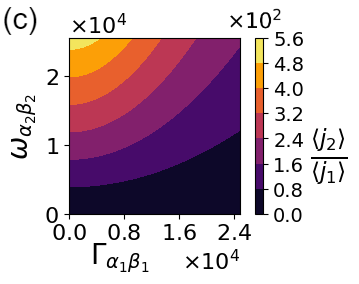}
    \end{minipage}
    \begin{minipage}{0.494\columnwidth}
        \centering
        \includegraphics[width=4cm, height=3.5cm]{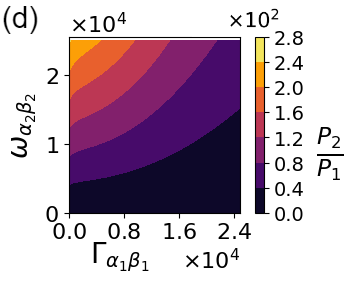}
    \end{minipage}

     \caption{\justifying Contour plots of ${\langle j_{2} \rangle}/{\langle j_{1} \rangle}$ and ${P_{2}}/{P_{1}}$ when all the couplings are weak and comparable, with $t_{I_{1}\alpha_{2}} \approx t_{I_{3}\alpha_{2}}, \Gamma_{\alpha_{2}\alpha_{1}} > \Gamma_{\alpha_{1}\alpha_{2}}$ and $E'_{\alpha_{2}} < E'_{\alpha_{1}}$. For (a), (b): $U = -1000 cm^{-1}$ and (c), (d): $U = 1000 cm^{-1}$.}
    \label{fig:5}
\end{figure}

Secondly, we consider a scenario where $E'_{\alpha_2} < E'_{\alpha_1}$, $t_{I_1 \alpha_2} \approx t_{I_3 \alpha_2}$ and $\Gamma_{\alpha_2 \alpha_1} > \Gamma_{\alpha_1 \alpha_2}$. From the contour plots in Figs.~\ref{fig:5} and \ref{fig: 11}, we observe that the cotunneling current and power dominate over the sequential counterparts in both negative and positive $U$ regimes. 
This observation is true in the range $|U|\le 2000  cm^{-1}$, where a constant 100-fold enhancement of the cotunneling current and power over the sequential ones is seen. Beyond this range, a drop in the cotunneling current and power is observed as depicted by the contour plots in Fig.~\ref{fig: 11} of Appendix E. In this scenario, we conclude that the enhancement in cotunneling contributions is primarily driven by the energy of the negatively charged state rather than interaction energy. This can be justified by the fact that the energy of the $\ket{\alpha'_{2}}$ state is relatively lower than that of the inherent $\ket{\alpha_{1}}$ state resulting in the electrons tunneling to the negatively charged state and enhancing the output current and power. Similar results are obtained upon reversing the bias in the shuttling rates, $\Gamma_{\alpha_2 \alpha_1} < \Gamma_{\alpha_1 \alpha_2}$.  The cotunneling flux and power remain dominant over the sequential processes until $U$ reaches approximately 1000 cm$^{-1}$ (see Fig.~\ref{fig: 13}). Beyond this, cotunneling current and power diminish in magnitude but still exceed their sequential counterparts.

\begin{figure}[h!]
    \centering
    \begin{minipage}{0.494\columnwidth}
        \centering
        \includegraphics[width=4cm, height=3.5cm]{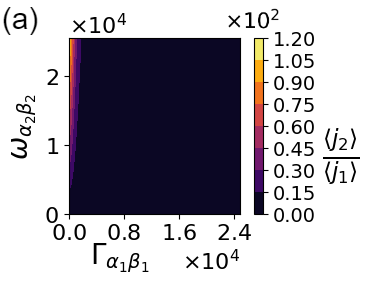} 
    \end{minipage}
    \begin{minipage}{0.494\columnwidth}
        \centering
        \includegraphics[width=4cm, height=3.5cm]{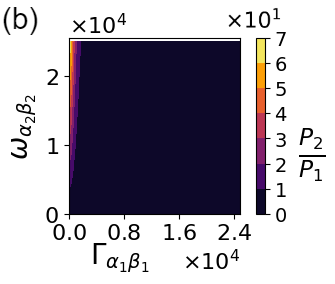}
    \end{minipage}

    \vspace{0.1cm} 

    \begin{minipage}{0.494\columnwidth}
        \centering
        \includegraphics[width=4cm, height=3.5cm]{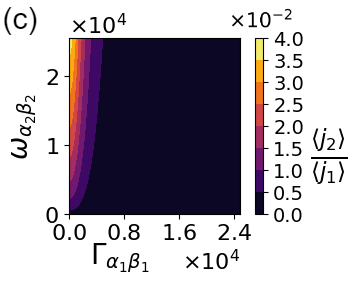}
    \end{minipage}
    \begin{minipage}{0.494\columnwidth}
        \centering
        \includegraphics[width=4cm, height=3.5cm]{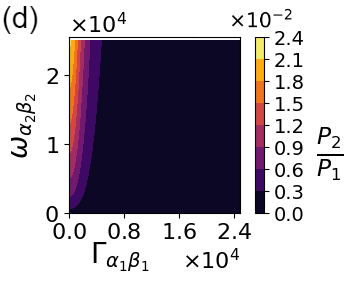}
    \end{minipage}
 
     \caption{\justifying Contour plots of ${\langle j_{2} \rangle}/{\langle j_{1} \rangle}$ and ${P_{2}}/{P_{1}}$ when $t_{I_{1}\alpha_{2}} > t_{I_{3}\alpha_{2}}, \Gamma_{\alpha_{2}\alpha_{1}} > \Gamma_{\alpha_{1}\alpha_{2}}$ and $E'_{\alpha_{2}} > E'_{\alpha_{1}}$. For (a), (b): U = -1000 cm$^{-1}$ and (c), (d): U = 1000 cm$^{-1}$.}
    \label{fig:6}
\end{figure}

Thirdly, we move on to a scenario where the cotunneling coupling coefficients are asymmetrically tuned ($t_{I_1 \alpha_2} > t_{I_3 \alpha_2}$) with $\Gamma_{\alpha_2 \alpha_1} > \Gamma_{\alpha_1 \alpha_2}$ and $E'_{\alpha_2} > E'_{\alpha_1}$. The interaction energy $U$ continues to affect the current and power ratios (Fig.~\ref{fig:6}). In the negative $U$ domain, both the cotunneling current and the power are enhanced (Figs.~\ref{fig:6}(a), \ref{fig:6}(b)), whereas these diminish in the positive $U$ domain (Figs.~\ref{fig:6}(c), \ref{fig:6}(d)). Interestingly maximum output flux and power are achieved only when $\Gamma_{\alpha_1\beta_{1}}$ is significantly smaller than $\omega_{\alpha_2\beta_{2}}$, suggesting that asymmetry in the strength of tunneling coefficients doesn't enhance the overall cotunneling flux and power even when the interaction energy is attractive. This behavior is contrary to what has been previously observed in quantum transport across junctions, where asymmetrically stronger tunneling amplitudes amplified the cotunneling contributions to current \cite{carmi2012enhanced}.

\begin{figure}[h!]
    \centering
    \begin{minipage}{0.494\columnwidth}
        \centering
        \includegraphics[width=4cm, height=3.5cm]{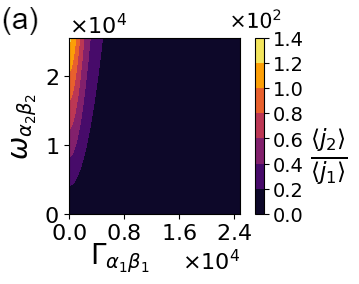} 
    \end{minipage}
    \begin{minipage}{0.494\columnwidth}
        \centering
        \includegraphics[width=4cm, height=3.5cm]{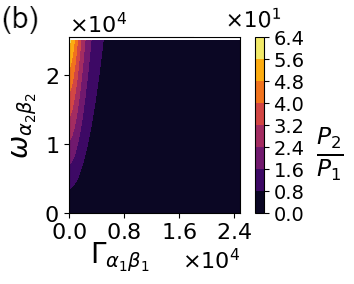}
    \end{minipage}

    \vspace{0.1cm} 

    \begin{minipage}{0.494\columnwidth}
        \centering
        \includegraphics[width=4cm, height=3.5cm]{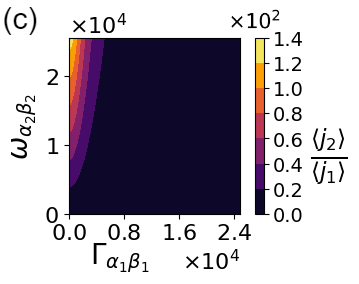}
    \end{minipage}
    \begin{minipage}{0.494\columnwidth}
        \centering
        \includegraphics[width=4cm, height=3.5cm]{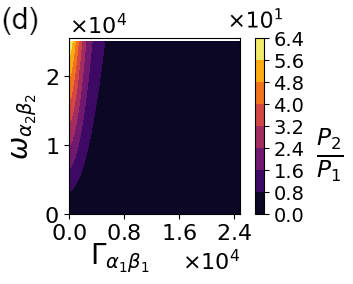}
    \end{minipage}

     \caption{\justifying Contour plots of ${\langle j_{2} \rangle}/{\langle j_{1} \rangle}$ and ${P_{2}}/{P_{1}}$ when $t_{I_{1}\alpha_{2}} > t_{I_{3}\alpha_{2}}, \Gamma_{\alpha_{2}\alpha_{1}} > \Gamma_{\alpha_{1}\alpha_{2}}$ and $E'_{\alpha_{1}} > E'_{\alpha_{2}}$. For (a), (b): $U = -1000 cm^{-1}$ and (c), (d): $U = 1000 cm^{-1}$.}
    \label{fig:7}
\end{figure}

Fourthly, for $E'_{\alpha_{2}} < E'_{\alpha_{1}}$ with $t_{I_{1}\alpha_{2}} > t_{I_{3}\alpha_{2}}$ and $\Gamma_{\alpha_{2}\alpha_{1}} > \Gamma_{\alpha_{1}\alpha_{2}}$ the cotunneling current and power dominates over their sequential counterparts throughout the negative range and extend into the positive domain before dropping at $U \approx 2000 \, \text{cm}^{-1}$ (Figs.~\ref{fig:7}, \ref{fig: 15}). However, similar to the third scenario, the maximum values of the cotunneling current and power are observed only at significantly smaller values of $\omega_{\alpha_{2}\beta_{2}}$.

\begin{figure}[h!]
    \centering
    \begin{minipage}{0.494\columnwidth}
        \centering
        \includegraphics[width=4cm, height=3.5cm]{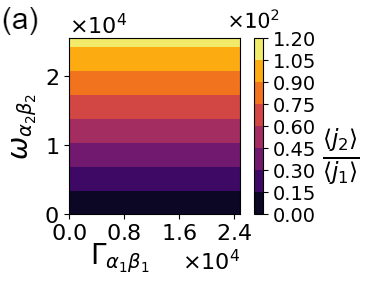} 
    \end{minipage}
    \begin{minipage}{0.494\columnwidth}
        \centering
        \includegraphics[width=4cm, height=3.5cm]{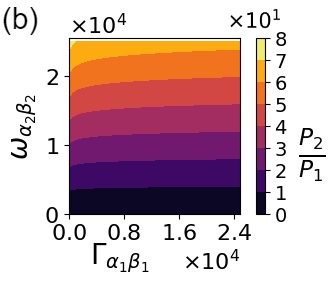}
    \end{minipage}

    \vspace{0.1cm} 

    \begin{minipage}{0.494\columnwidth}
        \centering
        \includegraphics[width=4cm, height=3.5cm]{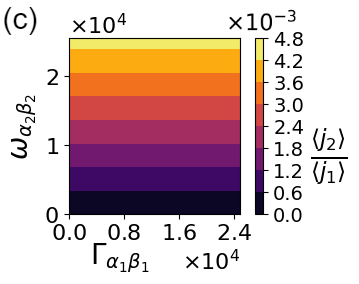}
    \end{minipage}
    \begin{minipage}{0.494\columnwidth}
        \centering
        \includegraphics[width=4cm, height=3.5cm]{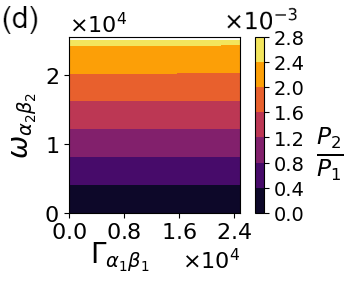}
    \end{minipage}

    \caption{\justifying Contour plots of ${\langle j_{2} \rangle}/{\langle j_{1} \rangle}$ and ${P_{2}}/{P_{1}}$ for $t_{I_{1}\alpha_{2}} > t_{I_{3}\alpha_{2}}, \Gamma_{\alpha_{2}\alpha_{1}} > \Gamma_{\alpha_{1}\alpha_{2}}$ with $E'_{\alpha_{2}} > E'_{\alpha_{1}}$. $t_{I_{1}\alpha_{2}}$ and $t_{I_{3}\alpha_{2}}$ are much larger in magnitude that other coupling parameters. For (a), (b): $U = -1000 cm^{-1}$ and (c), (d): $U = 1000 cm^{-1}$.}
    \label{fig:8}
\end{figure}

\begin{figure}[h!]
    \centering
    \begin{minipage}{0.494\columnwidth}
        \centering
        \includegraphics[width=4cm, height=3.5cm]{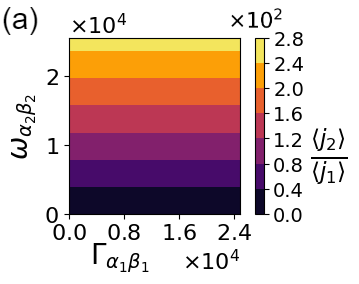} 
    \end{minipage}
    \begin{minipage}{0.494\columnwidth}
        \centering
        \includegraphics[width=4cm, height=3.5cm]{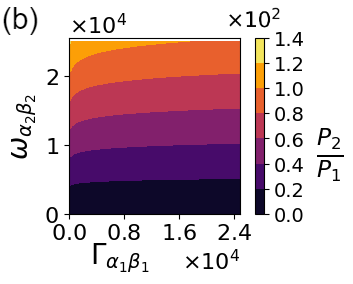}
    \end{minipage}

    \vspace{0.1cm} 

    \begin{minipage}{0.494\columnwidth}
        \centering
        \includegraphics[width=4cm, height=3.5cm]{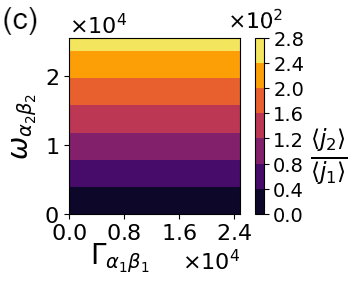}
    \end{minipage}
    \begin{minipage}{0.494\columnwidth}
        \centering
        \includegraphics[width=4cm, height=3.5cm]{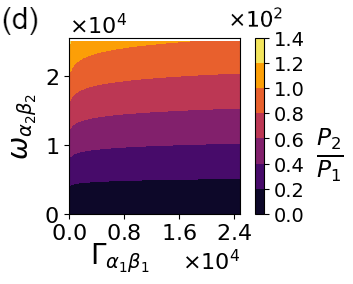}
    \end{minipage}
    
    \caption{\justifying Contour plots of ${\langle j_{2} \rangle}/{\langle j_{1} \rangle}$ and ${P_{2}}/{P_{1}}$ for $t_{I_{1}\alpha_{2}} > t_{I_{3}\alpha_{2}}, \Gamma_{\alpha_{2}\alpha_{1}} > \Gamma_{\alpha_{1}\alpha_{2}}$ with $E'_{\alpha_{1}} > E'_{\alpha_{2}}$. $t_{I_{1}\alpha_{2}}$ and $t_{I_{3}\alpha_{2}}$ are much larger in magnitude than the other coupling parameters. For (a), (b): $U = -1000 cm^{-1}$ and (c), (d): $U = 1000 cm^{-1}$.}
    \label{fig:9}
\end{figure}

When we reverse the asymmetry in the coupling strength, such that $t_{I_{1}\alpha_{2}} > t_{I_{3}\alpha_{2}}$, and the magnitude of these cotunneling coupling coefficients are relatively larger than other key parameters, the cotunneling current and power is seen to remain independent of $\Gamma_{\alpha_{1}\beta_{1}}$ for a fixed value of $\omega_{\alpha_{2}\beta}$. These are illustrated in Fig.~(\ref{fig:8}), where $E'_{\alpha_{2}} > E'_{\alpha_{1}}$ and Fig. (\ref{fig:9}), where $E'_{\alpha_{2}} < E'_{\alpha_{1}}$. However, when $E'_{\alpha_{2}} > E'_{\alpha_{1}}$ in the positive regime of $U$, the cotunneling contribution to the total current and power diminishes. On the other hand, the contributions show a significant increase when $E'_{\alpha_{1}} > E'_{\alpha_{2}}$.

\section{Conclusion}

We theoretically studied a photosystem II-based reaction center (PSIIRC) located between the oxygen evolution center and the quinone reduction center from a nonequilibrium source-system-drain perspective, i.e. by considering it as a biomolecular junction. The quantum elements describing this biomolecular junction were obtained from existing realistically simulated parameters. Inspired by the role of electroactive protein environment surrounding such pigment systems in introducing newer states into the quantum dynamics, we modify the Hamiltonian of the PSIIRC to include electron-electron interactions that allow cotunneling characteristics to be described in quantum mechanical rates in the master equation of the PSIIRC junction. Using the master equation, we investigated the role of cotunneling contributions in the thermodynamic quantifiers: current and power of the PSIIRC junction. We found that the cotunneling current can be amplified beyond the sequential current, but only within a smaller work or bias window at the junction terminal under study. We explored the interplay between key parameters such as cotunneling amplitudes, transition rates between the different manybody states, and the interaction energy ($U$).  Both attractive and repulsive interactions between the two electrons in the negatively charged state can lead to an enhancement in cotunneling current and power, achieved when there is an offset in the contributing energy states under equal cotunneling amplitudes.
Conversely, an asymmetry in the strength of the cotunneling amplitudes combined with an offset in the electron shuttling rate from the negative state allows a reduction in the cotunneling contributions even with attractive interactions. Notably, in the asymmetric regime of cotunneling amplitudes' strength and the shuttling rates, the current and power remained largely independent of the sequential electron ejection rate. Overall, in this study, we quantitatively highlighted the subtle interplay between the energy landscape, the tunneling amplitudes, and the electron-electron interaction in determining the PSIIRC junction's work, flux and power. In summary, our analysis revealed that cotunneling contributions can significantly alter the current and the power output in quantum systems, particularly under conditions where electron interactions and tunneling coefficients are finely tuned. These results provide important insights into the optimization of quantum transport devices, offering pathways to enhance performance by leveraging cotunneling processes. Future work could extend this analysis to more complex systems and explore the impact of strong coupling regimes, where the assumptions of the master equation approach may no longer hold.

\begin{acknowledgments}
DS acknowledges the support from the members of QuAInT research laboratory at GU, and MS thanks the Department of Chemistry at GU for its hospitality.
\end{acknowledgments}
  
\appendix

\section{PSIIRC Hamiltonian}\label{SD}

The considered PSIIRC model is assumed as a quantum system interacting with multiple environments whose Hamiltonian is taken to be of the form
\begin{equation}\label{A1}
    \hat H = \hat H_{q} + \hat H_{E} + \hat V_\nu, ~\hat V{_\nu} = \sum_{\nu = b, s, d} \hat{M}_{\nu}\hat{B}_{\nu} 
\end{equation}
with $\hat{H_{q}}$ being the PSIIRC molecular electronic Hamiltonian, $\hat{H}_{E}$ is the environment Hamiltonian and $\hat{V_\nu}$ is the coupling Hamiltonian between the molecule and the environment with $\nu = b,s,d$; representing the light-molecule ($b$), molecule-source ($s$) and molecule-drain ($d$) coupling respectively. $\hat{M}_\nu (\hat B_\nu)$ correspond to the molecular (environment) operator associated with $\nu$-th terminal. 
Assuming the initial density matrix describing the overall system to be separable into a molecular density matrix $\hat{\rho}_{q}$ and environment density matrix $\hat{\rho}_{E}$. $\hat\rho_q$ itself is a direct product of the system molecular components. We can write down an integro-differential equation describing the time evolution of the reduced density matrix (which is $\hat\rho_q =tr_E\{\hat \rho_q\otimes\hat\rho_E\})$ as follows
\begin{align}\label{eq-ide}
    \dot{\widetilde{\rho}}_{q}(t) &= \displaystyle\frac{i^2}{\hbar^{2}}\!\! \sum_{i.j = b,s,d}\! \int_{0}^{t}\Bigl[\widetilde{M}_{i}(t)\widetilde{M}_{j}(t')\widetilde{\rho}_{q}(t) \!-\! \widetilde{\rho}_{q}(t)\widetilde{M}_{j}(t')\widetilde{M}_{i}(t)\Bigr] \nonumber\\
    &\times
    tr_{E} \Bigl[\widetilde{B}_{i}(t)\widetilde{B}_{j}(t')\widetilde{\rho}_{E}(0) - \widetilde{\rho}_{E}(0)\widetilde{B}_{j}(t')\widetilde{B}_{i}(t) \Bigr]dt'
\end{align}
 with Dirac's interaction picture defined as
\begin{equation}\label{A3}
    \widetilde{O}(t) = e^{\frac{i}{\hbar}(\hat{H}_{q} + \hat{H}_{E})t} \hat{O} e^{-\frac{i}{\hbar}(\hat{H}_{q} + \hat{H}_{E})t}
\end{equation}
In Eq. (\ref{eq-ide}), the first term inside the square brackets of the integrand is the system correlation function while the second term is the bath correlation function \cite{ritschel2014analytic, timm2008tunneling}. Depending on the number of systems and bath states, these terms lead to a variety of terms upon explicit evaluation. To explicitly evaluate the time evolution of the molecular density matrix and the correlation functions, we define the following Hamiltonians for the PSIIRC. The molecular Hamiltonian is written as,
\begin{align}
    \label{eq-mole-ham}
    \hat H_q&=\hat H_o+U \hat c^\dag_{I_3}\hat c^\dag_{I_3}\hat c_{I_1}^{}\hat c_{I_1}^{},   \end{align}
where $\hat H_o$ is a diagonal Hamiltonian consisting of the energies of the considered states, $\ket{g}$, $\ket{X_n}$, $\ket{I_{1}}$, $\ket{I_{2}}$, $\ket{I_{3}}$, $\ket{\alpha_{1}}$, $\ket{\alpha_{2}}$, $\ket{\beta_{1}}$ and $\ket{\beta_{2}}$. All energies of the bare states except for $\ket{\alpha_2}$ and $\ket{\beta_{2}}$ are known. $U$ is the interaction energy associated when two electrons are simultaneously created on the N+2 manybodystate $|\alpha_2\rangle$ from $|I_1\rangle$ and $|I_3\rangle$ by the action of the Fermionic operators $\hat c$.  In this work, we kept the energy of $\ket{\alpha_2}$ (E$_{\alpha'_{1}}$ and E$_{\alpha'_{2}}$) and $U$ as free parameters, which we vary in our study. The overall environmental Hamiltonian is taken to be a sum of three local environments $\hat{H}_{E} = \hat{H}_{b} + \hat{H}_{s} + \hat{H}_{d}$ representing solar radiation, source electron supply (OEC) and the electron drain (QRC).  The concentrated sunlight is assumed to be a bosonic bath, whose  Hamiltonian is modeled as a non-interacting collection of harmonic modes given by 
\begin{equation}\label{A6}
\hat{H}_{b} = \sum_{k}\hbar\omega_{k}\hat{b}_{k}^{\dagger}\hat{b}_{k} 
\end{equation}
with $\hat{b}_{k}^{\dagger}(\hat{b}_{k})$ representing the bosonic creation (annihilation) operator of the k$^{th}$ harmonic mode of the energy $\hbar\omega_{k}$. The electronic reservoir Hamiltonian is defined as:
\begin{equation}\label{A7}
    \hat{H}_{e} = \sum_{k\in e}\hbar\widetilde{\omega}_{ke}\hat{c}_{ke}^{\dagger}\hat{c}_{ke}, e =s,d.
\end{equation}
 with $\hat{c}_{ke}^{\dagger}(\hat{c}_{ke})$ being the fermionic creation (annihilation) operator in the e$^{th}, e =s,d$ reservoir (source or drain) at the energy $\widetilde{\omega}_{kl}$. 

For the PSIIRC, the molecule-environment coupling Hamiltonian is composed of three components $(\hat{V} = \hat{V}_{b} + \hat{V}_{s} + \hat{V}_{d})$.  Mathematically,
\begin{align}\label{eq-exc-Ham}
    \hat{V}_{b} &= \hat M_{b} \hat B_{b} \equiv \sum_{ki} g_{ki}^{b} \ket{i}\bra{i}(\hat{b}_{k}^{\dagger} + \hat{b}_{k}), ~\ket{i} \in \{\ket {X_{n}},\ket{\beta_2} \}.
\end{align}
$g_{ki}^{b}$ represents the effective coupling of the i$^{th}$ molecular electronic state to the k$^{th}$ bosonic mode. The molecule-reservoir coupling Hamiltonian 
     $\hat{V}_{e} = \hat M_{e}\hat B_{e}, ~e =s, d$, based on the spectral evidence and existing reports \cite{Vladimir}, can be explicitly written as,
\begin{align}
\label{eq-fors-ham}
   \hat{V}_{e} &= \sum_{n}^{6}\sum_{i=1}^{2} \bigg[t_{I_{1}X_{n}}^{}\hat{c}_{I_{1}}^{\dagger}\hat{c}_{X_n} + \sum_{f=1,3,\beta} t_{\alpha_{1}I_{f}}\hat{c}_{\alpha_{1}}^{\dagger}\hat{c}_{I_{f}^{}} \nonumber\\
   &+ t_{I_{3}I_{2}}\hat{c}_{I_{3}}^{\dagger}\hat{c}^{}_{I_{2}} + t_{\alpha_{1}\alpha_{2}}\hat{c}_{\alpha_{1}}^{\dagger}\hat{c}_{\alpha_{2}} + t_{g\beta_1}^{}\hat{c}_{ks}^{\dagger}\hat{c}_{\beta_1} \nonumber\\
   &
   + 
    t_{\alpha_{i}\beta_i}^{}\hat{c}_{kd}^{\dagger}\hat{c}_{\alpha_{i}} + h.c\bigg]
\end{align}
with   $t_{mi}$ being the effective coupling term between the i$^{th}$ state to the m$^{th}$ state.  Substituting the explicit forms of the Hamiltonians in Eq. (\ref{eq-ide}, second and fourth order perturbation done on $\hat V_e$ for terms not involving $X_n$) and evaluating the elements of the molecular density matrix via the projection technique, $\hat{\rho}_{ij}=\bra i\hat\rho_q\ket j$, we define a vectorized density matrix $\ket {\rho_q}=\{\rho_{ij}\}$, we obtain a master equation in the Schrodinger picture, 
\begin{widetext}
\begin{equation} \label{A9}
\left[
    \begin{array}{c}
    \dot{\rho}_{g} \\[1.5mm]
    \dot{\rho}_{X_{1}}\\[1.5mm]
    \dot{\rho}_{X_{2}}\\[1.5mm]
    \dot{\rho}_{X_{3}}\\[1.5mm]
    \dot{\rho}_{X_{4}}\\[1.5mm]
    \dot{\rho}_{X_{5}}\\[1.5mm]
    \dot{\rho}_{X_{6}}\\[1.5mm]
    \dot{\rho}_{I_{1}}\\[1.5mm]
    \dot{\rho}_{I_{2}}\\[1.5mm]
    \dot{\rho}_{I_{3}}\\[1.5mm]
    \dot{\rho}_{\alpha_{1}}\\[1.5mm]
    \dot{\rho}_{\alpha_{2}}\\[1.5mm]
    \dot{\rho}_{\beta_{1}}\\[1.5mm]
    \dot{\rho}_{\beta_{2}}\\[1.5mm]
    \end{array}
\right]
=
\left[ 
    \begin{array}{cccccccccccccc}
    - \gamma_{ex} n & \gamma_{ex}(n+1) & 0 & 0 & 0 & 0 & 0 & 0 & 0 & 0 & 0 & 0 & \Gamma_{\beta_{1} g} & k_{\beta_{2} g}\\[1mm]
    \gamma_{ex} n & \breve{\cal L}_{11} & r_{21} & r_{31} & r_{41} & r_{51} & r_{61} & k_{I_{1}1} & k_{I_{2}1} & 0 & 0 & 0 & 0 & 0 \\[1mm]
    0 & r_{12} & \breve{\cal L}_{22} & r_{32} & r_{42} & r_{52} & r_{62} & k_{I_{1}2} & k_{I_{2}2} & 0 & 0 & 0 & 0 & 0 \\[1mm]
    0 & r_{13} & r_{23} & \breve{\cal L}_{33} & r_{43} & r_{53} & r_{63} & k_{I_{1}3} & k_{I_{2}3} & 0 & 0 & 0 & 0 & 0 \\[1mm]
    0 & r_{14} & r_{24} & r_{34} & \breve{\cal L}_{44} & r_{54} & r_{64} & k_{I_{1}4} & k_{I_{2}4} & 0 & 0 & 0 & 0 & 0 \\[1mm]
    0 & r_{15} & r_{25} & r_{35} & r_{45} & \breve{\cal L}_{55} & r_{65} & k_{I_{1}5} & k_{I_{2}5} & 0 & 0 & 0 & 0 & 0 \\[1mm]
    0 & r_{16} & r_{26} & r_{36} & r_{46} & r_{56} & \breve{\cal L}_{66} & k_{I_{1}6} & k_{I_{2}6} & 0 & 0 & 0 & 0 & 0 \\[1mm]
    0 & k_{1I_{1}} & k_{2I_{1}} & k_{3I_{1}} & k_{4I_{1}} & k_{5I_{1}} & k_{6I_{1}} & \breve{\cal L}_{I_{1}I_{1}} & 0 & 0 & k_{\alpha_{1}I_{1}} & \omega_{\alpha_{2}I_{1}} & 0 & 0 \\[1mm]
    0 & k_{1I_{2}} & k_{2I_{2}} & k_{3I_{2}} & k_{4I_{2}} & k_{5I_{2}} & k_{6I_{2}} & 0 & \breve{\cal L}_{I_{2}I_{2}} & k_{I_{3}I_{2}} & 0 & 0 & 0 & 0 \\[1mm]
    0 & 0 & 0 & 0 & 0 & 0 & 0 & 0 & k_{I_{2}I_{3}} & \breve{\cal L}_{I_{3}I_{3}} & k_{\alpha_{1}I_{3}} & \omega_{\alpha_{2}I_{3}} & 0 & 0 \\[1mm]
    0 & 0 & 0 & 0 & 0 & 0 & 0 & k_{I_{1}\alpha_{1}} & 0 & k_{I_{3}\alpha_{1}} & \breve{\cal L}_{\alpha_{1}\alpha_{1}} & \Gamma_{\alpha_{2}\alpha_{1}} & 0 & 0 \\[1mm]
    0 & 0 & 0 & 0 & 0 & 0 & 0 & \omega_{I_{1}\alpha_{2}} & 0 & \omega_{I_{3}\alpha_{2}} & \Gamma_{\alpha_{1}\alpha_{2}} & \breve{\cal L}_{\alpha_{2} \alpha_{2}} & 0 & 0 \\[1mm]
    0 & 0 & 0 & 0 & 0 & 0 & 0 & 0 & 0 & 0 & \Gamma_{\alpha_{1}\beta_{1}} & 0 & \breve{\cal L}_{\beta_{1}\beta_{1}} & \Gamma_{\beta_{2}\beta_{1}} \\[1mm]
    0 & 0 & 0 & 0 & 0 & 0 & 0 & 0 & 0 & 0 & 0 & \omega_{\alpha_{2}\beta_{2}} & \Gamma_{\beta_{1}\beta_{2}} & \breve{\cal L}_{\beta_{2}\beta_{2}} \\[1mm]
    \end{array} 
\right]
\left[ 
    \begin{array}{c}
    \rho_{g} \\[1.5mm]
    \rho_{X_{1}}\\[1.5mm]
    \rho_{X_{2}}\\[1.5mm]
    \rho_{X_{3}}\\[1.5mm]
    \rho_{X_{4}}\\[1.5mm]
    \rho_{X_{5}}\\[1.5mm]
    \rho_{X_{6}}\\[1.5mm]
    \rho_{I_{1}}\\[1.5mm]
    \rho_{I_{2}}\\[1.5mm]
    \rho_{I_{3}}\\[1.5mm]
    \rho_{\alpha_{1}}\\[1.5mm]
    \rho_{\alpha_{2}}\\[1.5mm]
    \rho_{\beta_{1}}\\[1.5mm]
    \rho_{\beta_{2}}\\[1.5mm]
    \end{array}
\right]
\end{equation}
\end{widetext}
which is of the form $\ket{\dot\rho_q}=\breve {\cal L}\ket{\rho_q}$, with the superoperator given by the $14\times 14 $ rate-matrix in Eq.~\ref{A9} and is of the Lindblad type akin to a Pauli master equation. Each diagonal element of the superoperator is a negative sum of its off-diagonal column ($\breve {\cal L}_{ii}=-\sum_{i\ne j}\breve{\cal L}_{ij})$.  Each rate (an element of the superoperator) is a direct result of the evaluation of the system and bath correlation functions of Eq. (\ref{eq-ide}). The rates $\gamma_{ex} n$ and $\gamma_{ex}(n+1)$ are a direct result of matter-field coupling with $\gamma_{ex}\propto|\sum_k g_{kX_1}|^2$ being the rate of photoexcitation involving the ground state $|g\rangle$ and the first excitonic state $|X_1\rangle$ populated as per the Bose-Einstein Distribution, $n$. 
The rates $r_{pq}, p\ne q$ representing transitions from the state $p$ to $q$ are the intra-exciton rates obtained by evaluating the bath correlation functions as per Redfield theory. The rates $k_{xy}$ are the charge transfer rates involving manybody states that differ by an electron which are obtained by calculating the bath correlation functions as per F$\ddot{o}$rster theory\cite{YANG2002355,runeson2024exciton}. The rates $\omega_{lm}$ are the cotunneling rates obtained by simplifying the bath correlation functions as per a perturbative regularisation procedure \cite{carmi2012enhanced}. The rates $\Gamma_{xy}$ are unidirectional electron transfer rates obtained using a standard perturbative technique \cite{harbola2006quantum}. While the unidirectional cotunneling rate $\omega_{\alpha_{2}\beta_{2}}$ is obtained using the Eq. ~\ref{D4} . More details shall follow in a later version. In the next section, we show explicitly, how we estimate each different type of rate.

\section{Redfield Theory of Excitonic Rates}\label{ED}

Exciton relaxation dynamics \cite{Vladimir} occur between the quantum states of the coupled chromophores when the number of electrons does not change in the manybody states. The dynamics are quantified by slightly modifying the Redfield theory describing the population-to-population transfer and the associated transfer rates between excitonic states $p$ and $q$ which belong to $\ket {X_n}$. Several rates between the excitonic states are obtained when Eq. (\ref{eq-exc-Ham}) enters Eq. (\ref{eq-ide}) through the bath correlation functions. A typical example of such a correlation function between two excitonic states $p$ and $q$ is,
\begin{align}
\langle p|tr_E\displaystyle\int_0^t \sum_kg_{kX_n}^b|\tilde X_n(t)\rangle\langle \tilde X_n(t')\tilde\rho_q(t)\tilde b_k^\dag(t')\tilde b_k(t)\rho_E|q\rangle,
\end{align}
with $\rho_q(t)=\rho_X(t)\rho_{q'}$, representing the density matrices of the excitons and the rest of the molecular components. In the above expression, assuming the source of excitation to be noisy and continuous, the relaxation to be dissipative happening over well separated timescales, it can be simplified to $r_{pq}\hat\rho_{p'q'}$ such that 
\begin{equation} \label{D1}
    r_{pq} = 2 \displaystyle\int_{0}^{\infty} e^{i(\omega_{pq} - \bar\lambda) t} e^{- \bar g(t))}\eta (t)dt
\end{equation}
with $\bar\lambda=(\lambda_{pppp} + \lambda_{qqqq} - 2\lambda_{qqpp}), \bar g(t)=g_{pppp}(t) - g_{qqqq}(t) + 2g_{qqpp}(t)$ and $\eta(t)=(\ddot{g}_{qpqp}(t) - (\dot{g}_{qpqq}(t) - \dot{g}_{qppp}(t) - 2i\lambda_{qpqq})^{2})$.    
 $\lambda_{pppp} = \sum_{i} c_{i}(p)c_{i}(p)c_{p}(s)c_{p}(t)\lambda_{i}$ and $g_{pppp} = \sum_{i}c_{i}(p)c_{i}(p)c_{i}(p)c_{i}(p)g_{i}(t)$ are the re-organization energies and the line broadening functions respectively. $c_{i}(p)$ is the amplitude of exciton $p$ on site $i$ while $\lambda_{i}$ is the re-organization energy of site $i$ and $g_{i}(t)$ is the line broadening function defined for site. The states $p, q \in \ket{X_n}, n= 1, 2, 3, 4, 5, 6$ such that while calculating the rates, $p \neq q$.

\section{F$\ddot{o}$rster Rates of Charge Transfer }\label{FD}
The transfer rates between the charge-separated states (among the $N$-electron manybody states) can be estimated using the F$\ddot{o}$rster theory\cite{YANG2002355}. It is equivalent to the Marcus rates and describes incoherent electron transfer rates. 
Similar to the previous case, several rate terms are obtained when Eq.~(\ref{eq-fors-ham})
enters Eq.~(\ref{eq-ide}) through the correlation functions. A typical example of such correlation functions between two charge-separated states $x$ and $y$ is,
\begin{align}
\langle x|tr_E\displaystyle\int_0^t |t_{I_1X_n}|^2\tilde c_{I_1}^\dag(t) \tilde c_{I_1}(t')\tilde\rho_q(t)|\tilde X_{1}(t')\tilde X_{1}(t)\rho_E|y\rangle, 
\end{align}
such that the above term represents $k_{xy}\rho_{x'y'}$. The general form of the F$\ddot{o}$rster rate expression is approximated to:
\begin{equation}\label{C2}
    k_{xy} = \left|t_{xy}\right|^{2}S_{xy}
\end{equation}
where $t_{xy}$ is the electronic coupling between states x and y, and $S_{xy}$ is the spectral overlap between the two states across which the transfer occurs. The indices x and y = 1, 2, 3, 4, 5, 6, $I_{1}$, $I_{2}$, $I_{3}$, $\alpha_{1}$ such that while calculating the F$\ddot{o}$rster like rates and the spectral overlap x $\neq$ y.
The spectral overlap $S_{xy}$ is expressed as:
\begin{equation} \label{C3}
    S_{xy}^{} = 2\mathbb{R}\int_{0}^{\infty}dte^{iw_{xy}t}e^{-i(\lambda_x + \lambda_y)t-(g_{x}(t)+g_{y}(t))}
\end{equation}
which in the frequency domain reads,
\begin{equation} \label{C4}
    S_{xy}^{} = \frac{1}{2\pi}\int_{-\infty}^{\infty}d\omega\overline{D}_{x}(\omega)D_{y}(\omega)
\end{equation}
where $\overline{D}_{x}(\omega) = 2\mathbb{R}\int_{0}^{\infty}dte^{i \omega t}e^{-i\omega_{xy}t + i \lambda_{x}t-g_{x}^{*}(t)}$ and $\overline{D}_{y}(\omega) = 2\mathbb{R}\int_{0}^{\infty}dte^{i \omega t}e^{-i\omega_{xy}t - i \lambda_{x}t-g_{x}(t)}$ are the respective fluorescence and absorption lineshapes. $k_{\beta_2g}$ can be obtained in a similar way if the spectral profile is know. In the current case, we simply treat it as a parameter and is fixed to be equal to $205cm^{-1}$.

\section{Unidirectional and Cotunneling Rates}\label{CT}
The presence of the negatively charged state allows for a cotunneling process in which there is a simultaneous electron transfer from states $\ket{I_{1}}$ and $\ket{I_{3}}$ to $\ket{\alpha_{2}}$. The dynamics of various tunneling \cite{harbola2006quantum, goswami2015electron} and cotunneling \cite{carmi2012enhanced,golovach2004transport,cabrera2023charge} electronic systems have been studied before using the quantum master framework based on second and fourth order perturbation theories respectively. A typical correlation function between two charge-separated states $l$ and $m$ ($l,m\in \ket{\beta_{1}}, \ket{\beta_{2}}, \ket{g}, \ket{\alpha_1}, \ket{\alpha_2}$) where we resort to second order perturbation theory is,
\begin{align}
\langle l|tr_E\displaystyle\int_0^t |t_{g\beta}|^2\tilde c_{g}^\dag(t) \tilde c_{\beta}(t')\tilde\rho_q(t)\tilde c_{d}(t')\tilde c_{d}(t)\rho_E|m\rangle
\end{align}
In the above expression, we do a second order perturbation on the coupling term $\hat V_e$ containing information only on the $l,m$ manybody states. Performing a standard Born-Markov approximation, assuming large bias and completely tracing over the reservoir degree of freedom, the above expression reduces to $\Gamma_{xy}\rho_{x'y'}$, with $\Gamma_{xy}=|t_{g\beta_{j}}|^2$. Likewise, when $l$ and $m$ $ \in  \ket{I_2}, \ket{I_3}, \ket{\alpha_2}$, we perform a fourth order perturbation on the Hamiltonian components of $\hat V_e$ which possess information only on these specific $l,m$ manybody states. 
In this case, the fourth order perturbative expansion reduces to $\omega_{lm}\rho_{l'm'}.$The rates, $\omega_{lm}$ connect the manybody states differing by two electrons and are
\begin{widetext}
\begin{small}
\begin{equation} \label{D2}
\begin{split}
    \omega_{I_{z}\alpha_{2}} = \frac{2\pi\nu^{2}}{\hbar}\int f(\epsilon - E_{I_{1}})f[(-\epsilon + E'_{\alpha_{1}} + E'_{\alpha_{2}} + U) - E_{I_{3}}]\left|\frac{t_{I_{1}\alpha_{2}}t_{I_{3}\alpha_{2}}}{\epsilon - E'_{\alpha_{1}}} - \frac{t_{I_{1}\alpha_{2}}t_{I_{3}\alpha_{2}}}{\epsilon - E'_{\alpha_{1}} + U} - \frac{t_{I_{1}\alpha_{2}}t_{I_{3}\alpha_{2}}}{\epsilon - E'_{\alpha_{2}}} + \frac{t_{I_{1}\alpha_{2}}t_{I_{3}\alpha_{2}}}{\epsilon - E'_{\alpha_{2}} + U} \right|^{2} d\epsilon\\ + \frac{2\pi\nu^{2}}{\hbar}\int f(\epsilon - E_{I_{3}})f[(-\epsilon + E'_{\alpha_{1}} + E'_{\alpha_{2}} + U) - E_{I_{1}}]\left|\frac{t_{I_{1}\alpha_{2}}t_{I_{3}\alpha_{2}}}{\epsilon - E'_{\alpha_{1}}} - \frac{t_{I_{1}\alpha_{2}}t_{I_{3}\alpha_{2}}}{\epsilon - E'_{\alpha_{1}} + U} - \frac{t_{I_{1}\alpha_{2}}t_{I_{3}\alpha_{2}}}{\epsilon - E'_{\alpha_{2}}} + \frac{t_{I_{1}\alpha_{2}}t_{I_{3}\alpha_{2}}}{\epsilon - E'_{\alpha_{2}} + U} \right|^{2} d\epsilon
\end{split}
\end{equation}

\begin{equation} \label{D3}
\begin{split}
     \omega_{\alpha_{2}I_{z}}^{} = \frac{2\pi\nu^{2}}{\hbar}\int f(E_{I_{1}} - \epsilon)f[E_{I_{3}}-(-\epsilon + E'_{\alpha_{1}} + E'_{\alpha_{2}} + U)]\left|\frac{t_{I_{1}\alpha_{2}}t_{I_{3}\alpha_{2}}}{\epsilon - E'_{\alpha_{1}}} - \frac{t_{I_{1}\alpha_{2}}t_{I_{3}\alpha_{2}}}{\epsilon - E'_{\alpha_{1}} - U} - \frac{t_{I_{1}\alpha_{2}}t_{I_{3}\alpha_{2}}}{\epsilon - E'_{\alpha_{2}}} + \frac{t_{I_{1}\alpha_{2}}t_{I_{3}\alpha_{2}}}{\epsilon - E'_{\alpha_{2}} - U} \right|^{2} d\epsilon\\ + \frac{2\pi\nu^{2}}{\hbar}\int f(E_{I_{3}} - \epsilon)f[E_{I_{1}} -(-\epsilon + E'_{\alpha_{1}} + E'_{\alpha_{2}} + U)]\left|\frac{t_{I_{1}\alpha_{2}}t_{I_{3}\alpha_{2}}}{\epsilon - E'_{\alpha_{1}}} - \frac{t_{I_{1}\alpha_{2}}t_{I_{3}\alpha_{2}}}{\epsilon - E'_{\alpha_{1}} - U} - \frac{t_{I_{1}\alpha_{2}}t_{I_{3}\alpha_{2}}}{\epsilon - E'_{\alpha_{2}}} + \frac{t_{I_{1}\alpha_{2}}t_{I_{3}\alpha_{2}}}{\epsilon - E'_{\alpha_{2}} - U} \right|^{2} d\epsilon
\end{split}
\end{equation}

\begin{equation} \label{D4}
\begin{split}
    \omega_{\alpha_{2}\beta_{2}} = \frac{2\pi\nu^{2}}{\hbar}\int \frac{1}{2}F_{FD}(E_{\beta_{2}} - \epsilon)F_{FD}[E_{\beta_{2}} - (-\epsilon + E_{\alpha'_{1}} + E_{\alpha'_{2}} + U)]\left|\frac{t_{\alpha_{2}\beta_{2}}t_{\alpha_{2}\beta_{2}}}{\epsilon - E_{\alpha'_{1}}} - \frac{t_{\alpha_{2}\beta_{2}}t_{\alpha_{2}\beta_{2}}}{\epsilon - E_{\alpha'_{1}} - U} - \frac{t_{\alpha_{2}\beta_{2}}t_{\alpha_{2}\beta_{2}}}{\epsilon - E_{\alpha'_{2}}} + \frac{t_{\alpha_{2}\beta_{2}}t_{\alpha_{2}\beta_{2}}}{\epsilon - E_{\alpha'_{2}} - U} \right|^{2} d\epsilon
    \end{split}
\end{equation}
\end{small}
\end{widetext}
where the index z = 1, 3, $f(x-y(t)) = (1 + e^{\beta(x - y(t))})^{-1}$ represents the Fermi function, $t_{I_{z}\alpha_{2}}$ represents the tunneling coefficients from states $I_{1}$ and $I_{3}$, U represents the additional charging energy due to Coulomb interaction, $E_{I_{1}}$ and $E_{I_{3}}$ represents the energies of the states $I_{1}$ and $I_{3}$ respectively and $\nu$ represents the density of states.

\section{Other Relevant Information}\label{CP}

The line-broadening functions are defined 
\begin{equation} \label{B4}
    g(t) = g_{D}(t) + \sum_{k}g_{k}(t)
\end{equation}
where $g_{D}(t)$ and $g_{k}(t)$ are the line broadening functions for the Drude mode and the $j^{th}$ underdamped mode respectively, which are defined as:
    \begin{align} \label{B5}
        g_{D}(t) &= \frac{c_{o}^{D}}{\Omega_{D}^{2}}(e^{-\Omega_{D}t} + \Omega_{D}t - 1) + \sum_{k=1}^{\infty}\frac{c_{k}^{D}}{\nu_{k}^{2}}(e^{-\nu_{k}t} + \nu_{k}t -1)\\
\label{B6}
        g_{k}(t) &= \sum_{+,-}\frac{c_{0j}^{\pm}}{\nu_{j\pm}^{2}}(e^{-\nu_{j \pm}t} \!+\! \nu_{j \pm}t \!-\! 1) + \sum_{k=1}^{\infty}\frac{c_{kj}}{\nu_{k}^{2}}(e^{-\nu_{k}t} \!+\! \nu_{k}t \!-\! 1)
    \end{align}
    where $c_{o}^{D} = \lambda_{D}\Omega_{D}(cot(\frac{\beta\Omega_D}{2})-i)$, $c_{k}^{D} = \frac{4\lambda_D\Omega_D}{\beta}(\frac{\nu_k}{\nu_{k}^{2}-\Omega_{D}^{2}})$, $c_{0j}^{\pm} = \pm i\frac{\lambda_{k} \omega_{k}^{2}}{2 \zeta_{k}}(cot(\frac{\beta \nu_{\pm}}{2}) - i)$, $c_{kj} = \frac{-4 \lambda_{k} \gamma_{k} \omega^{2}}{\beta}(\frac{\nu_{k}}{(\omega_{k}^{2} + \nu_{k}^{2})^{2} - (\gamma_{k}^{2} \nu_{k}^{2})})$, $\nu_{j \pm} = \frac{\gamma_{k}}{2} \pm i \zeta_{k}$ and $\zeta_{k} = \sqrt{\omega_{k}^{2} - \frac{\gamma_{k}^{2}}{4}}$. 
Also, $\nu_{k} = \frac{2 \pi}{\beta}k$, such that depending on the temperature the summations over $\nu_{k}$ can be truncated at a suitable value of k.

The primary charge transfer involves the transfer from any one of the exciton states ($\ket{X}$) to the charge-separated states $\ket{I_{1}}$ or $\ket{I_{2}}$. The electronic coupling between $\ket{X}$ and $\ket{I_{2}}$ is $V_{XI_{2}} = \sum_{n\in X}c_{n}(X)V_{nI_{2}}$ where $V_{nI_{2}}$ is the electronic coupling between site n and state $\ket{I_{2}}$, and $c_{n}(X)$ is the amplitude of site n in exciton $\ket{X}$. Similarly, the coupling between $\ket{X}$ and $\ket{I_{1}}$ is calculated by the same formula as above. The re-organization energies of the exciton can be defined as $\lambda_{X} = \sum_{i}|c_{i}(X)|^{4}\lambda$ and the line broadening functions as $g_{X}(t) = \sum_{i}|c_{i}(X)|^{4}g(t)$ where $c_{i}(X)$ represents the amplitudes of exciton $\ket{X_n}$ at site i; $\lambda$ and g(t) are the site-reorganisation energy and line broadening function. The line broadening function and the re-organization energy of the charge-separated states are $g_{I}(t) = \nu_{I}g(t)$ and $\lambda_{I} = \nu_{I}\lambda$, $I \in \{ \ket{I_{1}}, \ket{I_{2}} \}$. The rest of the charge transfer rates can be calculated similarly.
The values, $\lambda_{D} = 35 cm^{-1}$ and $\Omega_{D} = 40cm^{-1}$ are used for the calculation of $g_{D}(t)$. The frequencies and Huang-Rhys factors needed for the calculation of $g_{k}(t)$ are given in Table 2; the exciton states and associated energies are given in Table 3; the energies of the 6 co-factors along with the charge transfer states with their scaling factors are given in Table 4. The couplings between the 6 co-factors, couplings between the excitons and charge-separated states  in $cm^{-1}$ are given in Table 1.

\begin{table*}[t!]
\centering
\[
\begin{array}{c|cccccccccc}
    & P_{D1} & P_{D2} & Chl_{D2} & Phe_{D1} & Phe_{D2} & Chl_{D1} & P_{D2}^{+}P_{D1}^{-} & Chl_{D1}^{+}Phe_{D1}^{-} & P_{D1}^{+}Chl_{D1}^{-} & P_{D1}^{+}Phe_{D1}^{-} \\
    \hline
    P_{D1} &  & 150 & -55 & -6 & 17 & -42 & 45 & 0 & 0 & 0\\ 
    P_{D2} &  &  & -36 & 20 & -2 & -56 & 45 & 0 & 0 & 0\\
    Chl_{D2} &  &  &  & -5 & 37 & 7 & 0 & 0 & 0 & 0\\
    Phe_{D1} &  &  &  &  & -3 & 46 & 0 & 70 & 0 & 0\\
    Phe_{D2} &  &  &  &  &  & -4 & 0 & 0 & 0 & 0\\ 
    Chl_{D1} &  &  &  &  &  &  & 0 & 70 & 0 & 0\\ 
    P_{D2}^{+}P_{D1}^{-} &  &  &  &  &  &  &  & 0 & 70 & 0\\ 
    Chl_{D1}^{+}Phe_{D1}^{-} &  &  &  &  &  &  &  &  & 0 & 40\\ 
    P_{D1}^{+}Chl_{D1}^{-} &  &  &  &  &  &  &  &  &  & 40\\ 
    P_{D1}^{+}Phe_{D1}^{-} &  &  &  &  &  &  &  &  &  & \\
\end{array}
\]
\caption{Couplings among the 6 co-factors,  between the excitons and charge-separated states and charge-separated-charge-separated couplings in $cm^{-1}$(data from Raszewski et al. \cite{Raszewski2005TheoryOO}).}
\label{Table 1}
\end{table*}

\begin{table*}[t!]
\centering
\[
\begin{array}{c|c||c|c||c|c||c|c||c|c}
    \omega_{k} & s_{k} & \omega_{k} & s_{k} & \omega_{k} & s_{k} & \omega_{k} & s_{k} & \omega_{k} & s_{k} \\
    \hline \hline
    97 & 0.0371 & 573 & 0.0094 & 995 & 0.0293 & 1252 & 0.0051 & 1524 & 0.0067\\ 
    138 & 0.0455 & 585 & 0.0034 & 1052 & 0.0131 & 1260 & 0.0064 & 1537 & 0.0222\\
    213 & 0.0606 & 604 & 0.0034 & 1069 & 0.0064 & 1286 & 0.0047 & 1553 & 0.0091\\
    260 & 0.0539 & 700 & 0.005  & 1110 & 0.0192 & 1304 & 0.0057 & 1573 & 0.0044\\
    298 & 0.0488 & 722 & 0.0074 & 1143 & 0.0303 & 1322 & 0.0202 & 1580 & 0.0044\\
    342 & 0.038 & 742 & 0.0269 & 1181 & 0.0179 & 1338 & 0.0037 & 1612 & 0.0044\\
    388 & 0.0202 & 752 & 0.0219 & 1190 & 0.0084 & 1354 & 0.0057 & 1645 & 0.0034\\
    425 & 0.0168 & 795 & 0.0077 & 1208 & 0.0121 & 1382 & 0.0067 & 1673 & 0.001\\
    518 & 0.0303 & 916 & 0.0286 & 1216 & 0.0111 & 1439 & 0.0067 & &\\
    546 & 0.0030 & 986 & 0.0162 & 1235 & 0.0034 & 1487 & 0.0074 & &\\
\end{array}
\]
\caption{Frequencies (cm$^{-1}$) and Huang-Rhys factors for the 48 modes of PSIIRC (data taken from Stones et al. \cite{C7SC02983G}).}
\label{Table 2}
\end{table*}

\begin{table*}[t!]
\centering
\[
\begin{array}{c | c  c  c  c  c  c  c} 
  & P_{D1} & P_{D2} & Chl_{D1} & Chl_{D2} & Phe_{D1} & Phe_{D2}  & \textbf{E}(cm^{-1}) \\ [0.5ex] 
 \hline
 X_{1} & -0.029 & 0.251 & 0.8 & -0.018 & -0.541 & 0.048 & 14412.5\\ 
 X_{2} & -0.185 & 0.057 & -0.061 & -0.404 & 0.04 & 0.891 & 14459.7\\
 X_{3} & 0.265 & -0.145 & 0.56 & 0.112 & 0.754 & 0.12 & 14571.0\\
 X_{4} & -0.558 & 0.735 & -0.007 & 0.116  & 0.345 & -0.126 & 14540.8\\
 X_{5} & 0.134 & 0.096 & -0.121 & 0.876 & -0.134 & 0.416 & 14565.0\\
 X_{6} & -0.752 & -0.603 & 0.165 & 0.209 & -0.003 & -0.012 & 14864.0\\[1ex]
\end{array}
\]
\caption{Site amplitudes and associated energies of the six exciton states of the PSIIRC (data from Stones et al. \cite{C7SC02983G}).}
\label{Table 3}
\end{table*}

\begin{table*}[t!]
\centering
\[
\begin{array}{c | c c c c} 
  & \textbf{E}(cm^{-1}) & - \lambda (cm^{-1}) & (scaling factor)_{\lambda}& E_i\\ [0.5ex] 
 \hline
 Chl_{D1}^{+}Phe_{D1}^{-}(\ket{I_{1}}) & 15992 & 1620 & 3 & 14372\\
 P_{D2}^{+}P_{D1}^{-}(\ket{I_{2}}) & 15182 & 810 & 1.5 & 14372\\
 P_{D1}^{+}Chl_{D1}^{-}(\ket{I_{3}}) & 15842 & 1620 & 3 & 14222 \\
 P_{D1}^{+}Phe_{D1}^{-}(\alpha_{1}) & 16132 & 2160 & 4 & 13972\\
 \hline
 \textbf{E}_{\alpha'_{2}} > \textbf{E}_{\alpha'_{1}} \\
 \hline
 \ket{\alpha'_{1}} & - & - & - & 13972\\
 \ket{\alpha'_{2}} & - & - & - & 15670\\
 \hline
 \textbf{E}_{\alpha'_{2}} < \textbf{E}_{\alpha'_{1}} \\
 \hline
 \ket{\alpha'_{1}} & - & - & - & 13972\\
 \ket{\alpha'_{2}} & - & - & - & 11670\\
 [1ex]
\end{array}
\]
\caption{Electronic excitation energies and re-organization energies of the 6 charge-separated states along with the scaling factors (data from Novoderezhkin et al. \cite{Vladimir}).}
\label{Table 4}
\end{table*}

\FloatBarrier
\begin{figure}[h!]
\centering
\begin{minipage}{0.494\columnwidth}
    \centering
    \includegraphics[width=4cm, height=3.5cm]{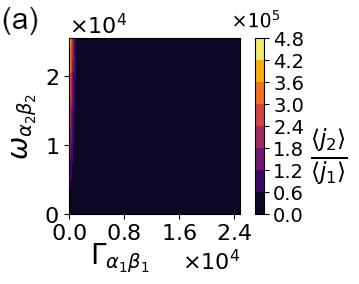} 
\end{minipage}
\begin{minipage}{0.494\columnwidth}
    \centering
    \includegraphics[width=4cm, height=3.5cm]{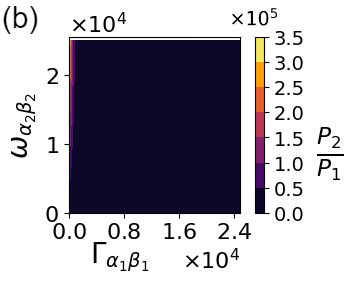}
\end{minipage}

\vspace{0.1cm} 

\begin{minipage}{0.494\columnwidth}
    \centering
    \includegraphics[width=4cm, height=3.5cm]{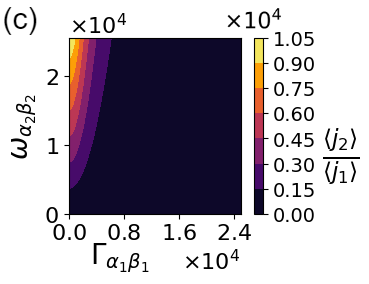}
\end{minipage}
\begin{minipage}{0.494\columnwidth}
    \centering
    \includegraphics[width=4cm, height=3.5cm]{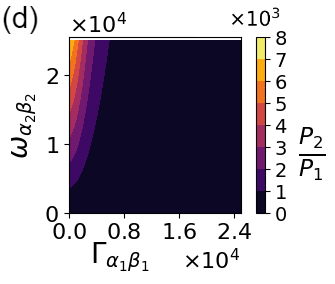}
\end{minipage}

\vspace{0.1cm}

\begin{minipage}{0.494\columnwidth}
    \centering
    \includegraphics[width=4cm, height=3.5cm]{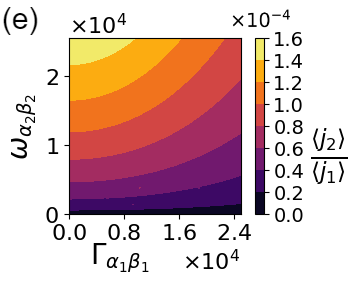} 
\end{minipage}
\begin{minipage}{0.494\columnwidth}
    \centering
    \includegraphics[width=4cm, height=3.5cm]{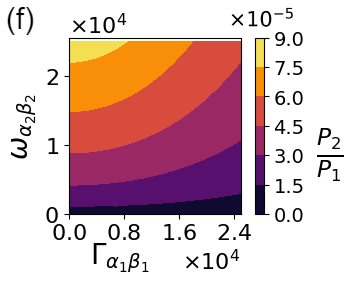}
\end{minipage}

\vspace{0.1cm} 

\begin{minipage}{0.494\columnwidth}
    \centering
    \includegraphics[width=4cm, height=3.5cm]{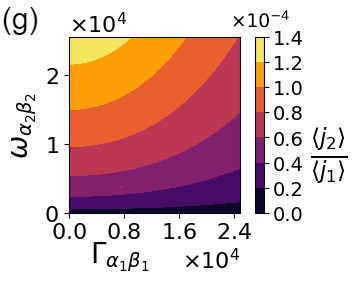}
\end{minipage}
\begin{minipage}{0.494\columnwidth}
    \centering
    \includegraphics[width=4cm, height=3.5cm]{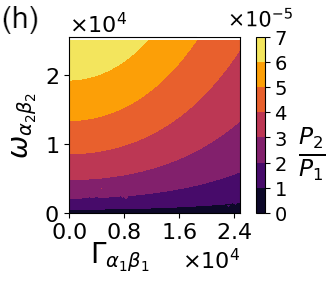}
\end{minipage}

\caption{\justifying Contour plots of ${\langle j_{2} \rangle}/{\langle j_{1} \rangle}$ and ${P_{2}}/{P_{1}}$ when $t_{I_{1}\alpha_{1}} \approx t_{I_{3}\alpha_{2}}$ with $E_{\alpha'_{2}} > E_{\alpha'_{1}}$ and $\Gamma_{\alpha_{2}\alpha_{1}} > \Gamma_{\alpha_{1}\alpha_{2}}$. (a), (b): $U = -3000 cm^{-1}$; (c), (d): $U = -2000 cm^{-1}$; (e), (f): $U = 2000 cm^{-1}$ and (g), (h): $U = 3000 cm^{-1}$.}
\label{fig: 10}
\end{figure}

\begin{figure}[h!]
\centering
\begin{minipage}{0.494\columnwidth}
    \centering
    \includegraphics[width=4cm, height=3.5cm]{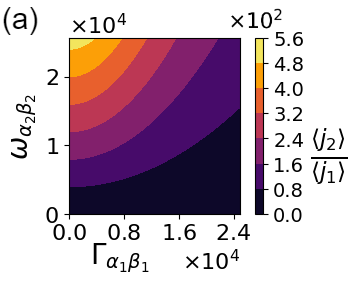} 
\end{minipage}
\begin{minipage}{0.494\columnwidth}
    \centering
    \includegraphics[width=4cm, height=3.5cm]{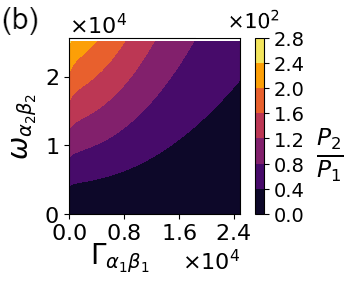}
\end{minipage}

\vspace{0.1cm} 

\begin{minipage}{0.494\columnwidth}
    \centering
    \includegraphics[width=4cm, height=3.5cm]{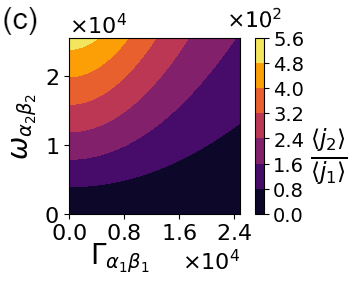}
\end{minipage}
\begin{minipage}{0.494\columnwidth}
    \centering
    \includegraphics[width=4cm, height=3.5cm]{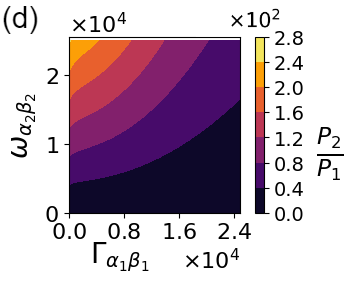}
\end{minipage}

\vspace{0.1cm}

\begin{minipage}{0.494\columnwidth}
    \centering
    \includegraphics[width=4cm, height=3.5cm]{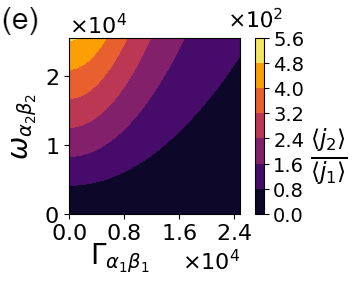} 
\end{minipage}
\begin{minipage}{0.494\columnwidth}
    \centering
    \includegraphics[width=4cm, height=3.5cm]{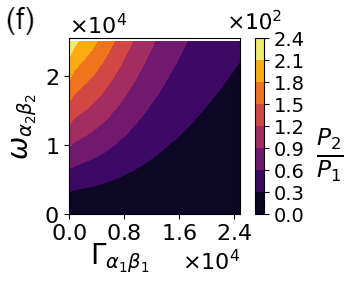}
\end{minipage}

\vspace{0.1cm} 

\begin{minipage}{0.494\columnwidth}
    \centering
    \includegraphics[width=4cm, height=3.5cm]{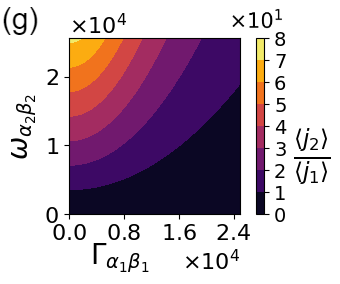}
\end{minipage}
\begin{minipage}{0.494\columnwidth}
    \centering
    \includegraphics[width=4cm, height=3.5cm]{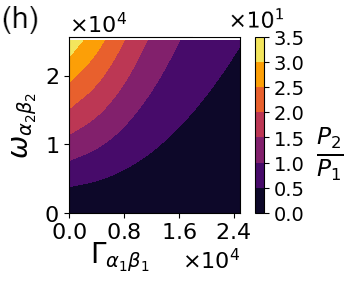}
\end{minipage}

\caption{\justifying Contour plots of ${\langle j_{2} \rangle}/{\langle j_{1} \rangle}$ and ${P_{2}}/{P_{1}}$ when $t_{I_{1}\alpha_{1}} \approx t_{I_{3}\alpha_{2}}$ with $E_{\alpha'_{2}} < E_{\alpha'_{1}}$ and $\Gamma_{\alpha_{2}\alpha_{1}} > \Gamma_{\alpha_{1}\alpha_{2}}$. (a), (b): $U = -3000 cm^{-1}$; (c), (d): $U = -2000 cm^{-1}$; (e), (f): $U = 2000 cm^{-1}$ and (g), (h): $U = 3000 cm^{-1}$.}
\label{fig: 11}
\end{figure}

\begin{figure}[h!]
\centering
\begin{minipage}{0.494\columnwidth}
    \centering
    \includegraphics[width=4cm, height=3.5cm]{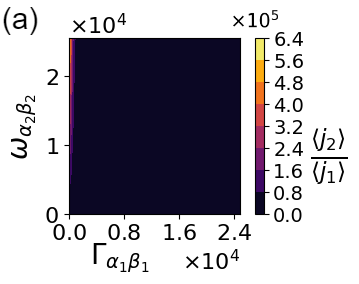}
\end{minipage}
\begin{minipage}{0.494\columnwidth}
    \centering
    \includegraphics[width=4cm, height=3.5cm]{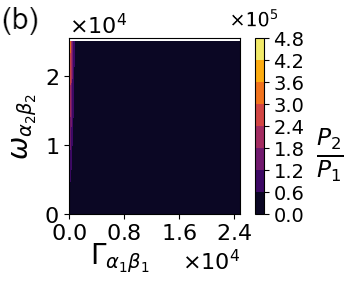}
\end{minipage}

\vspace{0.1cm}

\begin{minipage}{0.494\columnwidth}
    \centering
    \includegraphics[width=4cm, height=3.5cm]{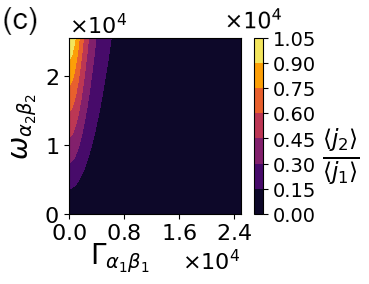} 
\end{minipage}
\begin{minipage}{0.494\columnwidth}
    \centering
    \includegraphics[width=4cm, height=3.5cm]{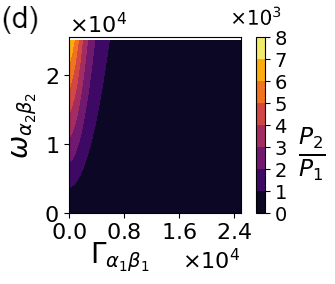}
\end{minipage}

\vspace{0.1cm} 

\begin{minipage}{0.494\columnwidth}
    \centering
    \includegraphics[width=4cm, height=3.5cm]{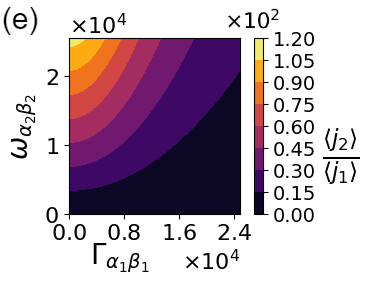}
\end{minipage}
\begin{minipage}{0.494\columnwidth}
    \centering
    \includegraphics[width=4cm, height=3.5cm]{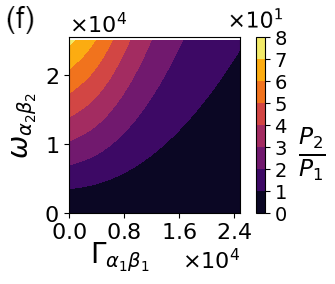}
\end{minipage}

\vspace{0.1cm} 

\begin{minipage}{0.494\columnwidth}
    \centering
    \includegraphics[width=4cm, height=3.5cm]{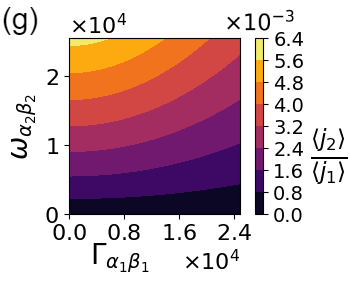}
\end{minipage}
\begin{minipage}{0.494\columnwidth}
    \centering
    \includegraphics[width=4cm, height=3.5cm]{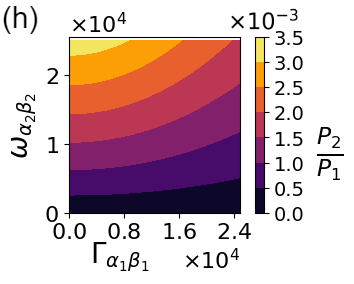}
\end{minipage}

\vspace{0.1cm}

\begin{minipage}{0.494\columnwidth}
    \centering
    \includegraphics[width=4cm, height=3.5cm]{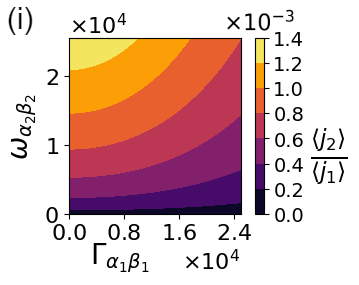} 
\end{minipage}
\begin{minipage}{0.494\columnwidth}
    \centering
    \includegraphics[width=4cm, height=3.5cm]{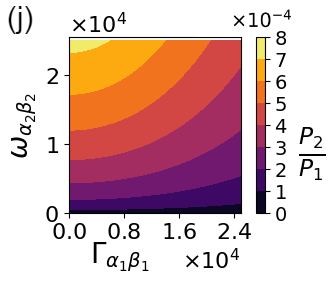}
\end{minipage}

\vspace{0.1cm} 

\begin{minipage}{0.494\columnwidth}
    \centering
    \includegraphics[width=4cm, height=3.5cm]{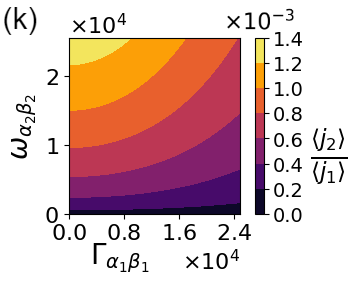}
\end{minipage}
\begin{minipage}{0.494\columnwidth}
    \centering
    \includegraphics[width=4cm, height=3.5cm]{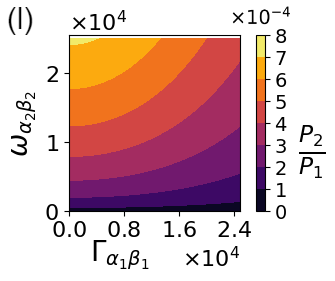}
\end{minipage}

\caption{\justifying Contour plots of ${\langle j_{2} \rangle}/{\langle j_{1} \rangle}$ and ${P_{2}}/{P_{1}}$ when $t_{I_{1}\alpha_{1}} \approx t_{I_{3}\alpha_{2}}$ with $E_{\alpha'_{2}} > E_{\alpha'_{1}}$ and $\Gamma_{\alpha_{2}\alpha_{1}} < \Gamma_{\alpha_{1}\alpha_{2}}$. (a), (b): $U = -3000 cm^{-1}$; (c), (d): $U = -2000 cm^{-1}$; (e), (f): $U = -1000 cm^{-1}$, (g), (h): $U = 1000 cm^{-1}$, (i), (j): $U = 2000 cm^{-1}$ and (k), (l): $U = 3000 cm^{-1}$.}
\label{fig: 12}
\end{figure}

\begin{figure}[h!] 
\centering
\begin{minipage}{0.494\columnwidth}
    \centering
    \includegraphics[width=4cm, height=3.5cm]{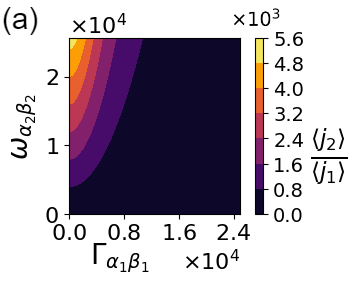}
\end{minipage}
\begin{minipage}{0.494\columnwidth}
    \centering
    \includegraphics[width=4cm, height=3.5cm]{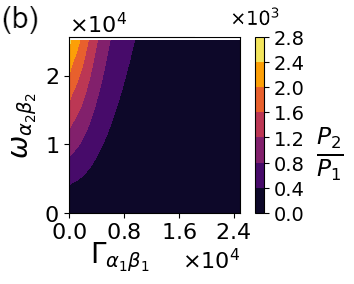}
\end{minipage}

\vspace{0.1cm}

\begin{minipage}{0.494\columnwidth}
    \centering
    \includegraphics[width=4cm, height=3.5cm]{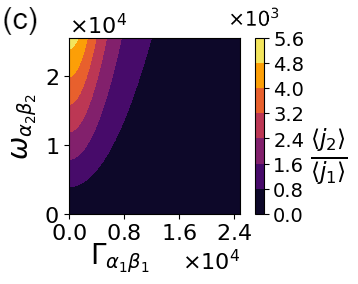} 
\end{minipage}
\begin{minipage}{0.494\columnwidth}
    \centering
    \includegraphics[width=4cm, height=3.5cm]{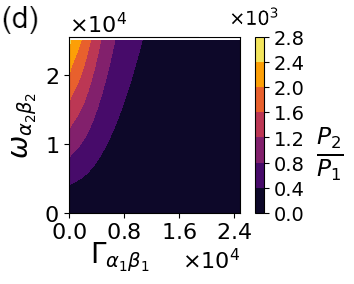}
\end{minipage}

\vspace{0.1cm} 

\begin{minipage}{0.494\columnwidth}
    \centering
    \includegraphics[width=4cm, height=3.5cm]{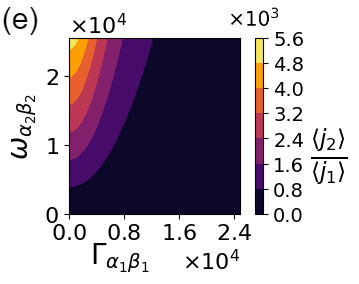}
\end{minipage}
\begin{minipage}{0.494\columnwidth}
    \centering
    \includegraphics[width=4cm, height=3.5cm]{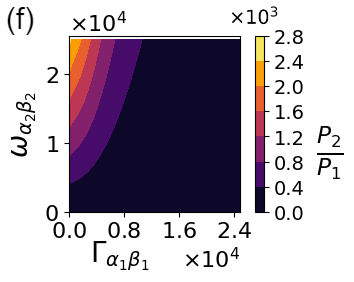}
\end{minipage}

\vspace{0.1cm} 

\begin{minipage}{0.494\columnwidth}
    \centering
    \includegraphics[width=4cm, height=3.5cm]{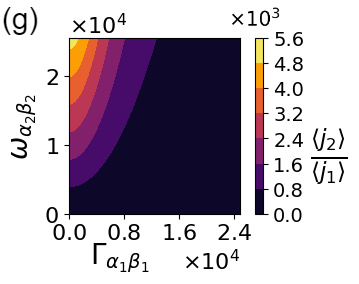}
\end{minipage}
\begin{minipage}{0.494\columnwidth}
    \centering
    \includegraphics[width=4cm, height=3.5cm]{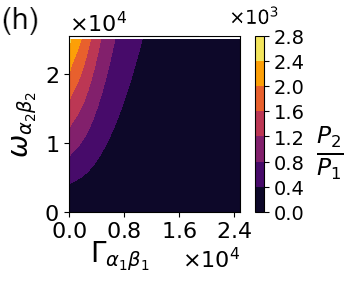}
\end{minipage}

\vspace{0.1cm}

\begin{minipage}{0.494\columnwidth}
    \centering
    \includegraphics[width=4cm, height=3.5cm]{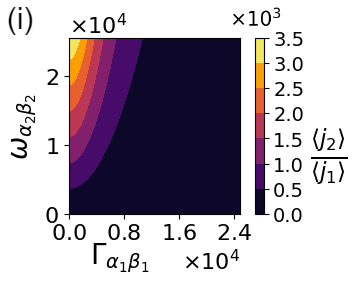} 
\end{minipage}
\begin{minipage}{0.494\columnwidth}
    \centering
    \includegraphics[width=4cm, height=3.5cm]{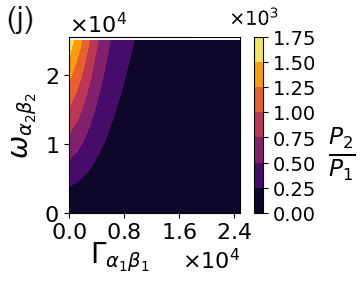}
\end{minipage}

\vspace{0.1cm} 

\begin{minipage}{0.494\columnwidth}
    \centering
    \includegraphics[width=4cm, height=3.5cm]{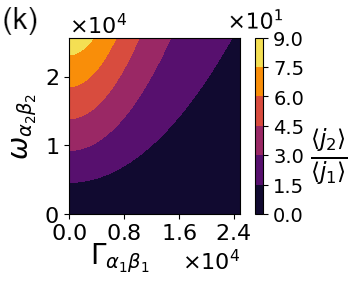}
\end{minipage}
\begin{minipage}{0.494\columnwidth}
    \centering
    \includegraphics[width=4cm, height=3.5cm]{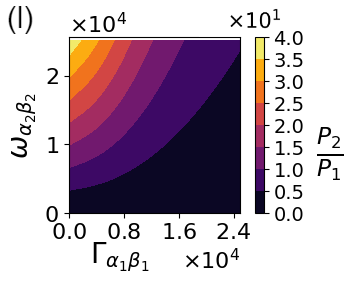}
\end{minipage}

\caption{\justifying Contour plots of ${\langle j_{2} \rangle}/{\langle j_{1} \rangle}$ and ${P_{2}}/{P_{1}}$ when $t_{I_{1}\alpha_{1}} \approx t_{I_{3}\alpha_{2}}$ with $E_{\alpha'_{2}} < E_{\alpha'_{1}}$ and $\Gamma_{\alpha_{2}\alpha_{1}} < \Gamma_{\alpha_{1}\alpha_{2}}$. (a), (b): $U = -3000 cm^{-1}$; (c), (d): $U = -2000 cm^{-1}$; (e), (f): $U = -1000 cm^{-1}$, (g), (h): $U = 1000 cm^{-1}$, (i), (j): $U = 2000 cm^{-1}$ and (k), (l): $U = 3000 cm^{-1}$.}
\label{fig: 13}
\end{figure}

\begin{figure}[h!]
\centering
\begin{minipage}{0.494\columnwidth}
    \centering
    \includegraphics[width=4cm, height=3.5cm]{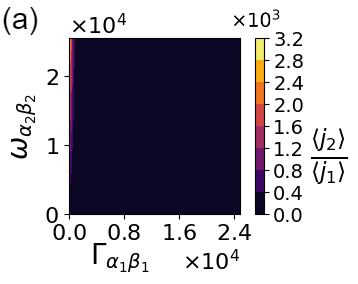} 
\end{minipage}
\begin{minipage}{0.494\columnwidth}
    \centering
    \includegraphics[width=4cm, height=3.5cm]{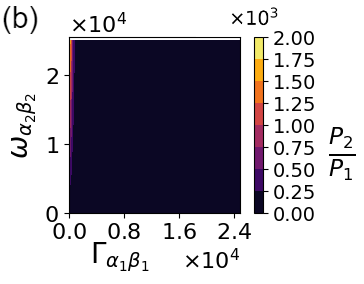}
\end{minipage}

\vspace{0.1cm} 

\begin{minipage}{0.494\columnwidth}
    \centering
    \includegraphics[width=4cm, height=3.5cm]{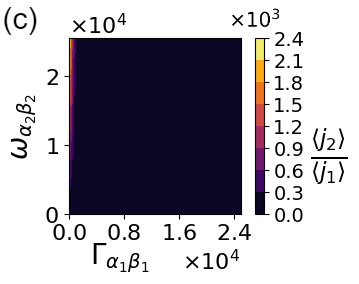}
\end{minipage}
\begin{minipage}{0.494\columnwidth}
    \centering
    \includegraphics[width=4cm, height=3.5cm]{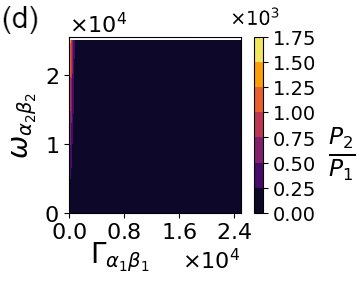}
\end{minipage}

\vspace{0.1cm}

\begin{minipage}{0.494\columnwidth}
    \centering
    \includegraphics[width=4cm, height=3.5cm]{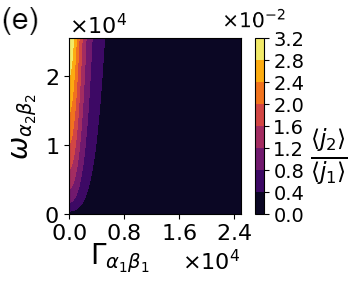} 
\end{minipage}
\begin{minipage}{0.494\columnwidth}
    \centering
    \includegraphics[width=4cm, height=3.5cm]{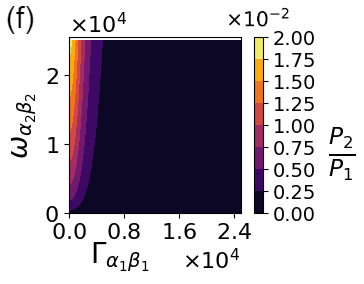}
\end{minipage}

\vspace{0.1cm} 

\begin{minipage}{0.494\columnwidth}
    \centering
    \includegraphics[width=4cm, height=3.5cm]{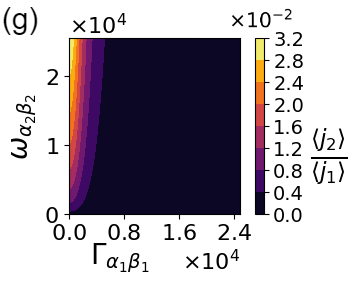}
\end{minipage}
\begin{minipage}{0.494\columnwidth}
    \centering
    \includegraphics[width=4cm, height=3.5cm]{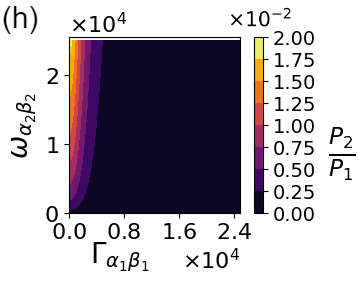}
\end{minipage}

\caption{\justifying Contour plots of ${\langle j_{2} \rangle}/{\langle j_{1} \rangle}$ and ${P_{2}}/{P_{1}}$ when $t_{I_{1}\alpha_{1}} > t_{I_{3}\alpha_{2}}$ with $E_{\alpha'_{2}} > E_{\alpha'_{1}}$ and $\Gamma_{\alpha_{2}\alpha_{1}} > \Gamma_{\alpha_{1}\alpha_{2}}$. (a), (b): $U = -3000 cm^{-1}$; (c), (d): $U = -2000 cm^{-1}$; (e), (f): $U = 2000 cm^{-1}$ and (g), (h): $U = 3000 cm^{-1}$.}
\label{fig: 14}
\end{figure}

\begin{figure}[h!]
\centering
\begin{minipage}{0.494\columnwidth}
    \centering
    \includegraphics[width=4cm, height=3.5cm]{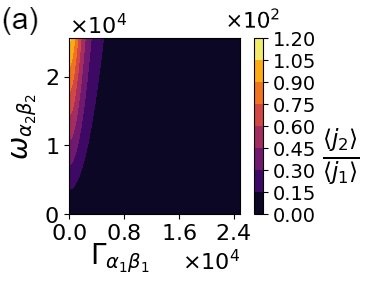} 
\end{minipage}
\begin{minipage}{0.494\columnwidth}
    \centering
    \includegraphics[width=4cm, height=3.5cm]{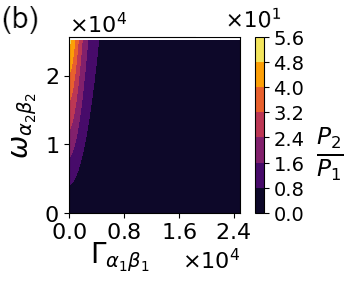}
\end{minipage}

\vspace{0.1cm} 

\begin{minipage}{0.494\columnwidth}
    \centering
    \includegraphics[width=4cm, height=3.5cm]{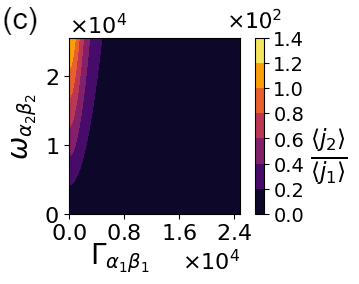}
\end{minipage}
\begin{minipage}{0.494\columnwidth}
    \centering
    \includegraphics[width=4cm, height=3.5cm]{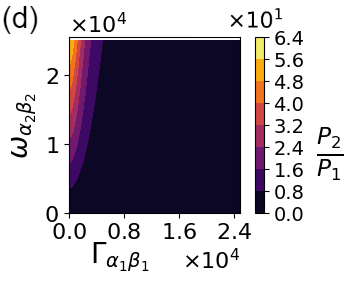}
\end{minipage}

\vspace{0.1cm}

\begin{minipage}{0.494\columnwidth}
    \centering
    \includegraphics[width=4cm, height=3.5cm]{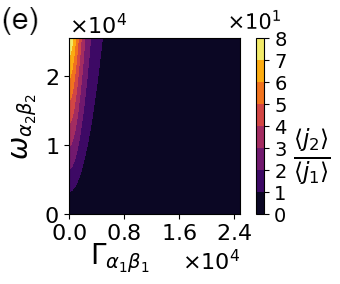} 
\end{minipage}
\begin{minipage}{0.494\columnwidth}
    \centering
    \includegraphics[width=4cm, height=3.5cm]{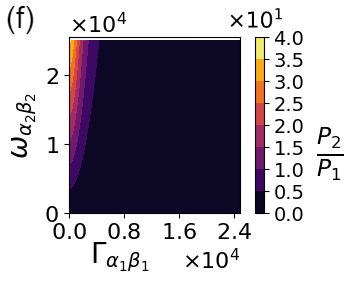}
\end{minipage}

\vspace{0.1cm} 

\begin{minipage}{0.494\columnwidth}
    \centering
    \includegraphics[width=4cm, height=3.5cm]{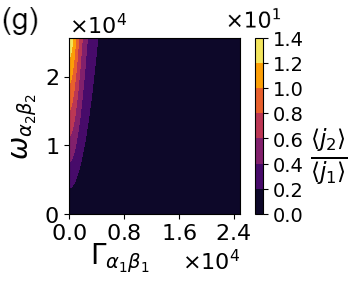}
\end{minipage}
\begin{minipage}{0.494\columnwidth}
    \centering
    \includegraphics[width=4cm, height=3.5cm]{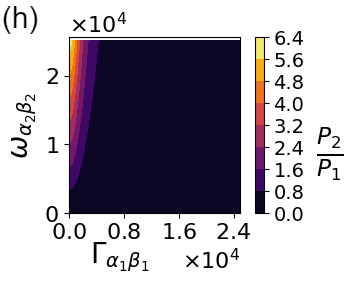}
\end{minipage}

\caption{\justifying Contour plots of ${\langle j_{2} \rangle}/{\langle j_{1} \rangle}$ and ${P_{2}}/{P_{1}}$ when $t_{I_{1}\alpha_{1}} > t_{I_{3}\alpha_{2}}$ with $E_{\alpha'_{2}} < E_{\alpha'_{1}}$ and $\Gamma_{\alpha_{2}\alpha_{1}} > \Gamma_{\alpha_{1}\alpha_{2}}$. (a), (b): $U = -3000 cm^{-1}$; (c), (d): $U = -2000 cm^{-1}$; (e), (f): $U = 2000 cm^{-1}$ and (g), (h): $U = 3000 cm^{-1}$.}
\label{fig: 15}
\end{figure}

\begin{figure}[h!] 
\centering
\begin{minipage}{0.494\columnwidth}
    \centering
    \includegraphics[width=4cm, height=3.5cm]{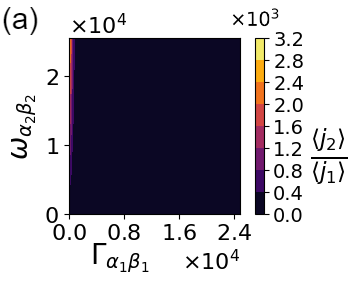}
\end{minipage}
\begin{minipage}{0.494\columnwidth}
    \centering
    \includegraphics[width=4cm, height=3.5cm]{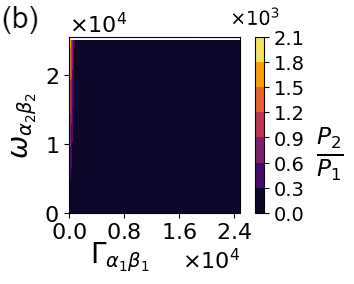}
\end{minipage}

\vspace{0.1cm}

\begin{minipage}{0.494\columnwidth}
    \centering
    \includegraphics[width=4cm, height=3.5cm]{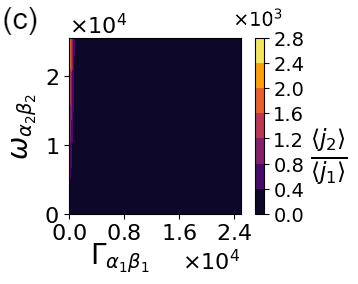} 
\end{minipage}
\begin{minipage}{0.494\columnwidth}
    \centering
    \includegraphics[width=4cm, height=3.5cm]{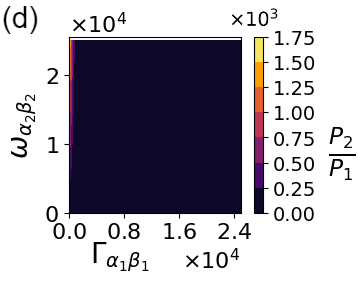}
\end{minipage}

\vspace{0.1cm} 

\begin{minipage}{0.494\columnwidth}
    \centering
    \includegraphics[width=4cm, height=3.5cm]{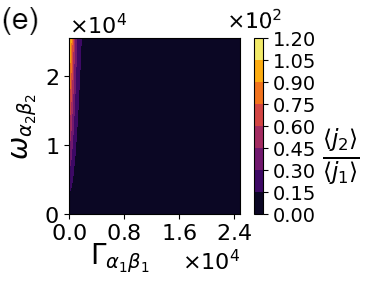}
\end{minipage}
\begin{minipage}{0.494\columnwidth}
    \centering
    \includegraphics[width=4cm, height=3.5cm]{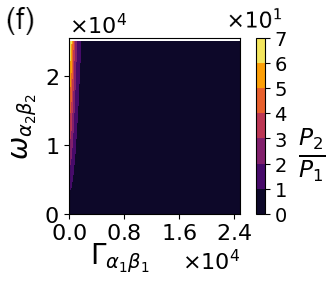}
\end{minipage}

\vspace{0.1cm} 

\begin{minipage}{0.494\columnwidth}
    \centering
    \includegraphics[width=4cm, height=3.5cm]{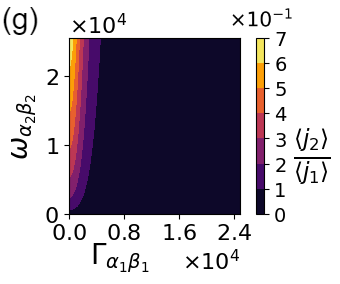}
\end{minipage}
\begin{minipage}{0.494\columnwidth}
    \centering
    \includegraphics[width=4cm, height=3.5cm]{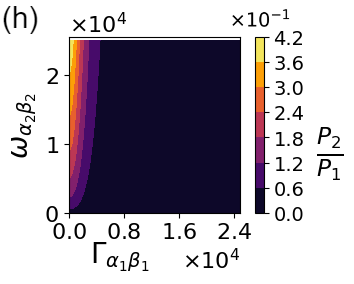}
\end{minipage}

\vspace{0.1cm}

\begin{minipage}{0.494\columnwidth}
    \centering
    \includegraphics[width=4cm, height=3.5cm]{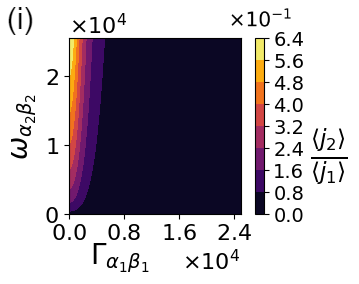}
\end{minipage}
\begin{minipage}{0.494\columnwidth}
    \centering
    \includegraphics[width=4cm, height=3.5cm]{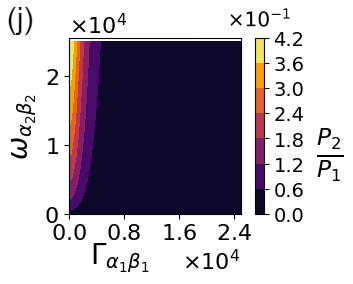}
\end{minipage}

\vspace{0.1cm} 

\begin{minipage}{0.494\columnwidth}
    \centering
    \includegraphics[width=4cm, height=3.5cm]{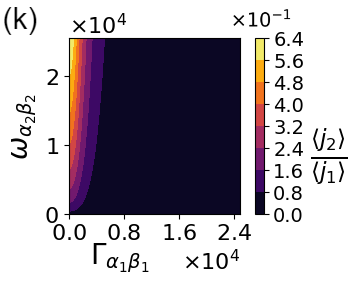}
\end{minipage}
\begin{minipage}{0.494\columnwidth}
    \centering
    \includegraphics[width=4cm, height=3.5cm]{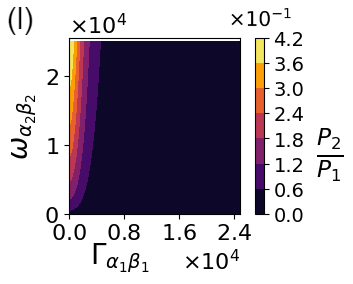}
\end{minipage}

\caption{\justifying Contour plots of ${\langle j_{2} \rangle}/{\langle j_{1} \rangle}$ and ${P_{2}}/{P_{1}}$ when $t_{I_{1}\alpha_{1}} > t_{I_{3}\alpha_{2}}$ with $E_{\alpha'_{2}} > E_{\alpha'_{1}}$ and $\Gamma_{\alpha_{2}\alpha_{1}} < \Gamma_{\alpha_{1}\alpha_{2}}$. (a), (b): $U = -3000 cm^{-1}$; (c), (d): $U = -2000 cm^{-1}$; (e), (f): $U = -1000 cm^{-1}$, (g), (h): $U = 1000 cm^{-1}$, (i), (j): $U = 2000 cm^{-1}$ and (k), (l): $U = 3000 cm^{-1}$.}
\label{fig: 16}
\end{figure}

\begin{figure}[h!] 
\centering
\begin{minipage}{0.494\columnwidth}
    \centering
    \includegraphics[width=4cm, height=3.5cm]{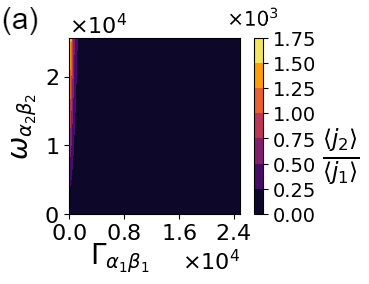}
\end{minipage}
\begin{minipage}{0.494\columnwidth}
    \centering
    \includegraphics[width=4cm, height=3.5cm]{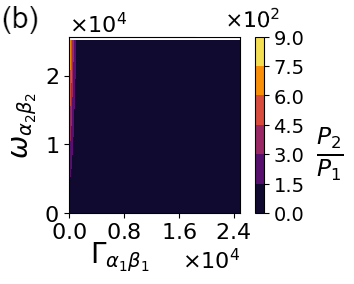}
\end{minipage}

\vspace{0.1cm}

\begin{minipage}{0.494\columnwidth}
    \centering
    \includegraphics[width=4cm, height=3.5cm]{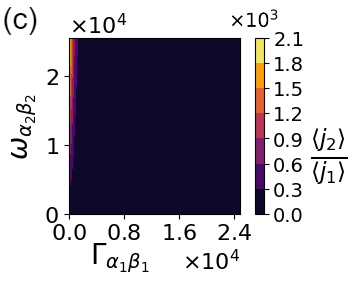} 
\end{minipage}
\begin{minipage}{0.494\columnwidth}
    \centering
    \includegraphics[width=4cm, height=3.5cm]{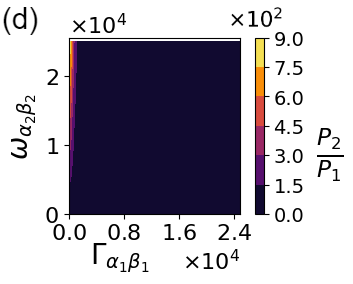}
\end{minipage}

\vspace{0.1cm} 

\begin{minipage}{0.494\columnwidth}
    \centering
    \includegraphics[width=4cm, height=3.5cm]{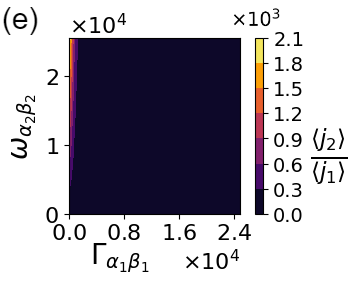}
\end{minipage}
\begin{minipage}{0.494\columnwidth}
    \centering
    \includegraphics[width=4cm, height=3.5cm]{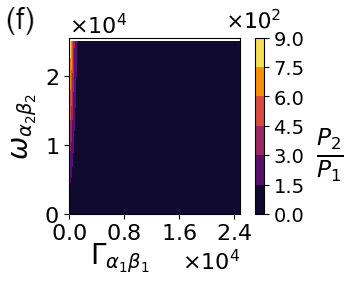}
\end{minipage}

\vspace{0.1cm} 

\begin{minipage}{0.494\columnwidth}
    \centering
    \includegraphics[width=4cm, height=3.5cm]{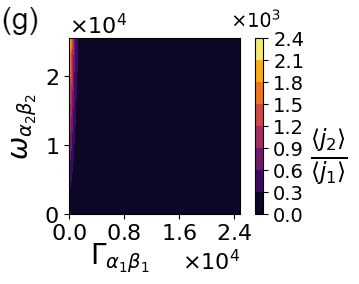}
\end{minipage}
\begin{minipage}{0.494\columnwidth}
    \centering
    \includegraphics[width=4cm, height=3.5cm]{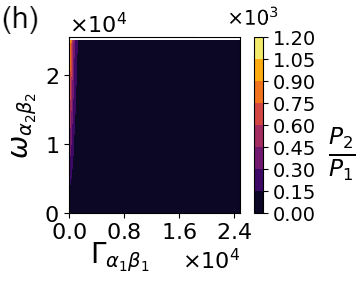}
\end{minipage}

\vspace{0.1cm}

\begin{minipage}{0.494\columnwidth}
    \centering
    \includegraphics[width=4cm, height=3.5cm]{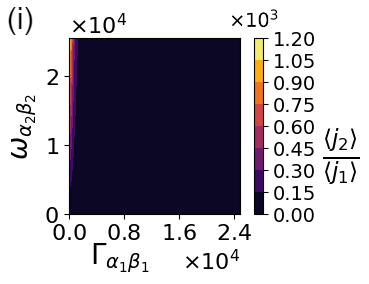} 
\end{minipage}
\begin{minipage}{0.494\columnwidth}
    \centering
    \includegraphics[width=4cm, height=3.5cm]{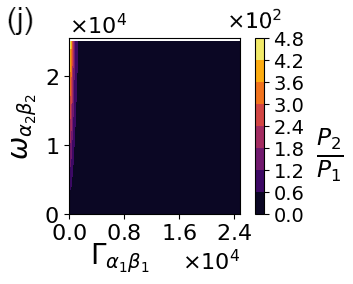}
\end{minipage}

\vspace{0.1cm} 

\begin{minipage}{0.494\columnwidth}
    \centering
    \includegraphics[width=4cm, height=3.5cm]{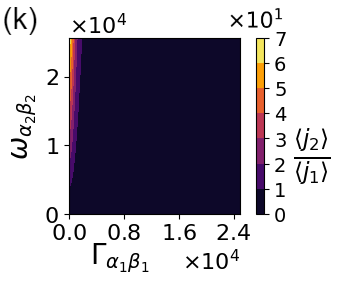}
\end{minipage}
\begin{minipage}{0.494\columnwidth}
    \centering
    \includegraphics[width=4cm, height=3.5cm]{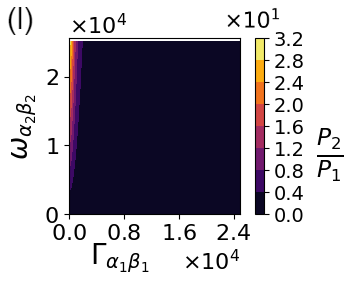}
\end{minipage}

\caption{\justifying Contour plots of ${\langle j_{2} \rangle}/{\langle j_{1} \rangle}$ and ${P_{2}}/{P_{1}}$ when $t_{I_{1}\alpha_{1}} > t_{I_{3}\alpha_{2}}$ with $E_{\alpha'_{2}} < E_{\alpha'_{1}}$ and $\Gamma_{\alpha_{2}\alpha_{1}} < \Gamma_{\alpha_{1}\alpha_{2}}$. (a), (b): $U = -3000 cm^{-1}$; (c), (d): $U = -2000 cm^{-1}$; (e), (f): $U = -1000 cm^{-1}$, (g), (h): $U = 1000 cm^{-1}$, (i), (j): $U = 2000 cm^{-1}$ and (k), (l): $U = 3000 cm^{-1}$.}
\label{fig: 17}
\end{figure}

\begin{figure}[h!]
\centering
\begin{minipage}{0.494\columnwidth}
    \centering
    \includegraphics[width=4cm, height=3.5cm]{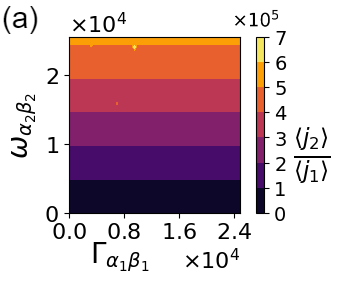} 
\end{minipage}
\begin{minipage}{0.494\columnwidth}
    \centering
    \includegraphics[width=4cm, height=3.5cm]{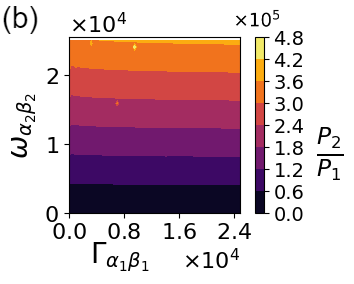}
\end{minipage}

\vspace{0.1cm} 

\begin{minipage}{0.494\columnwidth}
    \centering
    \includegraphics[width=4cm, height=3.5cm]{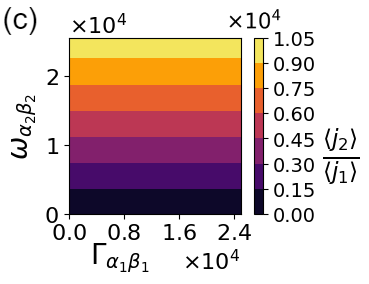}
\end{minipage}
\begin{minipage}{0.494\columnwidth}
    \centering
    \includegraphics[width=4cm, height=3.5cm]{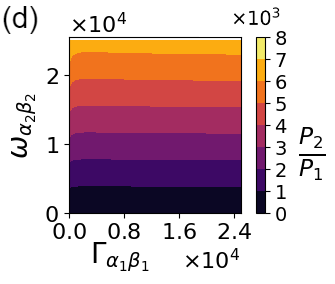}
\end{minipage}

\vspace{0.1cm}

\begin{minipage}{0.494\columnwidth}
    \centering
    \includegraphics[width=4cm, height=3.5cm]{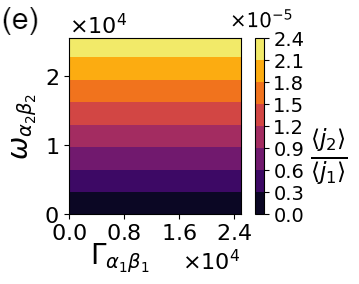} 
\end{minipage}
\begin{minipage}{0.494\columnwidth}
    \centering
    \includegraphics[width=4cm, height=3.5cm]{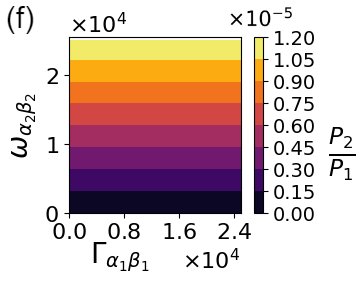}
\end{minipage}

\vspace{0.1cm} 

\begin{minipage}{0.494\columnwidth}
    \centering
    \includegraphics[width=4cm, height=3.5cm]{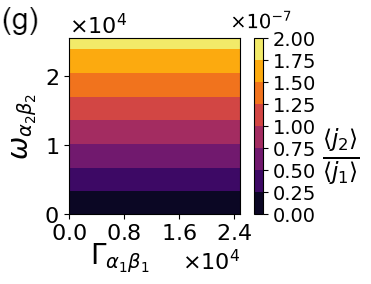}
\end{minipage}
\begin{minipage}{0.494\columnwidth}
    \centering
    \includegraphics[width=4cm, height=3.5cm]{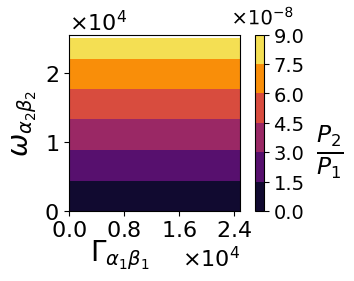}
\end{minipage}

\caption{\justifying Contour plots of ${\langle j_{2} \rangle}/{\langle j_{1} \rangle}$ and ${P_{2}}/{P_{1}}$  when $t_{I_{1}\alpha_{1}} > t_{I_{3}\alpha_{2}}$ (cotunneling coupling coefficients are relatively much larger than other parameters) with $E_{\alpha'_{2}} > E_{\alpha'_{1}}$ and $\Gamma_{\alpha_{2}\alpha_{1}} > \Gamma_{\alpha_{1}\alpha_{2}}$. (a), (b): $U = -3000 cm^{-1}$; (c), (d): $U = -2000 cm^{-1}$; (e), (f): $U = 2000 cm^{-1}$ and, (g), (h): $U = 3000 cm^{-1}$.}
\label{fig: 18}
\end{figure}

\begin{figure}[h!]
\centering
\begin{minipage}{0.494\columnwidth}
    \centering
    \includegraphics[width=4cm, height=3.5cm]{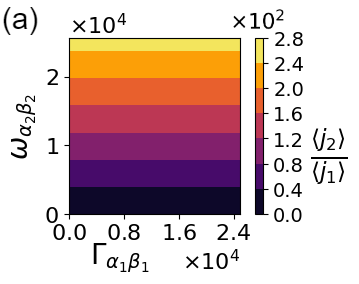} 
\end{minipage}
\begin{minipage}{0.494\columnwidth}
    \centering
    \includegraphics[width=4cm, height=3.5cm]{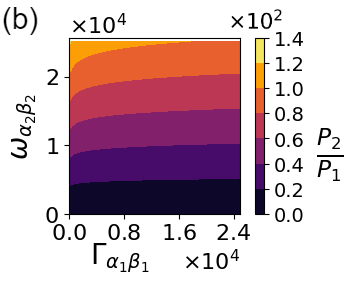}
\end{minipage}

\vspace{0.1cm} 

\begin{minipage}{0.494\columnwidth}
    \centering
    \includegraphics[width=4cm, height=3.5cm]{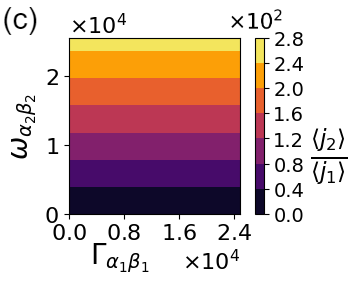}
\end{minipage}
\begin{minipage}{0.494\columnwidth}
    \centering
    \includegraphics[width=4cm, height=3.5cm]{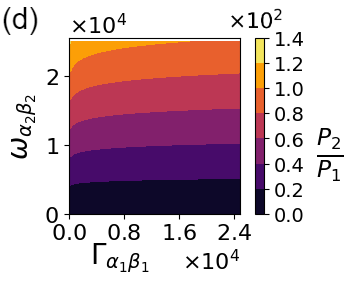}
\end{minipage}

\vspace{0.1cm}

\begin{minipage}{0.494\columnwidth}
    \centering
    \includegraphics[width=4cm, height=3.5cm]{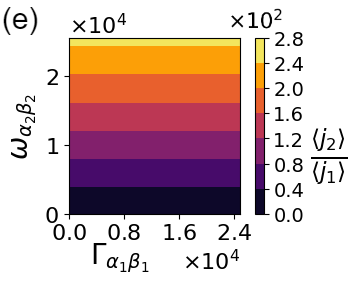} 
\end{minipage}
\begin{minipage}{0.494\columnwidth}
    \centering
    \includegraphics[width=4cm, height=3.5cm]{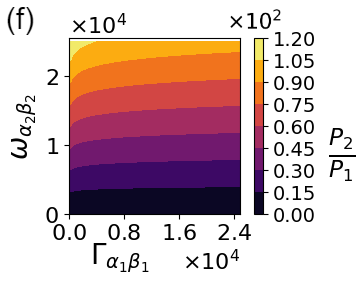}
\end{minipage}

\vspace{0.1cm} 

\begin{minipage}{0.494\columnwidth}
    \centering
    \includegraphics[width=4cm, height=3.5cm]{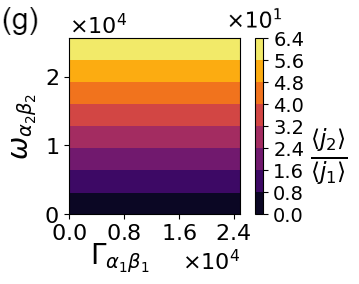}
\end{minipage}
\begin{minipage}{0.494\columnwidth}
    \centering
    \includegraphics[width=4cm, height=3.5cm]{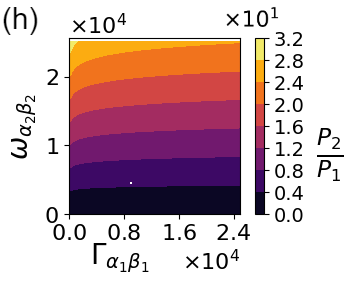}
\end{minipage}

\caption{\justifying Contour plots of ${\langle j_{2} \rangle}/{\langle j_{1} \rangle}$ and ${P_{2}}/{P_{1}}$ when $t_{I_{1}\alpha_{1}} > t_{I_{3}\alpha_{2}}$ (cotunneling coupling coefficients are relatively much larger than other parameters) with $E_{\alpha'_{2}} < E_{\alpha'_{1}}$ and $\Gamma_{\alpha_{2}\alpha_{1}} > \Gamma_{\alpha_{1}\alpha_{2}}$. (a), (b): $U = -3000 cm^{-1}$; (c), (d): $U = -2000 cm^{-1}$; (e), (f): $U = 2000 cm^{-1}$ and (g), (h): $U = 3000 cm^{-1}$.}
\label{fig: 19}
\end{figure}

\begin{figure}[h!] 
\centering
\begin{minipage}{0.494\columnwidth}
    \centering
    \includegraphics[width=4cm, height=3.5cm]{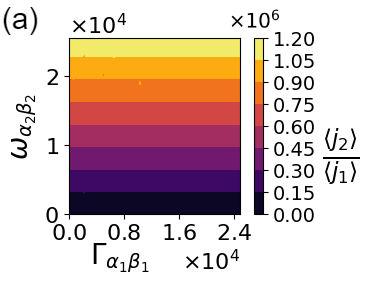}
\end{minipage}
\begin{minipage}{0.494\columnwidth}
    \centering
    \includegraphics[width=4cm, height=3.5cm]{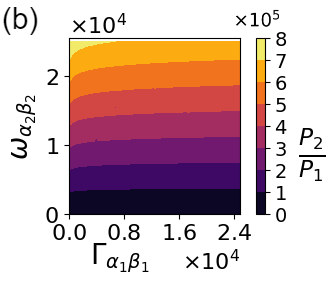}
\end{minipage}

\vspace{0.1cm}

\begin{minipage}{0.494\columnwidth}
    \centering
    \includegraphics[width=4cm, height=3.5cm]{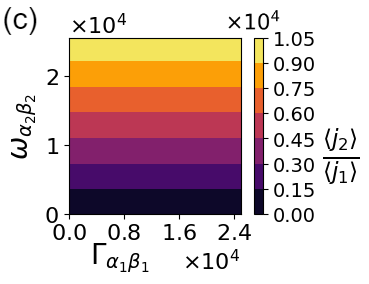} 
\end{minipage}
\begin{minipage}{0.494\columnwidth}
    \centering
    \includegraphics[width=4cm, height=3.5cm]{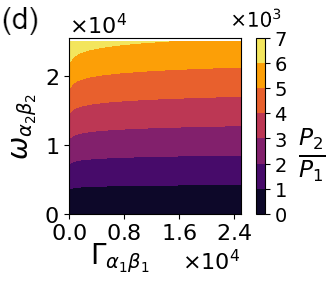}
\end{minipage}

\vspace{0.1cm} 

\begin{minipage}{0.494\columnwidth}
    \centering
    \includegraphics[width=4cm, height=3.5cm]{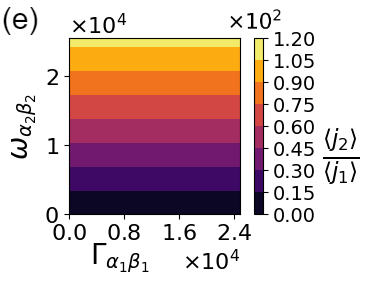}
\end{minipage}
\begin{minipage}{0.494\columnwidth}
    \centering
    \includegraphics[width=4cm, height=3.5cm]{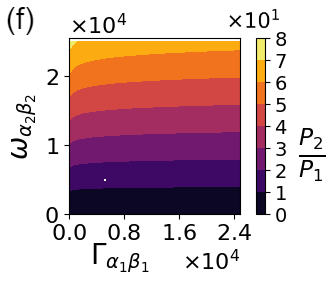}
\end{minipage}

\vspace{0.1cm} 

\begin{minipage}{0.494\columnwidth}
    \centering
    \includegraphics[width=4cm, height=3.5cm]{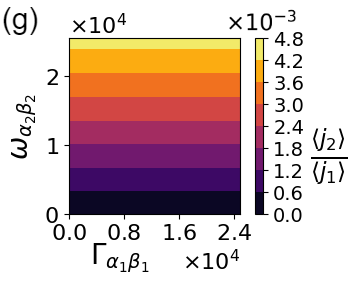}
\end{minipage}
\begin{minipage}{0.494\columnwidth}
    \centering
    \includegraphics[width=4cm, height=3.5cm]{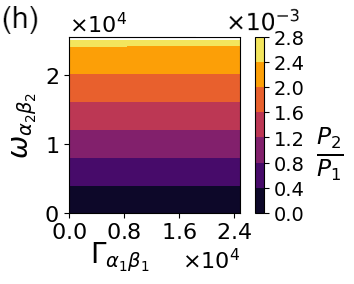}
\end{minipage}

\vspace{0.1cm}

\begin{minipage}{0.494\columnwidth}
    \centering
    \includegraphics[width=4cm, height=3.5cm]{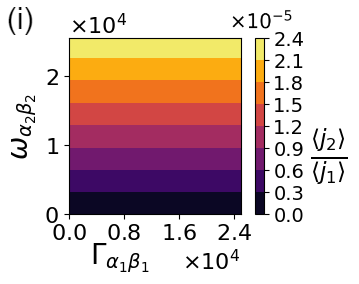} 
\end{minipage}
\begin{minipage}{0.494\columnwidth}
    \centering
    \includegraphics[width=4cm, height=3.5cm]{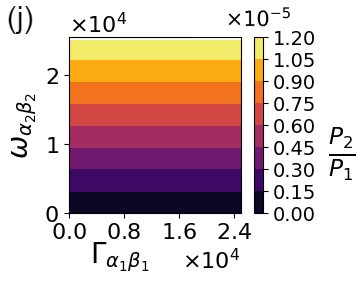}
\end{minipage}

\vspace{0.1cm} 

\begin{minipage}{0.494\columnwidth}
    \centering
    \includegraphics[width=4cm, height=3.5cm]{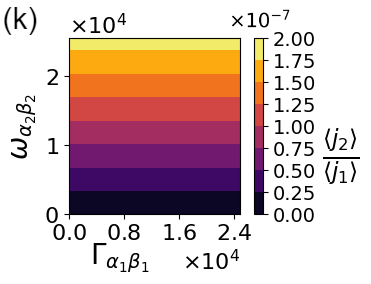}
\end{minipage}
\begin{minipage}{0.494\columnwidth}
    \centering
    \includegraphics[width=4cm, height=3.5cm]{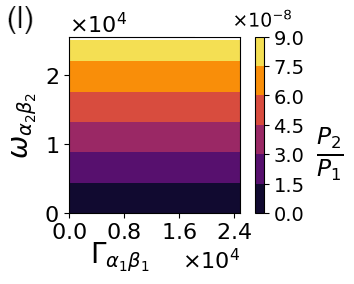}
\end{minipage}

\caption{\justifying Contour plots of ${\langle j_{2} \rangle}/{\langle j_{1} \rangle}$ and ${P_{2}}/{P_{1}}$  when $t_{I_{1}\alpha_{1}} > t_{I_{3}\alpha_{2}}$ (cotunneling coupling coefficients are relatively much larger than other parameters) with $E_{\alpha'_{2}} > E_{\alpha'_{1}}$ and $\Gamma_{\alpha_{2}\alpha_{1}} < \Gamma_{\alpha_{1}\alpha_{2}}$. (a), (b): $U = -3000 cm^{-1}$; (c), (d): $U = -2000 cm^{-1}$; (e), (f): $U = -1000 cm^{-1}$, (g), (h): $U = 1000 cm^{-1}$, (i), (j): $U = 2000 cm^{-1}$ and (k), (l): $U = 3000 cm^{-1}$.}
\label{fig: 20}
\end{figure}

\begin{figure}[h!] 
\centering
\begin{minipage}{0.494\columnwidth}
    \centering
    \includegraphics[width=4cm, height=3.5cm]{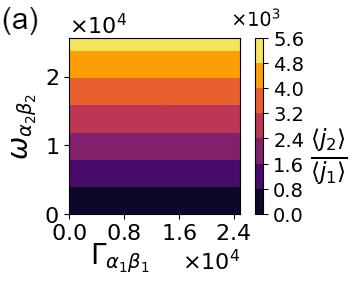}
\end{minipage}
\begin{minipage}{0.494\columnwidth}
    \centering
    \includegraphics[width=4cm, height=3.5cm]{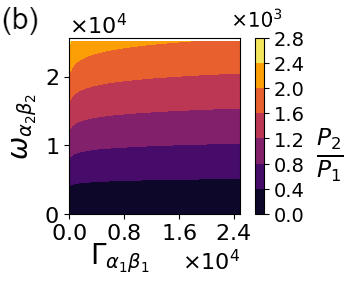}
\end{minipage}

\vspace{0.1cm}

\begin{minipage}{0.494\columnwidth}
    \centering
    \includegraphics[width=4cm, height=3.5cm]{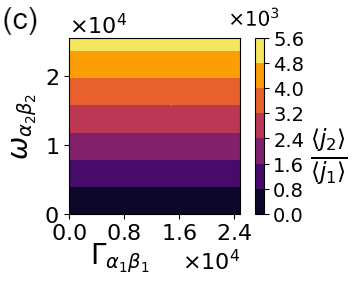} 
\end{minipage}
\begin{minipage}{0.494\columnwidth}
    \centering
    \includegraphics[width=4cm, height=3.5cm]{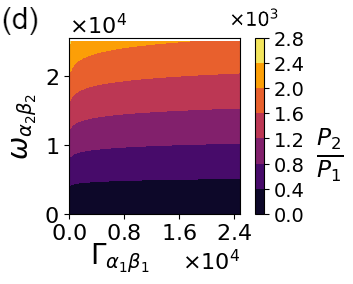}
\end{minipage}

\vspace{0.1cm} 

\begin{minipage}{0.494\columnwidth}
    \centering
    \includegraphics[width=4cm, height=3.5cm]{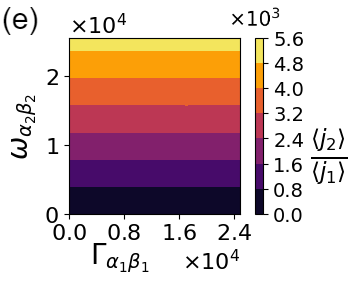}
\end{minipage}
\begin{minipage}{0.494\columnwidth}
    \centering
    \includegraphics[width=4cm, height=3.5cm]{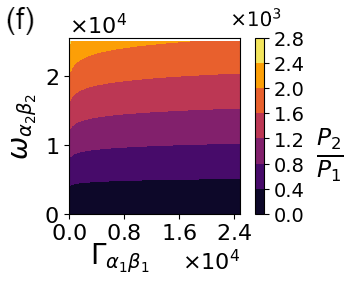}
\end{minipage}

\vspace{0.1cm} 

\begin{minipage}{0.494\columnwidth}
    \centering
    \includegraphics[width=4cm, height=3.5cm]{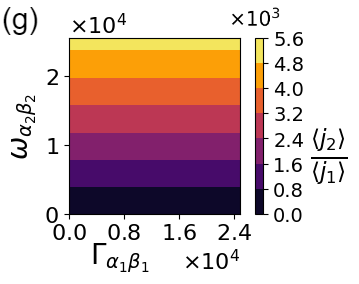}
\end{minipage}
\begin{minipage}{0.494\columnwidth}
    \centering
    \includegraphics[width=4cm, height=3.5cm]{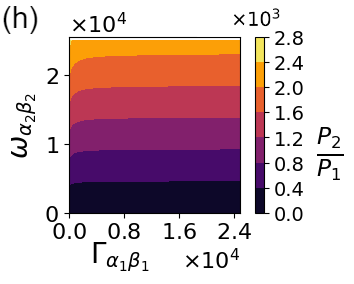}
\end{minipage}

\vspace{0.1cm}

\begin{minipage}{0.494\columnwidth}
    \centering
    \includegraphics[width=4cm, height=3.5cm]{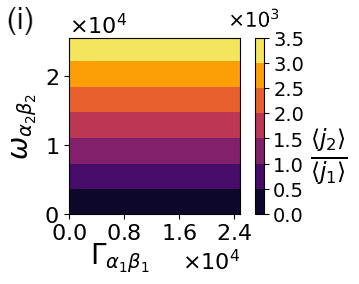} 
\end{minipage}
\begin{minipage}{0.494\columnwidth}
    \centering
    \includegraphics[width=4cm, height=3.5cm]{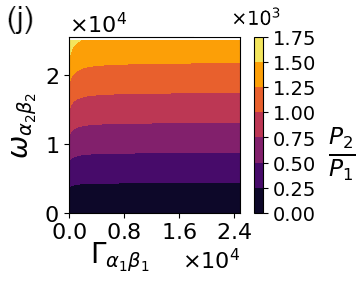}
\end{minipage}

\vspace{0.1cm} 

\begin{minipage}{0.494\columnwidth}
    \centering
    \includegraphics[width=4cm, height=3.5cm]{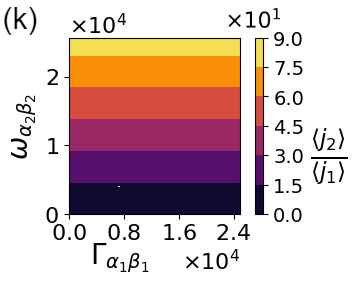}
\end{minipage}
\begin{minipage}{0.494\columnwidth}
    \centering
    \includegraphics[width=4cm, height=3.5cm]{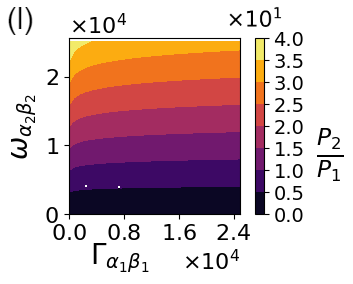}
\end{minipage}

\caption{\justifying Contour plots of ${\langle j_{2} \rangle}/{\langle j_{1} \rangle}$ and ${P_{2}}/{P_{1}}$ when $t_{I_{1}\alpha_{1}} > t_{I_{3}\alpha_{2}}$ (cotunneling coupling coefficients are relatively much larger than other parameters) with $E_{\alpha'_{2}} < E_{\alpha'_{1}}$ and $\Gamma_{\alpha_{2}\alpha_{1}} < \Gamma_{\alpha_{1}\alpha_{2}}$. (a), (b): $U = -3000 cm^{-1}$; (c), (d): $U = -2000 cm^{-1}$; (e), (f): $U = -1000 cm^{-1}$; (g), (h): $U = 1000 cm^{-1}$; (i), (j): $U = 2000 cm^{-1}$ and (k), (l): $U = 3000 cm^{-1}$.}
\label{fig: 21}
\end{figure}

\FloatBarrier
\newpage
\bibliography{Refernces}

\begin{thebibliography}{76}%
\makeatletter
\providecommand \@ifxundefined [1]{%
 \@ifx{#1\undefined}
}%
\providecommand \@ifnum [1]{%
 \ifnum #1\expandafter \@firstoftwo
 \else \expandafter \@secondoftwo
 \fi
}%
\providecommand \@ifx [1]{%
 \ifx #1\expandafter \@firstoftwo
 \else \expandafter \@secondoftwo
 \fi
}%
\providecommand \natexlab [1]{#1}%
\providecommand \enquote  [1]{``#1''}%
\providecommand \bibnamefont  [1]{#1}%
\providecommand \bibfnamefont [1]{#1}%
\providecommand \citenamefont [1]{#1}%
\providecommand \href@noop [0]{\@secondoftwo}%
\providecommand \href [0]{\begingroup \@sanitize@url \@href}%
\providecommand \@href[1]{\@@startlink{#1}\@@href}%
\providecommand \@@href[1]{\endgroup#1\@@endlink}%
\providecommand \@sanitize@url [0]{\catcode `\\12\catcode `\$12\catcode `\&12\catcode `\#12\catcode `\^12\catcode `\_12\catcode `\%12\relax}%
\providecommand \@@startlink[1]{}%
\providecommand \@@endlink[0]{}%
\providecommand \url  [0]{\begingroup\@sanitize@url \@url }%
\providecommand \@url [1]{\endgroup\@href {#1}{\urlprefix }}%
\providecommand \urlprefix  [0]{URL }%
\providecommand \Eprint [0]{\href }%
\providecommand \doibase [0]{http://dx.doi.org/}%
\providecommand \selectlanguage [0]{\@gobble}%
\providecommand \bibinfo  [0]{\@secondoftwo}%
\providecommand \bibfield  [0]{\@secondoftwo}%
\providecommand \translation [1]{[#1]}%
\providecommand \BibitemOpen [0]{}%
\providecommand \bibitemStop [0]{}%
\providecommand \bibitemNoStop [0]{.\EOS\space}%
\providecommand \EOS [0]{\spacefactor3000\relax}%
\providecommand \BibitemShut  [1]{\csname bibitem#1\endcsname}%
\let\auto@bib@innerbib\@empty
\bibitem [{\citenamefont {Panitchayangkoon}\ \emph {et~al.}(2010)\citenamefont {Panitchayangkoon}, \citenamefont {Hayes}, \citenamefont {Fransted}, \citenamefont {Caram}, \citenamefont {Harel}, \citenamefont {Wen}, \citenamefont {Blankenship},\ and\ \citenamefont {Engel}}]{Gitt}%
  \BibitemOpen
  \bibfield  {author} {\bibinfo {author} {\bibfnamefont {G.}~\bibnamefont {Panitchayangkoon}}, \bibinfo {author} {\bibfnamefont {D.}~\bibnamefont {Hayes}}, \bibinfo {author} {\bibfnamefont {K.~A.}\ \bibnamefont {Fransted}}, \bibinfo {author} {\bibfnamefont {J.~R.}\ \bibnamefont {Caram}}, \bibinfo {author} {\bibfnamefont {E.}~\bibnamefont {Harel}}, \bibinfo {author} {\bibfnamefont {J.}~\bibnamefont {Wen}}, \bibinfo {author} {\bibfnamefont {R.~E.}\ \bibnamefont {Blankenship}}, \ and\ \bibinfo {author} {\bibfnamefont {G.~S.}\ \bibnamefont {Engel}},\ }\href@noop {} {\bibfield  {journal} {\bibinfo  {journal} {Proceedings of the National Academy of Sciences}\ }\textbf {\bibinfo {volume} {107}},\ \bibinfo {pages} {12766} (\bibinfo {year} {2010})}\BibitemShut {NoStop}%
\bibitem [{\citenamefont {Moser}\ \emph {et~al.}(1992)\citenamefont {Moser}, \citenamefont {Keske}, \citenamefont {Warncke}, \citenamefont {Farid},\ and\ \citenamefont {Dutton}}]{Moser1992}%
  \BibitemOpen
  \bibfield  {author} {\bibinfo {author} {\bibfnamefont {C.~C.}\ \bibnamefont {Moser}}, \bibinfo {author} {\bibfnamefont {J.~M.}\ \bibnamefont {Keske}}, \bibinfo {author} {\bibfnamefont {K.}~\bibnamefont {Warncke}}, \bibinfo {author} {\bibfnamefont {R.~S.}\ \bibnamefont {Farid}}, \ and\ \bibinfo {author} {\bibfnamefont {P.~L.}\ \bibnamefont {Dutton}},\ }\href@noop {} {\bibfield  {journal} {\bibinfo  {journal} {Nature}\ }\textbf {\bibinfo {volume} {355}},\ \bibinfo {pages} {796} (\bibinfo {year} {1992})}\BibitemShut {NoStop}%
\bibitem [{\citenamefont {Engel}\ \emph {et~al.}(2007)\citenamefont {Engel}, \citenamefont {Calhoun}, \citenamefont {Read}, \citenamefont {Ahn}, \citenamefont {Man{\v{c}}al}, \citenamefont {Cheng}, \citenamefont {Blankenship},\ and\ \citenamefont {Fleming}}]{Engel2007}%
  \BibitemOpen
  \bibfield  {author} {\bibinfo {author} {\bibfnamefont {G.~S.}\ \bibnamefont {Engel}}, \bibinfo {author} {\bibfnamefont {T.~R.}\ \bibnamefont {Calhoun}}, \bibinfo {author} {\bibfnamefont {E.~L.}\ \bibnamefont {Read}}, \bibinfo {author} {\bibfnamefont {T.-K.}\ \bibnamefont {Ahn}}, \bibinfo {author} {\bibfnamefont {T.}~\bibnamefont {Man{\v{c}}al}}, \bibinfo {author} {\bibfnamefont {Y.-C.}\ \bibnamefont {Cheng}}, \bibinfo {author} {\bibfnamefont {R.~E.}\ \bibnamefont {Blankenship}}, \ and\ \bibinfo {author} {\bibfnamefont {G.~R.}\ \bibnamefont {Fleming}},\ }\href@noop {} {\bibfield  {journal} {\bibinfo  {journal} {Nature}\ }\textbf {\bibinfo {volume} {446}},\ \bibinfo {pages} {782} (\bibinfo {year} {2007})}\BibitemShut {NoStop}%
\bibitem [{\citenamefont {Marcus}\ and\ \citenamefont {Sutin}(1985)}]{MARCUS1985265}%
  \BibitemOpen
  \bibfield  {author} {\bibinfo {author} {\bibfnamefont {R.}~\bibnamefont {Marcus}}\ and\ \bibinfo {author} {\bibfnamefont {N.}~\bibnamefont {Sutin}},\ }\href@noop {} {\bibfield  {journal} {\bibinfo  {journal} {Biochimica et Biophysica Acta (BBA) - Reviews on Bioenergetics}\ }\textbf {\bibinfo {volume} {811}},\ \bibinfo {pages} {265} (\bibinfo {year} {1985})}\BibitemShut {NoStop}%
\bibitem [{\citenamefont {Renger}(2012)}]{RENGER20121164}%
  \BibitemOpen
  \bibfield  {author} {\bibinfo {author} {\bibfnamefont {G.}~\bibnamefont {Renger}},\ }\href@noop {} {\bibfield  {journal} {\bibinfo  {journal} {Biochimica et Biophysica Acta (BBA) - Bioenergetics}\ }\textbf {\bibinfo {volume} {1817}},\ \bibinfo {pages} {1164} (\bibinfo {year} {2012})},\ \bibinfo {note} {photosynthesis Research for Sustainability: From Natural to Artificial}\BibitemShut {NoStop}%
\bibitem [{\citenamefont {Siegbahn}\ and\ \citenamefont {Blomberg}(2008)}]{Siegbahn2008}%
  \BibitemOpen
  \bibfield  {author} {\bibinfo {author} {\bibfnamefont {P.~E.~M.}\ \bibnamefont {Siegbahn}}\ and\ \bibinfo {author} {\bibfnamefont {M.~R.~A.}\ \bibnamefont {Blomberg}},\ }\href@noop {} {\bibfield  {journal} {\bibinfo  {journal} {The Journal of Physical Chemistry A}\ }\textbf {\bibinfo {volume} {112}},\ \bibinfo {pages} {12772} (\bibinfo {year} {2008})}\BibitemShut {NoStop}%
\bibitem [{\citenamefont {Blankenship}(2002)}]{ch1}%
  \BibitemOpen
  \bibfield  {author} {\bibinfo {author} {\bibfnamefont {R.~E.}\ \bibnamefont {Blankenship}},\ }\enquote {\bibinfo {title} {Molecular mechanisms of photosynthesis},}\ \ (\bibinfo  {publisher} {John Wiley \& Sons, Ltd},\ \bibinfo {year} {2002})\BibitemShut {NoStop}%
\bibitem [{\citenamefont {Guskov}\ \emph {et~al.}(2009)\citenamefont {Guskov}, \citenamefont {Kern}, \citenamefont {Gabdulkhakov}, \citenamefont {Broser}, \citenamefont {Zouni},\ and\ \citenamefont {Saenger}}]{Guskov2009}%
  \BibitemOpen
  \bibfield  {author} {\bibinfo {author} {\bibfnamefont {A.}~\bibnamefont {Guskov}}, \bibinfo {author} {\bibfnamefont {J.}~\bibnamefont {Kern}}, \bibinfo {author} {\bibfnamefont {A.}~\bibnamefont {Gabdulkhakov}}, \bibinfo {author} {\bibfnamefont {M.}~\bibnamefont {Broser}}, \bibinfo {author} {\bibfnamefont {A.}~\bibnamefont {Zouni}}, \ and\ \bibinfo {author} {\bibfnamefont {W.}~\bibnamefont {Saenger}},\ }\href@noop {} {\bibfield  {journal} {\bibinfo  {journal} {Nature Structural {\&} Molecular Biology}\ }\textbf {\bibinfo {volume} {16}},\ \bibinfo {pages} {334} (\bibinfo {year} {2009})}\BibitemShut {NoStop}%
\bibitem [{\citenamefont {Tomo}\ and\ \citenamefont {Allakhverdiev}(2021)}]{Tomo2021}%
  \BibitemOpen
  \bibfield  {author} {\bibinfo {author} {\bibfnamefont {T.}~\bibnamefont {Tomo}}\ and\ \bibinfo {author} {\bibfnamefont {S.~I.}\ \bibnamefont {Allakhverdiev}},\ }\enquote {\bibinfo {title} {Chlorophyll species and their functions in the photosynthetic energy conversion},}\ \ (\bibinfo  {publisher} {Springer International Publishing},\ \bibinfo {address} {Cham},\ \bibinfo {year} {2021})\ pp.\ \bibinfo {pages} {133--161}\BibitemShut {NoStop}%
\bibitem [{\citenamefont {Panitchayangkoon}\ \emph {et~al.}(2011)\citenamefont {Panitchayangkoon}, \citenamefont {Voronine}, \citenamefont {Abramavicius}, \citenamefont {Caram}, \citenamefont {Lewis}, \citenamefont {Mukamel},\ and\ \citenamefont {Engel}}]{pnas.1105234108}%
  \BibitemOpen
  \bibfield  {author} {\bibinfo {author} {\bibfnamefont {G.}~\bibnamefont {Panitchayangkoon}}, \bibinfo {author} {\bibfnamefont {D.~V.}\ \bibnamefont {Voronine}}, \bibinfo {author} {\bibfnamefont {D.}~\bibnamefont {Abramavicius}}, \bibinfo {author} {\bibfnamefont {J.~R.}\ \bibnamefont {Caram}}, \bibinfo {author} {\bibfnamefont {N.~H.~C.}\ \bibnamefont {Lewis}}, \bibinfo {author} {\bibfnamefont {S.}~\bibnamefont {Mukamel}}, \ and\ \bibinfo {author} {\bibfnamefont {G.~S.}\ \bibnamefont {Engel}},\ }\href@noop {} {\bibfield  {journal} {\bibinfo  {journal} {Proceedings of the National Academy of Sciences}\ }\textbf {\bibinfo {volume} {108}},\ \bibinfo {pages} {20908} (\bibinfo {year} {2011})}\BibitemShut {NoStop}%
\bibitem [{\citenamefont {Novoderezhkin}\ \emph {et~al.}(2011)\citenamefont {Novoderezhkin}, \citenamefont {Romero}, \citenamefont {Dekker},\ and\ \citenamefont {van Grondelle}}]{Vladimir}%
  \BibitemOpen
  \bibfield  {author} {\bibinfo {author} {\bibfnamefont {V.~I.}\ \bibnamefont {Novoderezhkin}}, \bibinfo {author} {\bibfnamefont {E.}~\bibnamefont {Romero}}, \bibinfo {author} {\bibfnamefont {J.~P.}\ \bibnamefont {Dekker}}, \ and\ \bibinfo {author} {\bibfnamefont {R.}~\bibnamefont {van Grondelle}},\ }\href@noop {} {\bibfield  {journal} {\bibinfo  {journal} {ChemPhysChem}\ }\textbf {\bibinfo {volume} {12}},\ \bibinfo {pages} {681} (\bibinfo {year} {2011})}\BibitemShut {NoStop}%
\bibitem [{\citenamefont {Hayase}\ \emph {et~al.}(2023)\citenamefont {Hayase}, \citenamefont {Shimada}, \citenamefont {Mitomi}, \citenamefont {Nagao},\ and\ \citenamefont {Noguchi}}]{Hayase2023}%
  \BibitemOpen
  \bibfield  {author} {\bibinfo {author} {\bibfnamefont {T.}~\bibnamefont {Hayase}}, \bibinfo {author} {\bibfnamefont {Y.}~\bibnamefont {Shimada}}, \bibinfo {author} {\bibfnamefont {T.}~\bibnamefont {Mitomi}}, \bibinfo {author} {\bibfnamefont {R.}~\bibnamefont {Nagao}}, \ and\ \bibinfo {author} {\bibfnamefont {T.}~\bibnamefont {Noguchi}},\ }\href@noop {} {\bibfield  {journal} {\bibinfo  {journal} {The Journal of Physical Chemistry B}\ }\textbf {\bibinfo {volume} {127}},\ \bibinfo {pages} {1758} (\bibinfo {year} {2023})}\BibitemShut {NoStop}%
\bibitem [{\citenamefont {Kamlowski}\ \emph {et~al.}()\citenamefont {Kamlowski}, \citenamefont {Frankemöller}, \citenamefont {van~der Est}, \citenamefont {Stehlik},\ and\ \citenamefont {Holzwart}}]{Kamlowski}%
  \BibitemOpen
  \bibfield  {author} {\bibinfo {author} {\bibfnamefont {A.}~\bibnamefont {Kamlowski}}, \bibinfo {author} {\bibfnamefont {L.}~\bibnamefont {Frankemöller}}, \bibinfo {author} {\bibfnamefont {A.}~\bibnamefont {van~der Est}}, \bibinfo {author} {\bibfnamefont {D.}~\bibnamefont {Stehlik}}, \ and\ \bibinfo {author} {\bibfnamefont {A.~R.}\ \bibnamefont {Holzwart}},\ }\href@noop {} {\bibfield  {journal} {\bibinfo  {journal} {Berichte der Bunsengesellschaft für physikalische Chemie}\ }\textbf {\bibinfo {volume} {100}},\ \bibinfo {pages} {2045}}\BibitemShut {NoStop}%
\bibitem [{\citenamefont {Bhattacharjee}\ \emph {et~al.}(2023)\citenamefont {Bhattacharjee}, \citenamefont {Neese},\ and\ \citenamefont {Pantazis}}]{D3SC02985A}%
  \BibitemOpen
  \bibfield  {author} {\bibinfo {author} {\bibfnamefont {S.}~\bibnamefont {Bhattacharjee}}, \bibinfo {author} {\bibfnamefont {F.}~\bibnamefont {Neese}}, \ and\ \bibinfo {author} {\bibfnamefont {D.~A.}\ \bibnamefont {Pantazis}},\ }\href@noop {} {\bibfield  {journal} {\bibinfo  {journal} {Chemical Science}\ }\textbf {\bibinfo {volume} {14}},\ \bibinfo {pages} {9503} (\bibinfo {year} {2023})}\BibitemShut {NoStop}%
\bibitem [{\citenamefont {Diner}\ and\ \citenamefont {Rappaport}(2002)}]{annurev}%
  \BibitemOpen
  \bibfield  {author} {\bibinfo {author} {\bibfnamefont {B.~A.}\ \bibnamefont {Diner}}\ and\ \bibinfo {author} {\bibfnamefont {F.}~\bibnamefont {Rappaport}},\ }\href@noop {} {\bibfield  {journal} {\bibinfo  {journal} {Annual Review of Plant Biology}\ }\textbf {\bibinfo {volume} {53}},\ \bibinfo {pages} {551} (\bibinfo {year} {2002})}\BibitemShut {NoStop}%
\bibitem [{\citenamefont {Sirohiwal}\ \emph {et~al.}(2020)\citenamefont {Sirohiwal}, \citenamefont {Neese},\ and\ \citenamefont {Pantazis}}]{sirohiwal2020protein}%
  \BibitemOpen
  \bibfield  {author} {\bibinfo {author} {\bibfnamefont {A.}~\bibnamefont {Sirohiwal}}, \bibinfo {author} {\bibfnamefont {F.}~\bibnamefont {Neese}}, \ and\ \bibinfo {author} {\bibfnamefont {D.~A.}\ \bibnamefont {Pantazis}},\ }\href@noop {} {\bibfield  {journal} {\bibinfo  {journal} {Journal of the American Chemical Society}\ }\textbf {\bibinfo {volume} {142}},\ \bibinfo {pages} {18174} (\bibinfo {year} {2020})}\BibitemShut {NoStop}%
\bibitem [{\citenamefont {Mohseni}\ \emph {et~al.}(2008)\citenamefont {Mohseni}, \citenamefont {Rebentrost}, \citenamefont {Lloyd},\ and\ \citenamefont {Aspuru-Guzik}}]{Mohseni}%
  \BibitemOpen
  \bibfield  {author} {\bibinfo {author} {\bibfnamefont {M.}~\bibnamefont {Mohseni}}, \bibinfo {author} {\bibfnamefont {P.}~\bibnamefont {Rebentrost}}, \bibinfo {author} {\bibfnamefont {S.}~\bibnamefont {Lloyd}}, \ and\ \bibinfo {author} {\bibfnamefont {A.}~\bibnamefont {Aspuru-Guzik}},\ }\href@noop {} {\bibfield  {journal} {\bibinfo  {journal} {The Journal of Chemical Physics}\ }\textbf {\bibinfo {volume} {129}},\ \bibinfo {pages} {174106} (\bibinfo {year} {2008})}\BibitemShut {NoStop}%
\bibitem [{\citenamefont {Skourtis}\ \emph {et~al.}(2010)\citenamefont {Skourtis}, \citenamefont {Waldeck},\ and\ \citenamefont {Beratan}}]{skourtis2010fluctuations}%
  \BibitemOpen
  \bibfield  {author} {\bibinfo {author} {\bibfnamefont {S.~S.}\ \bibnamefont {Skourtis}}, \bibinfo {author} {\bibfnamefont {D.~H.}\ \bibnamefont {Waldeck}}, \ and\ \bibinfo {author} {\bibfnamefont {D.~N.}\ \bibnamefont {Beratan}},\ }\href@noop {} {\bibfield  {journal} {\bibinfo  {journal} {Annual Reviews of Physical Chemistry}\ }\textbf {\bibinfo {volume} {61}},\ \bibinfo {pages} {461} (\bibinfo {year} {2010})}\BibitemShut {NoStop}%
\bibitem [{\citenamefont {Dorfman}\ \emph {et~al.}(2013{\natexlab{a}})\citenamefont {Dorfman}, \citenamefont {Voronine}, \citenamefont {Mukamel},\ and\ \citenamefont {Scully}}]{dorfman2013photosynthetic}%
  \BibitemOpen
  \bibfield  {author} {\bibinfo {author} {\bibfnamefont {K.~E.}\ \bibnamefont {Dorfman}}, \bibinfo {author} {\bibfnamefont {D.~V.}\ \bibnamefont {Voronine}}, \bibinfo {author} {\bibfnamefont {S.}~\bibnamefont {Mukamel}}, \ and\ \bibinfo {author} {\bibfnamefont {M.~O.}\ \bibnamefont {Scully}},\ }\href@noop {} {\bibfield  {journal} {\bibinfo  {journal} {Proceedings of the National Academy of Sciences}\ }\textbf {\bibinfo {volume} {110}},\ \bibinfo {pages} {2746} (\bibinfo {year} {2013}{\natexlab{a}})}\BibitemShut {NoStop}%
\bibitem [{\citenamefont {Rouse}\ \emph {et~al.}(2024)\citenamefont {Rouse}, \citenamefont {Kushwaha}, \citenamefont {Tomasi}, \citenamefont {Lovett}, \citenamefont {Gauger},\ and\ \citenamefont {Kassal}}]{rouse2024light}%
  \BibitemOpen
  \bibfield  {author} {\bibinfo {author} {\bibfnamefont {D.~M.}\ \bibnamefont {Rouse}}, \bibinfo {author} {\bibfnamefont {A.}~\bibnamefont {Kushwaha}}, \bibinfo {author} {\bibfnamefont {S.}~\bibnamefont {Tomasi}}, \bibinfo {author} {\bibfnamefont {B.~W.}\ \bibnamefont {Lovett}}, \bibinfo {author} {\bibfnamefont {E.~M.}\ \bibnamefont {Gauger}}, \ and\ \bibinfo {author} {\bibfnamefont {I.}~\bibnamefont {Kassal}},\ }\href@noop {} {\bibfield  {journal} {\bibinfo  {journal} {The Journal of Physical Chemistry Letters}\ }\textbf {\bibinfo {volume} {15}},\ \bibinfo {pages} {254} (\bibinfo {year} {2024})}\BibitemShut {NoStop}%
\bibitem [{\citenamefont {Werren}\ \emph {et~al.}(2023)\citenamefont {Werren}, \citenamefont {Brown},\ and\ \citenamefont {Gauger}}]{PRXEnergy.2.013002}%
  \BibitemOpen
  \bibfield  {author} {\bibinfo {author} {\bibfnamefont {N.}~\bibnamefont {Werren}}, \bibinfo {author} {\bibfnamefont {W.}~\bibnamefont {Brown}}, \ and\ \bibinfo {author} {\bibfnamefont {E.~M.}\ \bibnamefont {Gauger}},\ }\href {\doibase 10.1103/PRXEnergy.2.013002} {\bibfield  {journal} {\bibinfo  {journal} {PRX Energy}\ }\textbf {\bibinfo {volume} {2}},\ \bibinfo {pages} {013002} (\bibinfo {year} {2023})}\BibitemShut {NoStop}%
\bibitem [{\citenamefont {Wang}\ and\ \citenamefont {Mirza}(2020)}]{wang2020dissipative}%
  \BibitemOpen
  \bibfield  {author} {\bibinfo {author} {\bibfnamefont {Z.}~\bibnamefont {Wang}}\ and\ \bibinfo {author} {\bibfnamefont {I.}~\bibnamefont {Mirza}}\ }(\bibinfo {organization} {Optica Publishing Group},\ \bibinfo {year} {2020})\ pp.\ \bibinfo {pages} {JM6B--26}\BibitemShut {NoStop}%
\bibitem [{\citenamefont {Dodin}\ and\ \citenamefont {Brumer}(2022)}]{dodin2022noise}%
  \BibitemOpen
  \bibfield  {author} {\bibinfo {author} {\bibfnamefont {A.}~\bibnamefont {Dodin}}\ and\ \bibinfo {author} {\bibfnamefont {P.}~\bibnamefont {Brumer}},\ }\href@noop {} {\bibfield  {journal} {\bibinfo  {journal} {Journal of Physics B: Atomic, Molecular and Optical Physics}\ }\textbf {\bibinfo {volume} {54}},\ \bibinfo {pages} {223001} (\bibinfo {year} {2022})}\BibitemShut {NoStop}%
\bibitem [{\citenamefont {Poteshman}\ \emph {et~al.}(2023)\citenamefont {Poteshman}, \citenamefont {Ouellet}, \citenamefont {Bassett},\ and\ \citenamefont {Bassett}}]{poteshman2023network}%
  \BibitemOpen
  \bibfield  {author} {\bibinfo {author} {\bibfnamefont {A.~N.}\ \bibnamefont {Poteshman}}, \bibinfo {author} {\bibfnamefont {M.}~\bibnamefont {Ouellet}}, \bibinfo {author} {\bibfnamefont {L.~C.}\ \bibnamefont {Bassett}}, \ and\ \bibinfo {author} {\bibfnamefont {D.~S.}\ \bibnamefont {Bassett}},\ }\href@noop {} {\bibfield  {journal} {\bibinfo  {journal} {Physical Review Research}\ }\textbf {\bibinfo {volume} {5}},\ \bibinfo {pages} {023125} (\bibinfo {year} {2023})}\BibitemShut {NoStop}%
\bibitem [{\citenamefont {Fang}\ \emph {et~al.}(2019)\citenamefont {Fang}, \citenamefont {Kruse}, \citenamefont {Lu},\ and\ \citenamefont {Wang}}]{fang2019nonequilibrium}%
  \BibitemOpen
  \bibfield  {author} {\bibinfo {author} {\bibfnamefont {X.}~\bibnamefont {Fang}}, \bibinfo {author} {\bibfnamefont {K.}~\bibnamefont {Kruse}}, \bibinfo {author} {\bibfnamefont {T.}~\bibnamefont {Lu}}, \ and\ \bibinfo {author} {\bibfnamefont {J.}~\bibnamefont {Wang}},\ }\href@noop {} {\bibfield  {journal} {\bibinfo  {journal} {Reviews of Modern Physics}\ }\textbf {\bibinfo {volume} {91}},\ \bibinfo {pages} {045004} (\bibinfo {year} {2019})}\BibitemShut {NoStop}%
\bibitem [{\citenamefont {Joubert-Doriol}\ \emph {et~al.}(2023)\citenamefont {Joubert-Doriol}, \citenamefont {Jung}, \citenamefont {Izmaylov},\ and\ \citenamefont {Brumer}}]{joubert2023quantum}%
  \BibitemOpen
  \bibfield  {author} {\bibinfo {author} {\bibfnamefont {L.}~\bibnamefont {Joubert-Doriol}}, \bibinfo {author} {\bibfnamefont {K.~A.}\ \bibnamefont {Jung}}, \bibinfo {author} {\bibfnamefont {A.~F.}\ \bibnamefont {Izmaylov}}, \ and\ \bibinfo {author} {\bibfnamefont {P.}~\bibnamefont {Brumer}},\ }\href@noop {} {\bibfield  {journal} {\bibinfo  {journal} {Journal of Chemical Theory and Computation}\ }\textbf {\bibinfo {volume} {19}},\ \bibinfo {pages} {1130} (\bibinfo {year} {2023})}\BibitemShut {NoStop}%
\bibitem [{\citenamefont {Singh}\ and\ \citenamefont {Brumer}(2011)}]{singh2011electronic}%
  \BibitemOpen
  \bibfield  {author} {\bibinfo {author} {\bibfnamefont {N.}~\bibnamefont {Singh}}\ and\ \bibinfo {author} {\bibfnamefont {P.}~\bibnamefont {Brumer}},\ }\href@noop {} {\bibfield  {journal} {\bibinfo  {journal} {Faraday Discussions}\ }\textbf {\bibinfo {volume} {153}},\ \bibinfo {pages} {41} (\bibinfo {year} {2011})}\BibitemShut {NoStop}%
\bibitem [{\citenamefont {Yang}\ and\ \citenamefont {Cao}(2020)}]{yang2020steady}%
  \BibitemOpen
  \bibfield  {author} {\bibinfo {author} {\bibfnamefont {P.-Y.}\ \bibnamefont {Yang}}\ and\ \bibinfo {author} {\bibfnamefont {J.}~\bibnamefont {Cao}},\ }\href@noop {} {\bibfield  {journal} {\bibinfo  {journal} {The Journal of Physical Chemistry Letters}\ }\textbf {\bibinfo {volume} {11}},\ \bibinfo {pages} {7204} (\bibinfo {year} {2020})}\BibitemShut {NoStop}%
\bibitem [{\citenamefont {Chen}\ \emph {et~al.}(2015)\citenamefont {Chen}, \citenamefont {Lambert}, \citenamefont {Cheng}, \citenamefont {Chen},\ and\ \citenamefont {Nori}}]{chen2015using}%
  \BibitemOpen
  \bibfield  {author} {\bibinfo {author} {\bibfnamefont {H.-B.}\ \bibnamefont {Chen}}, \bibinfo {author} {\bibfnamefont {N.}~\bibnamefont {Lambert}}, \bibinfo {author} {\bibfnamefont {Y.-C.}\ \bibnamefont {Cheng}}, \bibinfo {author} {\bibfnamefont {Y.-N.}\ \bibnamefont {Chen}}, \ and\ \bibinfo {author} {\bibfnamefont {F.}~\bibnamefont {Nori}},\ }\href@noop {} {\bibfield  {journal} {\bibinfo  {journal} {Scientific reports}\ }\textbf {\bibinfo {volume} {5}},\ \bibinfo {pages} {12753} (\bibinfo {year} {2015})}\BibitemShut {NoStop}%
\bibitem [{\citenamefont {Lambert}\ \emph {et~al.}(2023)\citenamefont {Lambert}, \citenamefont {Raheja}, \citenamefont {Cross}, \citenamefont {Menczel}, \citenamefont {Ahmed}, \citenamefont {Pitchford}, \citenamefont {Burgarth},\ and\ \citenamefont {Nori}}]{PhysRevResearch.5.013181}%
  \BibitemOpen
  \bibfield  {author} {\bibinfo {author} {\bibfnamefont {N.}~\bibnamefont {Lambert}}, \bibinfo {author} {\bibfnamefont {T.}~\bibnamefont {Raheja}}, \bibinfo {author} {\bibfnamefont {S.}~\bibnamefont {Cross}}, \bibinfo {author} {\bibfnamefont {P.}~\bibnamefont {Menczel}}, \bibinfo {author} {\bibfnamefont {S.}~\bibnamefont {Ahmed}}, \bibinfo {author} {\bibfnamefont {A.}~\bibnamefont {Pitchford}}, \bibinfo {author} {\bibfnamefont {D.}~\bibnamefont {Burgarth}}, \ and\ \bibinfo {author} {\bibfnamefont {F.}~\bibnamefont {Nori}},\ }\href {\doibase 10.1103/PhysRevResearch.5.013181} {\bibfield  {journal} {\bibinfo  {journal} {Phys. Rev. Res.}\ }\textbf {\bibinfo {volume} {5}},\ \bibinfo {pages} {013181} (\bibinfo {year} {2023})}\BibitemShut {NoStop}%
\bibitem [{\citenamefont {Zhang}\ \emph {et~al.}(2023)\citenamefont {Zhang}, \citenamefont {Liu}, \citenamefont {Jiang},\ and\ \citenamefont {Ma}}]{zhang2023many}%
  \BibitemOpen
  \bibfield  {author} {\bibinfo {author} {\bibfnamefont {M.}~\bibnamefont {Zhang}}, \bibinfo {author} {\bibfnamefont {Y.}~\bibnamefont {Liu}}, \bibinfo {author} {\bibfnamefont {Y.-n.}\ \bibnamefont {Jiang}}, \ and\ \bibinfo {author} {\bibfnamefont {Y.}~\bibnamefont {Ma}},\ }\href@noop {} {\bibfield  {journal} {\bibinfo  {journal} {The Journal of Physical Chemistry Letters}\ }\textbf {\bibinfo {volume} {14}},\ \bibinfo {pages} {5267} (\bibinfo {year} {2023})}\BibitemShut {NoStop}%
\bibitem [{\citenamefont {Harbola}\ and\ \citenamefont {Mukamel}(2008)}]{harbola2008superoperator}%
  \BibitemOpen
  \bibfield  {author} {\bibinfo {author} {\bibfnamefont {U.}~\bibnamefont {Harbola}}\ and\ \bibinfo {author} {\bibfnamefont {S.}~\bibnamefont {Mukamel}},\ }\href@noop {} {\bibfield  {journal} {\bibinfo  {journal} {Physics Reports}\ }\textbf {\bibinfo {volume} {465}},\ \bibinfo {pages} {191} (\bibinfo {year} {2008})}\BibitemShut {NoStop}%
\bibitem [{\citenamefont {Levi}\ \emph {et~al.}(2015)\citenamefont {Levi}, \citenamefont {Mostarda}, \citenamefont {Rao},\ and\ \citenamefont {Mintert}}]{levi2015quantum}%
  \BibitemOpen
  \bibfield  {author} {\bibinfo {author} {\bibfnamefont {F.}~\bibnamefont {Levi}}, \bibinfo {author} {\bibfnamefont {S.}~\bibnamefont {Mostarda}}, \bibinfo {author} {\bibfnamefont {F.}~\bibnamefont {Rao}}, \ and\ \bibinfo {author} {\bibfnamefont {F.}~\bibnamefont {Mintert}},\ }\href@noop {} {\bibfield  {journal} {\bibinfo  {journal} {Reports on Progress in Physics}\ }\textbf {\bibinfo {volume} {78}},\ \bibinfo {pages} {082001} (\bibinfo {year} {2015})}\BibitemShut {NoStop}%
\bibitem [{\citenamefont {Karafyllidis}(2017)}]{karafyllidis2017quantum}%
  \BibitemOpen
  \bibfield  {author} {\bibinfo {author} {\bibfnamefont {I.~G.}\ \bibnamefont {Karafyllidis}},\ }\href@noop {} {\bibfield  {journal} {\bibinfo  {journal} {Journal of Biological Physics}\ }\textbf {\bibinfo {volume} {43}},\ \bibinfo {pages} {239} (\bibinfo {year} {2017})}\BibitemShut {NoStop}%
\bibitem [{\citenamefont {Suess}\ \emph {et~al.}(2014)\citenamefont {Suess}, \citenamefont {Eisfeld},\ and\ \citenamefont {Strunz}}]{suess2014hierarchy}%
  \BibitemOpen
  \bibfield  {author} {\bibinfo {author} {\bibfnamefont {D.}~\bibnamefont {Suess}}, \bibinfo {author} {\bibfnamefont {A.}~\bibnamefont {Eisfeld}}, \ and\ \bibinfo {author} {\bibfnamefont {W.}~\bibnamefont {Strunz}},\ }\href@noop {} {\bibfield  {journal} {\bibinfo  {journal} {Physical Review Letters}\ }\textbf {\bibinfo {volume} {113}},\ \bibinfo {pages} {150403} (\bibinfo {year} {2014})}\BibitemShut {NoStop}%
\bibitem [{\citenamefont {Timm}(2008)}]{timm2008tunneling}%
  \BibitemOpen
  \bibfield  {author} {\bibinfo {author} {\bibfnamefont {C.}~\bibnamefont {Timm}},\ }\href@noop {} {\bibfield  {journal} {\bibinfo  {journal} {Physical Review B—Condensed Matter and Materials Physics}\ }\textbf {\bibinfo {volume} {77}},\ \bibinfo {pages} {195416} (\bibinfo {year} {2008})}\BibitemShut {NoStop}%
\bibitem [{\citenamefont {Welack}\ \emph {et~al.}(2008)\citenamefont {Welack}, \citenamefont {Maddox}, \citenamefont {Esposito}, \citenamefont {Harbola},\ and\ \citenamefont {Mukamel}}]{welack2008single}%
  \BibitemOpen
  \bibfield  {author} {\bibinfo {author} {\bibfnamefont {S.}~\bibnamefont {Welack}}, \bibinfo {author} {\bibfnamefont {J.~B.}\ \bibnamefont {Maddox}}, \bibinfo {author} {\bibfnamefont {M.}~\bibnamefont {Esposito}}, \bibinfo {author} {\bibfnamefont {U.}~\bibnamefont {Harbola}}, \ and\ \bibinfo {author} {\bibfnamefont {S.}~\bibnamefont {Mukamel}},\ }\href@noop {} {\bibfield  {journal} {\bibinfo  {journal} {Nano letters}\ }\textbf {\bibinfo {volume} {8}},\ \bibinfo {pages} {1137} (\bibinfo {year} {2008})}\BibitemShut {NoStop}%
\bibitem [{\citenamefont {Papp}\ and\ \citenamefont {Vattay}(2024)}]{papp2024computation}%
  \BibitemOpen
  \bibfield  {author} {\bibinfo {author} {\bibfnamefont {E.}~\bibnamefont {Papp}}\ and\ \bibinfo {author} {\bibfnamefont {G.}~\bibnamefont {Vattay}},\ }\href@noop {} {\bibfield  {journal} {\bibinfo  {journal} {Scientific Reports}\ }\textbf {\bibinfo {volume} {14}},\ \bibinfo {pages} {19571} (\bibinfo {year} {2024})}\BibitemShut {NoStop}%
\bibitem [{\citenamefont {Duan}\ \emph {et~al.}(2022)\citenamefont {Duan}, \citenamefont {Jha}, \citenamefont {Chen}, \citenamefont {Tiwari}, \citenamefont {Cogdell}, \citenamefont {Ashraf}, \citenamefont {Prokhorenko}, \citenamefont {Thorwart},\ and\ \citenamefont {Miller}}]{Hong-Guang}%
  \BibitemOpen
  \bibfield  {author} {\bibinfo {author} {\bibfnamefont {H.-G.}\ \bibnamefont {Duan}}, \bibinfo {author} {\bibfnamefont {A.}~\bibnamefont {Jha}}, \bibinfo {author} {\bibfnamefont {L.}~\bibnamefont {Chen}}, \bibinfo {author} {\bibfnamefont {V.}~\bibnamefont {Tiwari}}, \bibinfo {author} {\bibfnamefont {R.~J.}\ \bibnamefont {Cogdell}}, \bibinfo {author} {\bibfnamefont {K.}~\bibnamefont {Ashraf}}, \bibinfo {author} {\bibfnamefont {V.~I.}\ \bibnamefont {Prokhorenko}}, \bibinfo {author} {\bibfnamefont {M.}~\bibnamefont {Thorwart}}, \ and\ \bibinfo {author} {\bibfnamefont {R.~J.~D.}\ \bibnamefont {Miller}},\ }\href@noop {} {\bibfield  {journal} {\bibinfo  {journal} {Proceedings of the National Academy of Sciences}\ }\textbf {\bibinfo {volume} {119}},\ \bibinfo {pages} {e2212630119} (\bibinfo {year} {2022})}\BibitemShut {NoStop}%
\bibitem [{\citenamefont {Zhou}\ \emph {et~al.}(2021)\citenamefont {Zhou}, \citenamefont {Park}, \citenamefont {Timrov}, \citenamefont {Floris}, \citenamefont {Cococcioni}, \citenamefont {Marzari},\ and\ \citenamefont {Bernardi}}]{Zhou}%
  \BibitemOpen
  \bibfield  {author} {\bibinfo {author} {\bibfnamefont {J.-J.}\ \bibnamefont {Zhou}}, \bibinfo {author} {\bibfnamefont {J.}~\bibnamefont {Park}}, \bibinfo {author} {\bibfnamefont {I.}~\bibnamefont {Timrov}}, \bibinfo {author} {\bibfnamefont {A.}~\bibnamefont {Floris}}, \bibinfo {author} {\bibfnamefont {M.}~\bibnamefont {Cococcioni}}, \bibinfo {author} {\bibfnamefont {N.}~\bibnamefont {Marzari}}, \ and\ \bibinfo {author} {\bibfnamefont {M.}~\bibnamefont {Bernardi}},\ }\href@noop {} {\bibfield  {journal} {\bibinfo  {journal} {Phys. Rev. Lett.}\ }\textbf {\bibinfo {volume} {127}},\ \bibinfo {pages} {126404} (\bibinfo {year} {2021})}\BibitemShut {NoStop}%
\bibitem [{\citenamefont {Jha}\ \emph {et~al.}(2024)\citenamefont {Jha}, \citenamefont {Zhang}, \citenamefont {Tiwari}, \citenamefont {Chen}, \citenamefont {Thorwart}, \citenamefont {Miller},\ and\ \citenamefont {Duan}}]{jha2024unraveling}%
  \BibitemOpen
  \bibfield  {author} {\bibinfo {author} {\bibfnamefont {A.}~\bibnamefont {Jha}}, \bibinfo {author} {\bibfnamefont {P.-P.}\ \bibnamefont {Zhang}}, \bibinfo {author} {\bibfnamefont {V.}~\bibnamefont {Tiwari}}, \bibinfo {author} {\bibfnamefont {L.}~\bibnamefont {Chen}}, \bibinfo {author} {\bibfnamefont {M.}~\bibnamefont {Thorwart}}, \bibinfo {author} {\bibfnamefont {R.~D.}\ \bibnamefont {Miller}}, \ and\ \bibinfo {author} {\bibfnamefont {H.-G.}\ \bibnamefont {Duan}},\ }\href@noop {} {\bibfield  {journal} {\bibinfo  {journal} {Science Advances}\ }\textbf {\bibinfo {volume} {10}},\ \bibinfo {pages} {eadk1312} (\bibinfo {year} {2024})}\BibitemShut {NoStop}%
\bibitem [{\citenamefont {Gerster}\ \emph {et~al.}(2012)\citenamefont {Gerster}, \citenamefont {Reichert}, \citenamefont {Bi}, \citenamefont {Barth}, \citenamefont {Kaniber}, \citenamefont {Holleitner}, \citenamefont {Visoly-Fisher}, \citenamefont {Sergani},\ and\ \citenamefont {Carmeli}}]{Gerster2012}%
  \BibitemOpen
  \bibfield  {author} {\bibinfo {author} {\bibfnamefont {D.}~\bibnamefont {Gerster}}, \bibinfo {author} {\bibfnamefont {J.}~\bibnamefont {Reichert}}, \bibinfo {author} {\bibfnamefont {H.}~\bibnamefont {Bi}}, \bibinfo {author} {\bibfnamefont {J.~V.}\ \bibnamefont {Barth}}, \bibinfo {author} {\bibfnamefont {S.~M.}\ \bibnamefont {Kaniber}}, \bibinfo {author} {\bibfnamefont {A.~W.}\ \bibnamefont {Holleitner}}, \bibinfo {author} {\bibfnamefont {I.}~\bibnamefont {Visoly-Fisher}}, \bibinfo {author} {\bibfnamefont {S.}~\bibnamefont {Sergani}}, \ and\ \bibinfo {author} {\bibfnamefont {I.}~\bibnamefont {Carmeli}},\ }\href@noop {} {\bibfield  {journal} {\bibinfo  {journal} {Nature Nanotechnology}\ }\textbf {\bibinfo {volume} {7}},\ \bibinfo {pages} {673} (\bibinfo {year} {2012})}\BibitemShut {NoStop}%
\bibitem [{\citenamefont {Pillai}\ \emph {et~al.}(2007)\citenamefont {Pillai}, \citenamefont {Catchpole}, \citenamefont {Trupke},\ and\ \citenamefont {Green}}]{Pillai}%
  \BibitemOpen
  \bibfield  {author} {\bibinfo {author} {\bibfnamefont {S.}~\bibnamefont {Pillai}}, \bibinfo {author} {\bibfnamefont {K.~R.}\ \bibnamefont {Catchpole}}, \bibinfo {author} {\bibfnamefont {T.}~\bibnamefont {Trupke}}, \ and\ \bibinfo {author} {\bibfnamefont {M.~A.}\ \bibnamefont {Green}},\ }\href@noop {} {\bibfield  {journal} {\bibinfo  {journal} {Journal of Applied Physics}\ }\textbf {\bibinfo {volume} {101}},\ \bibinfo {pages} {093105} (\bibinfo {year} {2007})}\BibitemShut {NoStop}%
\bibitem [{\citenamefont {Cabrera-Tinoco}\ \emph {et~al.}(2023)\citenamefont {Cabrera-Tinoco}, \citenamefont {Moreira}, \citenamefont {Borja-Castro}, \citenamefont {Valencia-Bedregal}, \citenamefont {Barnes},\ and\ \citenamefont {Santos~Valladares}}]{cabrera2023charge}%
  \BibitemOpen
  \bibfield  {author} {\bibinfo {author} {\bibfnamefont {H.}~\bibnamefont {Cabrera-Tinoco}}, \bibinfo {author} {\bibfnamefont {A.~C.}\ \bibnamefont {Moreira}}, \bibinfo {author} {\bibfnamefont {L.}~\bibnamefont {Borja-Castro}}, \bibinfo {author} {\bibfnamefont {R.}~\bibnamefont {Valencia-Bedregal}}, \bibinfo {author} {\bibfnamefont {C.~H.}\ \bibnamefont {Barnes}}, \ and\ \bibinfo {author} {\bibfnamefont {L.~d.~l.}\ \bibnamefont {Santos~Valladares}},\ }\href@noop {} {\bibfield  {journal} {\bibinfo  {journal} {The Journal of Physical Chemistry A}\ }\textbf {\bibinfo {volume} {127}},\ \bibinfo {pages} {10828} (\bibinfo {year} {2023})}\BibitemShut {NoStop}%
\bibitem [{\citenamefont {Weymann}\ \emph {et~al.}(2011)\citenamefont {Weymann}, \citenamefont {Bu{\l}ka},\ and\ \citenamefont {Barna{\'s}}}]{weymann2011dark}%
  \BibitemOpen
  \bibfield  {author} {\bibinfo {author} {\bibfnamefont {I.}~\bibnamefont {Weymann}}, \bibinfo {author} {\bibfnamefont {B.}~\bibnamefont {Bu{\l}ka}}, \ and\ \bibinfo {author} {\bibfnamefont {J.}~\bibnamefont {Barna{\'s}}},\ }\href@noop {} {\bibfield  {journal} {\bibinfo  {journal} {Physical Review B—Condensed Matter and Materials Physics}\ }\textbf {\bibinfo {volume} {83}},\ \bibinfo {pages} {195302} (\bibinfo {year} {2011})}\BibitemShut {NoStop}%
\bibitem [{\citenamefont {Bian}\ \emph {et~al.}(2022)\citenamefont {Bian}, \citenamefont {Chen}, \citenamefont {Sowa}, \citenamefont {Evangeli}, \citenamefont {Limburg}, \citenamefont {Swett}, \citenamefont {Baugh}, \citenamefont {Briggs}, \citenamefont {Anderson}, \citenamefont {Mol} \emph {et~al.}}]{bian2022charge}%
  \BibitemOpen
  \bibfield  {author} {\bibinfo {author} {\bibfnamefont {X.}~\bibnamefont {Bian}}, \bibinfo {author} {\bibfnamefont {Z.}~\bibnamefont {Chen}}, \bibinfo {author} {\bibfnamefont {J.~K.}\ \bibnamefont {Sowa}}, \bibinfo {author} {\bibfnamefont {C.}~\bibnamefont {Evangeli}}, \bibinfo {author} {\bibfnamefont {B.}~\bibnamefont {Limburg}}, \bibinfo {author} {\bibfnamefont {J.~L.}\ \bibnamefont {Swett}}, \bibinfo {author} {\bibfnamefont {J.}~\bibnamefont {Baugh}}, \bibinfo {author} {\bibfnamefont {G.~A.~D.}\ \bibnamefont {Briggs}}, \bibinfo {author} {\bibfnamefont {H.~L.}\ \bibnamefont {Anderson}}, \bibinfo {author} {\bibfnamefont {J.~A.}\ \bibnamefont {Mol}},  \emph {et~al.},\ }\href@noop {} {\bibfield  {journal} {\bibinfo  {journal} {Physical Review Letters}\ }\textbf {\bibinfo {volume} {129}},\ \bibinfo {pages} {207702} (\bibinfo {year} {2022})}\BibitemShut {NoStop}%
\bibitem [{\citenamefont {Donarini}\ and\ \citenamefont {Grifoni}(2024)}]{donarini2024transport}%
  \BibitemOpen
  \bibfield  {author} {\bibinfo {author} {\bibfnamefont {A.}~\bibnamefont {Donarini}}\ and\ \bibinfo {author} {\bibfnamefont {M.}~\bibnamefont {Grifoni}}\ }(\bibinfo  {publisher} {Springer},\ \bibinfo {year} {2024})\ pp.\ \bibinfo {pages} {365--410}\BibitemShut {NoStop}%
\bibitem [{\citenamefont {Hsiao}\ \emph {et~al.}(2024)\citenamefont {Hsiao}, \citenamefont {Cova~Fari{\~n}a}, \citenamefont {Oosterhout}, \citenamefont {Jirovec}, \citenamefont {Zhang}, \citenamefont {van Diepen}, \citenamefont {Lawrie}, \citenamefont {Wang}, \citenamefont {Sammak}, \citenamefont {Scappucci} \emph {et~al.}}]{hsiao2024exciton}%
  \BibitemOpen
  \bibfield  {author} {\bibinfo {author} {\bibfnamefont {T.-K.}\ \bibnamefont {Hsiao}}, \bibinfo {author} {\bibfnamefont {P.}~\bibnamefont {Cova~Fari{\~n}a}}, \bibinfo {author} {\bibfnamefont {S.~D.}\ \bibnamefont {Oosterhout}}, \bibinfo {author} {\bibfnamefont {D.}~\bibnamefont {Jirovec}}, \bibinfo {author} {\bibfnamefont {X.}~\bibnamefont {Zhang}}, \bibinfo {author} {\bibfnamefont {C.~J.}\ \bibnamefont {van Diepen}}, \bibinfo {author} {\bibfnamefont {W.}~\bibnamefont {Lawrie}}, \bibinfo {author} {\bibfnamefont {C.-A.}\ \bibnamefont {Wang}}, \bibinfo {author} {\bibfnamefont {A.}~\bibnamefont {Sammak}}, \bibinfo {author} {\bibfnamefont {G.}~\bibnamefont {Scappucci}},  \emph {et~al.},\ }\href@noop {} {\bibfield  {journal} {\bibinfo  {journal} {Physical Review X}\ }\textbf {\bibinfo {volume} {14}},\ \bibinfo {pages} {011048} (\bibinfo {year} {2024})}\BibitemShut {NoStop}%
\bibitem [{\citenamefont {Carmi}\ and\ \citenamefont {Oreg}(2012)}]{carmi2012enhanced}%
  \BibitemOpen
  \bibfield  {author} {\bibinfo {author} {\bibfnamefont {A.}~\bibnamefont {Carmi}}\ and\ \bibinfo {author} {\bibfnamefont {Y.}~\bibnamefont {Oreg}},\ }\href@noop {} {\bibfield  {journal} {\bibinfo  {journal} {Physical Review B}\ }\textbf {\bibinfo {volume} {85}},\ \bibinfo {pages} {045325} (\bibinfo {year} {2012})}\BibitemShut {NoStop}%
\bibitem [{\citenamefont {Sandilya}\ \emph {et~al.}(2024)\citenamefont {Sandilya}, \citenamefont {Akhtar}, \citenamefont {Sarmah},\ and\ \citenamefont {Goswami}}]{sandilya2024cotunneling}%
  \BibitemOpen
  \bibfield  {author} {\bibinfo {author} {\bibfnamefont {M.}~\bibnamefont {Sandilya}}, \bibinfo {author} {\bibfnamefont {J.}~\bibnamefont {Akhtar}}, \bibinfo {author} {\bibfnamefont {M.~J.}\ \bibnamefont {Sarmah}}, \ and\ \bibinfo {author} {\bibfnamefont {H.~P.}\ \bibnamefont {Goswami}},\ }\href {https://doi.org/10.1002/andp.202400143} {\bibfield  {journal} {\bibinfo  {journal} {Annalen der Physik}\ }\textbf {\bibinfo {volume} {536}},\ \bibinfo {pages} {2400143} (\bibinfo {year} {2024})}\BibitemShut {NoStop}%
\bibitem [{\citenamefont {Ferreira}\ \emph {et~al.}(2004)\citenamefont {Ferreira}, \citenamefont {Iverson}, \citenamefont {Maghlaoui}, \citenamefont {Barber},\ and\ \citenamefont {Iwata}}]{ferreira2004architecture}%
  \BibitemOpen
  \bibfield  {author} {\bibinfo {author} {\bibfnamefont {K.~N.}\ \bibnamefont {Ferreira}}, \bibinfo {author} {\bibfnamefont {T.~M.}\ \bibnamefont {Iverson}}, \bibinfo {author} {\bibfnamefont {K.}~\bibnamefont {Maghlaoui}}, \bibinfo {author} {\bibfnamefont {J.}~\bibnamefont {Barber}}, \ and\ \bibinfo {author} {\bibfnamefont {S.}~\bibnamefont {Iwata}},\ }\href@noop {} {\bibfield  {journal} {\bibinfo  {journal} {Science}\ }\textbf {\bibinfo {volume} {303}},\ \bibinfo {pages} {1831} (\bibinfo {year} {2004})}\BibitemShut {NoStop}%
\bibitem [{\citenamefont {Umena}\ \emph {et~al.}(2011)\citenamefont {Umena}, \citenamefont {Kawakami}, \citenamefont {Shen},\ and\ \citenamefont {Kamiya}}]{Umena2011}%
  \BibitemOpen
  \bibfield  {author} {\bibinfo {author} {\bibfnamefont {Y.}~\bibnamefont {Umena}}, \bibinfo {author} {\bibfnamefont {K.}~\bibnamefont {Kawakami}}, \bibinfo {author} {\bibfnamefont {J.-R.}\ \bibnamefont {Shen}}, \ and\ \bibinfo {author} {\bibfnamefont {N.}~\bibnamefont {Kamiya}},\ }\href@noop {} {\bibfield  {journal} {\bibinfo  {journal} {Nature}\ }\textbf {\bibinfo {volume} {473}},\ \bibinfo {pages} {55} (\bibinfo {year} {2011})}\BibitemShut {NoStop}%
\bibitem [{\citenamefont {Zouni}\ \emph {et~al.}(2001)\citenamefont {Zouni}, \citenamefont {Witt}, \citenamefont {Kern}, \citenamefont {Fromme}, \citenamefont {Krauss}, \citenamefont {Saenger},\ and\ \citenamefont {Orth}}]{Zouni2001}%
  \BibitemOpen
  \bibfield  {author} {\bibinfo {author} {\bibfnamefont {A.}~\bibnamefont {Zouni}}, \bibinfo {author} {\bibfnamefont {H.-T.}\ \bibnamefont {Witt}}, \bibinfo {author} {\bibfnamefont {J.}~\bibnamefont {Kern}}, \bibinfo {author} {\bibfnamefont {P.}~\bibnamefont {Fromme}}, \bibinfo {author} {\bibfnamefont {N.}~\bibnamefont {Krauss}}, \bibinfo {author} {\bibfnamefont {W.}~\bibnamefont {Saenger}}, \ and\ \bibinfo {author} {\bibfnamefont {P.}~\bibnamefont {Orth}},\ }\href@noop {} {\bibfield  {journal} {\bibinfo  {journal} {Nature}\ }\textbf {\bibinfo {volume} {409}},\ \bibinfo {pages} {739} (\bibinfo {year} {2001})}\BibitemShut {NoStop}%
\bibitem [{\citenamefont {Tao}(2006)}]{Tao2006}%
  \BibitemOpen
  \bibfield  {author} {\bibinfo {author} {\bibfnamefont {N.~J.}\ \bibnamefont {Tao}},\ }\href@noop {} {\bibfield  {journal} {\bibinfo  {journal} {Nature Nanotechnology}\ }\textbf {\bibinfo {volume} {1}},\ \bibinfo {pages} {173} (\bibinfo {year} {2006})}\BibitemShut {NoStop}%
\bibitem [{\citenamefont {Brixner}\ \emph {et~al.}(2005)\citenamefont {Brixner}, \citenamefont {Stenger}, \citenamefont {Vaswani}, \citenamefont {Cho}, \citenamefont {Blankenship},\ and\ \citenamefont {Fleming}}]{Brixner2005}%
  \BibitemOpen
  \bibfield  {author} {\bibinfo {author} {\bibfnamefont {T.}~\bibnamefont {Brixner}}, \bibinfo {author} {\bibfnamefont {J.}~\bibnamefont {Stenger}}, \bibinfo {author} {\bibfnamefont {H.~M.}\ \bibnamefont {Vaswani}}, \bibinfo {author} {\bibfnamefont {M.}~\bibnamefont {Cho}}, \bibinfo {author} {\bibfnamefont {R.~E.}\ \bibnamefont {Blankenship}}, \ and\ \bibinfo {author} {\bibfnamefont {G.~R.}\ \bibnamefont {Fleming}},\ }\href@noop {} {\bibfield  {journal} {\bibinfo  {journal} {Nature}\ }\textbf {\bibinfo {volume} {434}},\ \bibinfo {pages} {625} (\bibinfo {year} {2005})}\BibitemShut {NoStop}%
\bibitem [{\citenamefont {Duan}\ \emph {et~al.}(2017)\citenamefont {Duan}, \citenamefont {Prokhorenko}, \citenamefont {Wientjes}, \citenamefont {Croce}, \citenamefont {Thorwart},\ and\ \citenamefont {Miller}}]{Duan2017}%
  \BibitemOpen
  \bibfield  {author} {\bibinfo {author} {\bibfnamefont {H.-G.}\ \bibnamefont {Duan}}, \bibinfo {author} {\bibfnamefont {V.~I.}\ \bibnamefont {Prokhorenko}}, \bibinfo {author} {\bibfnamefont {E.}~\bibnamefont {Wientjes}}, \bibinfo {author} {\bibfnamefont {R.}~\bibnamefont {Croce}}, \bibinfo {author} {\bibfnamefont {M.}~\bibnamefont {Thorwart}}, \ and\ \bibinfo {author} {\bibfnamefont {R.~J.~D.}\ \bibnamefont {Miller}},\ }\href@noop {} {\bibfield  {journal} {\bibinfo  {journal} {Scientific Reports}\ }\textbf {\bibinfo {volume} {7}},\ \bibinfo {pages} {12347} (\bibinfo {year} {2017})}\BibitemShut {NoStop}%
\bibitem [{\citenamefont {Novoderezhkin}\ \emph {et~al.}(2007)\citenamefont {Novoderezhkin}, \citenamefont {Dekker},\ and\ \citenamefont {Van~Grondelle}}]{novoderezhkin2007mixing}%
  \BibitemOpen
  \bibfield  {author} {\bibinfo {author} {\bibfnamefont {V.~I.}\ \bibnamefont {Novoderezhkin}}, \bibinfo {author} {\bibfnamefont {J.~P.}\ \bibnamefont {Dekker}}, \ and\ \bibinfo {author} {\bibfnamefont {R.}~\bibnamefont {Van~Grondelle}},\ }\href@noop {} {\bibfield  {journal} {\bibinfo  {journal} {Biophysical Journal}\ }\textbf {\bibinfo {volume} {93}},\ \bibinfo {pages} {1293} (\bibinfo {year} {2007})}\BibitemShut {NoStop}%
\bibitem [{\citenamefont {Goswami}\ \emph {et~al.}(2015)\citenamefont {Goswami}, \citenamefont {Hua}, \citenamefont {Zhang}, \citenamefont {Mukamel},\ and\ \citenamefont {Harbola}}]{goswami2015electroluminescence}%
  \BibitemOpen
  \bibfield  {author} {\bibinfo {author} {\bibfnamefont {H.~P.}\ \bibnamefont {Goswami}}, \bibinfo {author} {\bibfnamefont {W.}~\bibnamefont {Hua}}, \bibinfo {author} {\bibfnamefont {Y.}~\bibnamefont {Zhang}}, \bibinfo {author} {\bibfnamefont {S.}~\bibnamefont {Mukamel}}, \ and\ \bibinfo {author} {\bibfnamefont {U.}~\bibnamefont {Harbola}},\ }\href {https://pubs.acs.org/doi/abs/10.1021/acs.jctc.5b00500} {\bibfield  {journal} {\bibinfo  {journal} {Journal of Chemical Theory and Computation}\ }\textbf {\bibinfo {volume} {11}},\ \bibinfo {pages} {4304} (\bibinfo {year} {2015})}\BibitemShut {NoStop}%
\bibitem [{\citenamefont {Brunk}\ and\ \citenamefont {Rothlisberger}(2015)}]{Brunk}%
  \BibitemOpen
  \bibfield  {author} {\bibinfo {author} {\bibfnamefont {E.}~\bibnamefont {Brunk}}\ and\ \bibinfo {author} {\bibfnamefont {U.}~\bibnamefont {Rothlisberger}},\ }\href@noop {} {\bibfield  {journal} {\bibinfo  {journal} {Chemical reviews}\ }\textbf {\bibinfo {volume} {115}},\ \bibinfo {pages} {6217} (\bibinfo {year} {2015})}\BibitemShut {NoStop}%
\bibitem [{\citenamefont {Aghassi}\ \emph {et~al.}(2008)\citenamefont {Aghassi}, \citenamefont {Hettler},\ and\ \citenamefont {Schön}}]{Aghassi}%
  \BibitemOpen
  \bibfield  {author} {\bibinfo {author} {\bibfnamefont {J.}~\bibnamefont {Aghassi}}, \bibinfo {author} {\bibfnamefont {M.~H.}\ \bibnamefont {Hettler}}, \ and\ \bibinfo {author} {\bibfnamefont {G.}~\bibnamefont {Schön}},\ }\href@noop {} {\bibfield  {journal} {\bibinfo  {journal} {Applied Physics Letters}\ }\textbf {\bibinfo {volume} {92}},\ \bibinfo {pages} {202101} (\bibinfo {year} {2008})}\BibitemShut {NoStop}%
\bibitem [{\citenamefont {Dorfman}\ \emph {et~al.}(2013{\natexlab{b}})\citenamefont {Dorfman}, \citenamefont {Voronine}, \citenamefont {Mukamel},\ and\ \citenamefont {Scully}}]{Konstantin}%
  \BibitemOpen
  \bibfield  {author} {\bibinfo {author} {\bibfnamefont {K.~E.}\ \bibnamefont {Dorfman}}, \bibinfo {author} {\bibfnamefont {D.~V.}\ \bibnamefont {Voronine}}, \bibinfo {author} {\bibfnamefont {S.}~\bibnamefont {Mukamel}}, \ and\ \bibinfo {author} {\bibfnamefont {M.~O.}\ \bibnamefont {Scully}},\ }\href@noop {} {\bibfield  {journal} {\bibinfo  {journal} {Proceedings of the National Academy of Sciences}\ }\textbf {\bibinfo {volume} {110}},\ \bibinfo {pages} {2746} (\bibinfo {year} {2013}{\natexlab{b}})}\BibitemShut {NoStop}%
\bibitem [{\citenamefont {Creatore}\ \emph {et~al.}(2013)\citenamefont {Creatore}, \citenamefont {Parker}, \citenamefont {Emmott},\ and\ \citenamefont {Chin}}]{Creatore}%
  \BibitemOpen
  \bibfield  {author} {\bibinfo {author} {\bibfnamefont {C.}~\bibnamefont {Creatore}}, \bibinfo {author} {\bibfnamefont {M.~A.}\ \bibnamefont {Parker}}, \bibinfo {author} {\bibfnamefont {S.}~\bibnamefont {Emmott}}, \ and\ \bibinfo {author} {\bibfnamefont {A.~W.}\ \bibnamefont {Chin}},\ }\href@noop {} {\bibfield  {journal} {\bibinfo  {journal} {Phys. Rev. Lett.}\ }\textbf {\bibinfo {volume} {111}},\ \bibinfo {pages} {253601} (\bibinfo {year} {2013})}\BibitemShut {NoStop}%
\bibitem [{\citenamefont {Yang}\ and\ \citenamefont {Fleming}(2002)}]{YANG2002355}%
  \BibitemOpen
  \bibfield  {author} {\bibinfo {author} {\bibfnamefont {M.}~\bibnamefont {Yang}}\ and\ \bibinfo {author} {\bibfnamefont {G.~R.}\ \bibnamefont {Fleming}},\ }\href@noop {} {\bibfield  {journal} {\bibinfo  {journal} {Chemical Physics}\ }\textbf {\bibinfo {volume} {275}},\ \bibinfo {pages} {355} (\bibinfo {year} {2002})}\BibitemShut {NoStop}%
\bibitem [{\citenamefont {Peterman}\ \emph {et~al.}(1998)\citenamefont {Peterman}, \citenamefont {van Amerongen}, \citenamefont {van Grondelle},\ and\ \citenamefont {Dekker}}]{Peterman1998TheNO}%
  \BibitemOpen
  \bibfield  {author} {\bibinfo {author} {\bibfnamefont {E.~J.}\ \bibnamefont {Peterman}}, \bibinfo {author} {\bibfnamefont {H.}~\bibnamefont {van Amerongen}}, \bibinfo {author} {\bibfnamefont {R.}~\bibnamefont {van Grondelle}}, \ and\ \bibinfo {author} {\bibfnamefont {J.~P.}\ \bibnamefont {Dekker}},\ }\href@noop {} {\bibfield  {journal} {\bibinfo  {journal} {Proceedings of the National Academy of Sciences of the United States of America}\ }\textbf {\bibinfo {volume} {95 11}},\ \bibinfo {pages} {6128} (\bibinfo {year} {1998})}\BibitemShut {NoStop}%
\bibitem [{\citenamefont {Stones}\ \emph {et~al.}(2017)\citenamefont {Stones}, \citenamefont {Hossein-Nejad}, \citenamefont {van Grondelle},\ and\ \citenamefont {Olaya-Castro}}]{C7SC02983G}%
  \BibitemOpen
  \bibfield  {author} {\bibinfo {author} {\bibfnamefont {R.}~\bibnamefont {Stones}}, \bibinfo {author} {\bibfnamefont {H.}~\bibnamefont {Hossein-Nejad}}, \bibinfo {author} {\bibfnamefont {R.}~\bibnamefont {van Grondelle}}, \ and\ \bibinfo {author} {\bibfnamefont {A.}~\bibnamefont {Olaya-Castro}},\ }\href@noop {} {\bibfield  {journal} {\bibinfo  {journal} {Chemical Science}\ }\textbf {\bibinfo {volume} {8}},\ \bibinfo {pages} {6871} (\bibinfo {year} {2017})}\BibitemShut {NoStop}%
\bibitem [{\citenamefont {Esposito}\ \emph {et~al.}(2009)\citenamefont {Esposito}, \citenamefont {Harbola},\ and\ \citenamefont {Mukamel}}]{esposito2009nonequilibrium}%
  \BibitemOpen
  \bibfield  {author} {\bibinfo {author} {\bibfnamefont {M.}~\bibnamefont {Esposito}}, \bibinfo {author} {\bibfnamefont {U.}~\bibnamefont {Harbola}}, \ and\ \bibinfo {author} {\bibfnamefont {S.}~\bibnamefont {Mukamel}},\ }\href@noop {} {\bibfield  {journal} {\bibinfo  {journal} {Reviews of modern physics}\ }\textbf {\bibinfo {volume} {81}},\ \bibinfo {pages} {1665} (\bibinfo {year} {2009})}\BibitemShut {NoStop}%
\bibitem [{\citenamefont {Ross}\ and\ \citenamefont {Calvin}(1967)}]{Ross}%
  \BibitemOpen
  \bibfield  {author} {\bibinfo {author} {\bibfnamefont {R.~T.}\ \bibnamefont {Ross}}\ and\ \bibinfo {author} {\bibfnamefont {M.}~\bibnamefont {Calvin}},\ }\href@noop {} {\bibfield  {journal} {\bibinfo  {journal} {Biophysical journal}\ }\textbf {\bibinfo {volume} {7}},\ \bibinfo {pages} {595} (\bibinfo {year} {1967})}\BibitemShut {NoStop}%
\bibitem [{\citenamefont {Shockley}\ and\ \citenamefont {Queisser}(1961)}]{Shockley}%
  \BibitemOpen
  \bibfield  {author} {\bibinfo {author} {\bibfnamefont {W.}~\bibnamefont {Shockley}}\ and\ \bibinfo {author} {\bibfnamefont {H.~J.}\ \bibnamefont {Queisser}},\ }\href@noop {} {\bibfield  {journal} {\bibinfo  {journal} {Journal of Applied Physics}\ }\textbf {\bibinfo {volume} {32}},\ \bibinfo {pages} {510} (\bibinfo {year} {1961})}\BibitemShut {NoStop}%
\bibitem [{\citenamefont {Huang}\ \emph {et~al.}(2023)\citenamefont {Huang}, \citenamefont {Xu},\ and\ \citenamefont {Markides}}]{huang2023high}%
  \BibitemOpen
  \bibfield  {author} {\bibinfo {author} {\bibfnamefont {G.}~\bibnamefont {Huang}}, \bibinfo {author} {\bibfnamefont {J.}~\bibnamefont {Xu}}, \ and\ \bibinfo {author} {\bibfnamefont {C.~N.}\ \bibnamefont {Markides}},\ }\href@noop {} {\bibfield  {journal} {\bibinfo  {journal} {Nature Communications}\ }\textbf {\bibinfo {volume} {14}},\ \bibinfo {pages} {3344} (\bibinfo {year} {2023})}\BibitemShut {NoStop}%
\bibitem [{\citenamefont {Takekuma}\ \emph {et~al.}(2020)\citenamefont {Takekuma}, \citenamefont {Ikeda}, \citenamefont {Kawakami}, \citenamefont {Kamiya}, \citenamefont {Nango},\ and\ \citenamefont {Nagata}}]{Photocurrent}%
  \BibitemOpen
  \bibfield  {author} {\bibinfo {author} {\bibfnamefont {Y.}~\bibnamefont {Takekuma}}, \bibinfo {author} {\bibfnamefont {N.}~\bibnamefont {Ikeda}}, \bibinfo {author} {\bibfnamefont {K.}~\bibnamefont {Kawakami}}, \bibinfo {author} {\bibfnamefont {N.}~\bibnamefont {Kamiya}}, \bibinfo {author} {\bibfnamefont {M.}~\bibnamefont {Nango}}, \ and\ \bibinfo {author} {\bibfnamefont {M.}~\bibnamefont {Nagata}},\ }\href@noop {} {\bibfield  {journal} {\bibinfo  {journal} {RSC Adv.}\ }\textbf {\bibinfo {volume} {10}},\ \bibinfo {pages} {15734} (\bibinfo {year} {2020})}\BibitemShut {NoStop}%
\bibitem [{\citenamefont {Ritschel}\ and\ \citenamefont {Eisfeld}(2014)}]{ritschel2014analytic}%
  \BibitemOpen
  \bibfield  {author} {\bibinfo {author} {\bibfnamefont {G.}~\bibnamefont {Ritschel}}\ and\ \bibinfo {author} {\bibfnamefont {A.}~\bibnamefont {Eisfeld}},\ }\href@noop {} {\bibfield  {journal} {\bibinfo  {journal} {The Journal of Chemical Physics}\ }\textbf {\bibinfo {volume} {141}} (\bibinfo {year} {2014})}\BibitemShut {NoStop}%
\bibitem [{\citenamefont {Runeson}\ \emph {et~al.}(2024)\citenamefont {Runeson}, \citenamefont {Fay},\ and\ \citenamefont {Manolopoulos}}]{runeson2024exciton}%
  \BibitemOpen
  \bibfield  {author} {\bibinfo {author} {\bibfnamefont {J.~E.}\ \bibnamefont {Runeson}}, \bibinfo {author} {\bibfnamefont {T.~P.}\ \bibnamefont {Fay}}, \ and\ \bibinfo {author} {\bibfnamefont {D.~E.}\ \bibnamefont {Manolopoulos}},\ }\href@noop {} {\bibfield  {journal} {\bibinfo  {journal} {Physical Chemistry Chemical Physics}\ }\textbf {\bibinfo {volume} {26}},\ \bibinfo {pages} {4929} (\bibinfo {year} {2024})}\BibitemShut {NoStop}%
\bibitem [{\citenamefont {Harbola}\ \emph {et~al.}(2006)\citenamefont {Harbola}, \citenamefont {Esposito},\ and\ \citenamefont {Mukamel}}]{harbola2006quantum}%
  \BibitemOpen
  \bibfield  {author} {\bibinfo {author} {\bibfnamefont {U.}~\bibnamefont {Harbola}}, \bibinfo {author} {\bibfnamefont {M.}~\bibnamefont {Esposito}}, \ and\ \bibinfo {author} {\bibfnamefont {S.}~\bibnamefont {Mukamel}},\ }\href@noop {} {\bibfield  {journal} {\bibinfo  {journal} {Physical Review B—Condensed Matter and Materials Physics}\ }\textbf {\bibinfo {volume} {74}},\ \bibinfo {pages} {235309} (\bibinfo {year} {2006})}\BibitemShut {NoStop}%
\bibitem [{\citenamefont {Goswami}\ and\ \citenamefont {Harbola}(2015)}]{goswami2015electron}%
  \BibitemOpen
  \bibfield  {author} {\bibinfo {author} {\bibfnamefont {H.~P.}\ \bibnamefont {Goswami}}\ and\ \bibinfo {author} {\bibfnamefont {U.}~\bibnamefont {Harbola}},\ }\href {https://doi.org/10.1063/1.4908230} {\bibfield  {journal} {\bibinfo  {journal} {The Journal of Chemical Physics}\ }\textbf {\bibinfo {volume} {142}} (\bibinfo {year} {2015})}\BibitemShut {NoStop}%
\bibitem [{\citenamefont {Golovach}\ and\ \citenamefont {Loss}(2004)}]{golovach2004transport}%
  \BibitemOpen
  \bibfield  {author} {\bibinfo {author} {\bibfnamefont {V.~N.}\ \bibnamefont {Golovach}}\ and\ \bibinfo {author} {\bibfnamefont {D.}~\bibnamefont {Loss}},\ }\href@noop {} {\bibfield  {journal} {\bibinfo  {journal} {Physical Review B—Condensed Matter and Materials Physics}\ }\textbf {\bibinfo {volume} {69}},\ \bibinfo {pages} {245327} (\bibinfo {year} {2004})}\BibitemShut {NoStop}%
\bibitem [{\citenamefont {Raszewski}\ \emph {et~al.}(2005)\citenamefont {Raszewski}, \citenamefont {Saenger},\ and\ \citenamefont {Renger}}]{Raszewski2005TheoryOO}%
  \BibitemOpen
  \bibfield  {author} {\bibinfo {author} {\bibfnamefont {G.}~\bibnamefont {Raszewski}}, \bibinfo {author} {\bibfnamefont {W.}~\bibnamefont {Saenger}}, \ and\ \bibinfo {author} {\bibfnamefont {T.}~\bibnamefont {Renger}},\ }\href@noop {} {\bibfield  {journal} {\bibinfo  {journal} {Biophysical Journal}\ }\textbf {\bibinfo {volume} {88 2}},\ \bibinfo {pages} {986} (\bibinfo {year} {2005})}\BibitemShut {NoStop}%
\end{thebibliography}%

\end{document}